\documentclass{aa}
\usepackage{txfonts}
\usepackage{graphics}
\usepackage{epsfig}
\newcommand{\JminK}{\mbox{$(J-K_{\mathrm s})$}}

\newcommand{\etal}{et al.~}
\newcommand{\eg}{e.g.~}

\begin{document}

   \title{AGB stars as tracers of metallicity and mean age across M33}

   \author{M.-R.L. Cioni\inst{1,2}
          \and M. Irwin\inst{3}
          \and A.M.N. Ferguson\inst{1} 
          \and A. McConnachie\inst{4}
          \and B.C. Conn\inst{5}
          \and A. Huxor\inst{1,6}
          \and R. Ibata\inst{7}
          \and G. Lewis\inst{8}
          \and N. Tanvir\inst{9}
          }

   \offprints{M.Cioni@herts.ac.uk}

   \institute{
     SUPA, School of Physics, University of Edinburgh,
     IfA, Blackford Hill, Edinburgh EH9 3HJ, UK
     \and
     Centre for Astrophysics Research, University of
     Hertfordshire, Hatfield AL10 9AB, UK
     \and
     Institute of Astronomy, University of Cambridge, 
     Madingley Road, Cambridge CB3 0HA, UK
     \and
     Deptartment of Physics \& Astronomy, University of Victoria, PO Box
     3055, STN CSC, Victoria, BC, V8W 3P6 Canada
     \and
     European Southern Observatory, Alonso de Cordova 3107, Vitacura,
     Santiago, Chile 
     \and
     Department of Physics, University of Bristol, Tyndall Avenue,
     Bristol BS8 1TL, UK 
     \and
     Observatoire de Strasbourg, 11 rue de l'Université, F-67000
     Strasbourg, France 
     \and
     Institute of Astronomy, School of Physics, A29, University of
     Sydney, NSW 2006, Australia 
     \and
     Department of Physics and Astronomy, University of Leicester,
     Leicester LE1 7RH, UK 
     }

   \date{Received 8 January 2008 / Accepted 5 May 2008}

   \titlerunning{Metallicity and age across M33}

   \authorrunning{Cioni et al.}

   \abstract{Wide-field $JHK_{\mathrm s}$ near-infrared observations
     covering an area of $1.8^{\circ}\times 1.8^{\circ}$ centred on
     M33 were obtained using WFCAM at UKIRT. These data show a large
     population of intermediate-age asymptotic giant branch stars
     (AGB).}{We have used both C-type and M-type AGB stars to
     determine spatial variations in metallicity and mean age across
     the galaxy.}{We distinguished between C-type and M-type AGB stars
     from their location in the colour-magnitude diagram
     ($J-K_{\mathrm s}$, $K_{\mathrm s}$). The distribution of these
     stars is supported by a cross-identification between our sample
     and a catalogue of optically confirmed, long-period variable
     stars, as well as with the list of sources detected by Spitzer in
     the mid-infrared. We calculated the C/M ratio and the orientation
     of the galaxy in the sky, and compared the $K_{\mathrm s}$
     magnitude distribution with theoretical distributions spanning a
     range of metallicities and star formation rates (SFRs).}  {The
     C/M ratio surface map confirms a metallicity gradient in the
     galaxy corresponding to a spread in [Fe/H] $=0.6$ dex with
     substructures in the inner and outer galaxy. Magnitude and colour
     variations suggest orientation and extinction effects on the
     galaxy disc. Maps showing the distribution of mean age and
     metallicity obtained from the $K_s$ method suggest that: the
     outer galaxy disc/halo is metal poorer than the nuclear region
     and metal-rich clumps in the inner galaxy change location with
     time.  The average outer ring and nuclear stellar population is
     $\sim 6$ Gyr old while central regions are a few Gyr younger.}{}

\maketitle

\section{Introduction}
\label{intro}
M33 (NGC 598 or the Triangulum galaxy) is located at about
$15^{\circ}$ from M31 (Andromeda) and is the third brightest member of
the Local Group. It is a spiral galaxy of type Sc II-III intermediate
between large spirals and dwarf irregulars in the Local Group. A
detailed review of studies of this galaxy until the year 1999,
included, is given by van den Berg (\cite{vdb}). These studies
concentrated on: the determination of the distance to the galaxy, the
characterisation of its structural components (nucleus, disc and
halo), the detection of variable stars although their numbers were far
from complete, the determination of the age and metallicity of
globular clusters which, contrary to the Large Magellanic Clouds
clusters, formed regularly throughout time, and the spatial
distribution and chemical analysis of HII regions, associations,
supernova and supernova remnants. Abundance gradients were clearly
detected.  The only $JK_{\mathrm s}$ near-infrared study at the time
of the resolved stellar population was that of McLean \& Liu
(\cite{mcli}) which failed to reveal the presence of a bulge but
showed a clearly numerous intermediate-age component in the disc as
well as in the centre of the galaxy.  Their observations covered a
field of $7.6^{\prime}\times7.6^{\prime}$ centred on M33 and barely
reached $K_{\mathrm s}=18$; sources fainter than $K_{\mathrm s}=17$
were affected by severe photometric errors and incompleteness.

A few years later Stephens \& Frogel (\cite{stfr}) presented
near-infrared diagrams as deep as $K_{\mathrm s}=22$ which outlined
young, intermediate-age as well as old stellar populations in the very
centre ($22^{\prime\prime}\times22^{\prime\prime}$) of the galaxy.
The outer region was explored by Davidge (\cite{da03}) who confirmed
that an intermediate-age population occurs well outside the young star
forming disc. Wide-area optical observations were analysed by
McConnachie et al. (\cite{co04}) to refine the distance modulus of the
galaxy using the tip of the red giant branch (RGB) method (see also
Kim et al. \cite{km02}). Similar observations extending over the whole
body of M33 were analysed by Li et al. (\cite{li04}). The authors
computed theoretical spectral energy distributions (SEDs) for three
different star formation rates (SFRs).  A comparison with their
observations suggests that for a constant SFR there is an age gradient
between stellar populations of the central regions ($10$ Gyr old) and
of the outer regions ($7$ Gyr old) with the youngest component ($5$
Gyr old) in the spiral arms. An exponentially decreasing SFR equally
reproduces the observations but ages are about $2$ Gyr lower. The SE
region of the galaxy, explored by Barker et al.~(\cite{ba07b}) using
deep images from the Hubble Space Telescope, shows that the mean age
increases from $6$ (inner) to $8$ (outer) Gyr assuming that star
formation began $14$ Gyr ago. This result was obtained from
observations of three fields located within $20^{\prime}-30^{\prime}$
from the nucleus. On the contrary, Li et al. (\cite{li04}), by
observing the entire galaxy, found a decrease in the age of the
stellar population from the central to the outer regions. Spiral arms,
however, appear younger than the outer galaxy. Block et
al. (\cite{bl04}) obtained deep 2MASS near-infrared photometry across
the whole galaxy suggesting that carbon stars delineate arcs which are
the signature of accretion of low metallicity gas in the outer
disc. Recent spectra confirm the existence of a few C stars at this
location (Block et al. \cite{bl07}). However, these arcs were not
found by Rowe et al. (\cite{ro05}) who obtained wide-area observations
in $VI$ broad-band filters but also in narrow-band filters,
distinguishing between C- and M-type asymptotic giant branch (AGB)
stars and investigated the metallicity gradient via the C/M
ratio. These authors suggest that the arcs are simply an extension of
the M33 disc.

The present work aims to recover the spatially resolved metallicity
and mean age of the stellar population of M33 by interpreting the
observed $K_{\mathrm s}$ band distribution of AGB stars from new
wide-field near-infrared data. Section \ref{obs} describes the
observations and the data reduction while Section \ref{ana}
concentrates on the analysis of the data: the comparison with
theoretical distributions and the production of final maps showing the
distribution of mean age and metallicity across the galaxy. Section
\ref{dis} compares the results with the information available from the
literature while Sect.~\ref{con} concludes this paper.

\section{Observations and Data Reduction}
\label{obs}

The observations of M33 were obtained as part of a program to survey
the luminous red stellar populations of the Local Group galaxies using
the Wide-Field CAMera (WFCAM) on the $3.8$m United Kingdom InfraRed
Telescope (UKIRT) in Hawaii. WFCAM utilises $4$ Rockwell Hawaii-II
(HgCdTe $2048\times2048$) arrays, such that $4$ separately pointed
observations can be tiled together to cover a filled square of sky (a
tile) of $0.75$ deg$^2$ with $0.4^{\prime\prime}$ pixels. A mosaic of
$4$ tiles was obtained in three broad-band filters ($J$, $H$ and
$K_{\mathrm s}$; Tokunaga et al. \cite{toku}) covering a $3$ deg$^2$
of sky centred on M33. Observations in the $J$ band were obtained
using a $5$ point jitter mode, using $10$s$\times3$ coaveraged
exposures at each position, without
microstepping\footnote{Microstepping is used to attempt to recover
some of the lost resolution due to under sampling good seeing
conditions with the $0.4^{\prime\prime}$ pixels.  The microstepped
frames are interwoven at the pixel level to give, in this case, an
effective sampling of $0.133^{\prime\prime}$ per interleaved pixel.},
to give a total exposure time of 150s per pointing. For the $H$ and
$K_{\mathrm s}$ band images a $3$ point jitter with $10$s per position
and a $3\times3$ microstepping was used, effectively giving $270$s
total exposure per pointing.

Target frames of M33 were acquired throughout the semester on
September $29^{\mathrm{th}}$ and $30^{\mathrm{th}}$, October
$24^{\mathrm{th}}$, November $5^{\mathrm{th}}$ and December
$16^{\mathrm{th}}$ 2005, and were processed, together with the rest of
the WFCAM data for those nights, using the WFCAM pipeline provided by
the Cambridge Astronomy Survey Unit (http://casu.ast.cam.ac.uk/).  The
pipeline performs all the standard near-infrared reduction steps for
instrumental signature removal: dark-correction, flatfielding,
crosstalk removal, sky-correction, and systematic noise removal, all
at the frame level.  In addition frames forming part of a dither or
interleave sequence are co-aligned and stacked/and or interleaved
prior to performing source extraction.  After source extraction the
pipeline then produces object morphological classifications from which
assorted quality control measures are computed, and also does full
astrometric and photometric calibration based on the 2MASS point
source catalogue (see Irwin \etal 2004; and the CASU web pages for
more details).  The quality control measures are used to monitor
parameters such as sky brightness, seeing, limiting depth etc., and
are used to select the best set of frames for further post-pipeline
analysis.

In the case of M33, since the average seeing on the frames varied
between $\approx 0.7-1.1$ arcsec, the $H$ and $K_s$ stacked
interleaved images were subsequently resampled using $2\times2$
binning to form images with a resolution of
$0.267^{\prime\prime}$/pix, which is better matched to the seeing and
helps to minimise the effects of seeing variations in the $3\times3$
interleaved stacked data.  Source extraction and calibration were then
re-run on the resampled images to give the final object catalogues per
pass band, again calibrated with respect to 2MASS (Skrutski \etal
2006).

The $J$, $H$, and $K_{\mathrm s}$ object catalogues were then matched
across the different passbands on a per pointing basis. To be
considered a good match, objects had to align to better than
$1^{\prime\prime}$.  This is significantly worse than the average
positional match error between the same object in the different
passbands which is better than 100 mas.  The band-merged products for
all pointings were then simply ``glued'' together to make the first
pass large area complete catalogue.  Duplicate entries in this
catalogue, due to the overlap between pointings within a tile and
between tiles, were then removed using the photometic flux error
combined with the morphological classification to choose a unique
``best'' entry per object.  This unique catalogue forms the basis for
all subsequent analysis.  We note that the accuracy of the 2MASS
photometric calibration of $1-2$\% (\eg Hodgkin \etal 2007) greatly
simplifies the process.

The observing conditions corresponded to an average seeing of
$1.07\pm0.06^{\prime\prime}$ which, combined with the photon-noise limited
science products, allowed us to reach magnitudes of $J=19.30\pm0.18$, 
$H=18.70\pm0.12$ and $K=18.32\pm0.10$ with a signal:to:noise better than 
10:1 in all wavebands.  We discuss completeness and crowding issues
separately in section 3.3. 

\begin{figure}
\resizebox{\hsize}{!}{\includegraphics{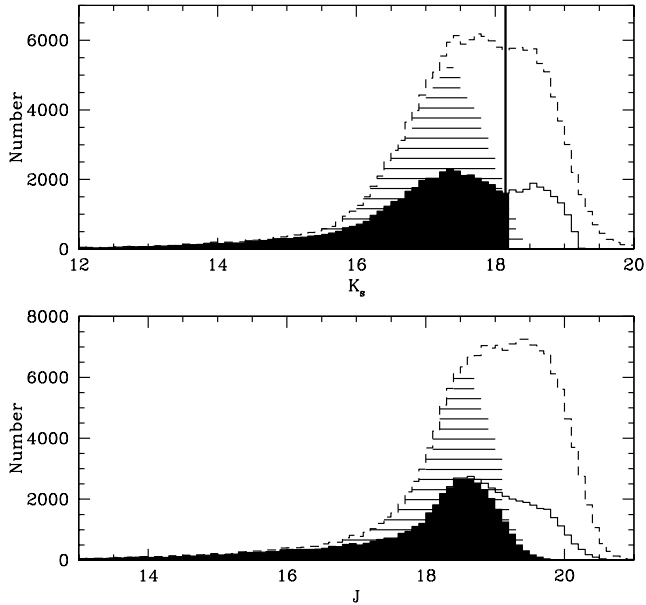}}
\caption{Distribution of all sources listed in the catalogue as a
  function of magnitude ({\it dashed-line histogram}), those with a
  photometric accuracy better than $10$\% ({\it dashed-filled
  histogram}) and of selected stellar as well as probably stellar
  objects ({\it continuous histogram}). $K_{\mathrm s}$ magnitudes are
  shown on the top panel and $J$ magnitudes on the bottom one. Bins
  correspond to $0.1$ mag and magnitudes are apparent. The vertical
  line at $K_s=18.15$ indicates the position of the tip of the RGB
  while the {\it continuous-filled} histograms show sources with
  $K_s<18.15$, most of them are AGB stars, and their corresponding $J$
  band distribution; this is the sample used in the paper.}
\label{compl}
\end{figure}

\section{Analysis and Results}
\label{ana}

The first step in the data analysis was to select from the catalogue
only stellar and probably stellar objects, from a pipeline stellarity
index, in both $J$ and $K_s$ bands. No selection was performed on the
$H$ band but it is unlikely that an object will look like a star in
two but not in three bands. A total of $60367$ objects satisfy this
selection criteriom (are within $1-2\sigma$ of the stellar
locus). Their histogram is shown in Figure \ref{compl} where the dip
at $K_s=18.15$ corresponds to the discontinuity marking the transition
between RGB and AGB stars. RGB stars reach their brightest magnitude
at the tip of the RGB when He burning begins in their core.  It is
interesting to note that this dip is also present in the distribution
of all sources in the catalogue for both the $K_s$ and $J$ band. The
latter occurs at $J=19.15$, but it is not clearly visible for
stellar-like objects. The difference between the position of the dip
in the two wave bands corresponds to the difference between the tip of
the RGB position observed, in the same wave-bands, for the Magellanic
Clouds (Cioni et al. \cite{tip}). In general, using the calibrations
by Bellazzini et al. (\cite{bel04}) this difference depends on the
total metallicity of the galaxy and for [M/H]$\sim-0.75$ (McConnachie
et al. \cite{co04}) it is $\sim 1.1$ mag, in agreement with what is
observed.

 The sample analyzed in this paper, for which a detailed selection
is presented in the next section, contains only sources brighter than
the tip of the RGB in the $K_{\mathrm s}$ band. Figure \ref{compl}
shows that all these objects, with $K_{\mathrm s}<18.15$, have a
photometric accuracy better than $10$\% in both the $J$ and
$K_{\mathrm s}$ bands (filled histograms). These data were then
dereddened for galactic foreground extinction assuming $E(B-V)=0.07$
(van den Bergh \cite{vdb}) and adopting the reddening law from Glass
et al. (\cite{glass}) which gives absorptions of: $J=0.06$, $H=0.04$
and $K_{\mathrm s}=0.02$ mag.

\subsection{Selection of AGB stars}
\label{sel}
\begin{figure}
\resizebox{\hsize}{!}{\includegraphics{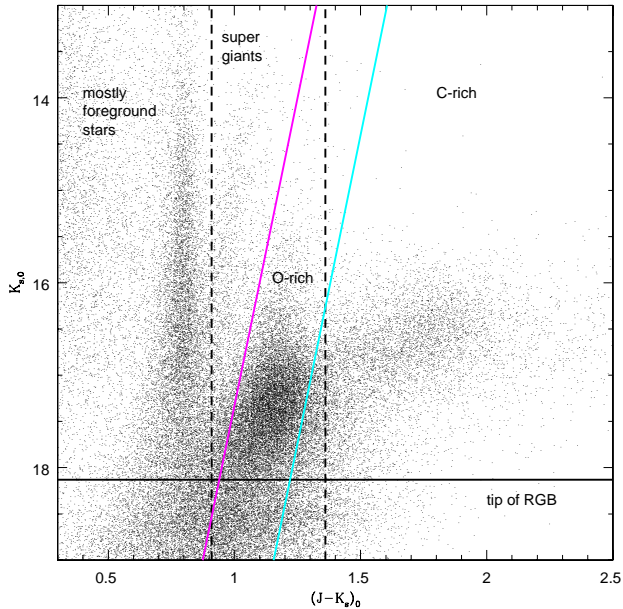}}
\caption{Colour-magnitude diagram of the stellar sources detected in
$J$ and $K_{\mathrm s}$ across M33.  Dashed {\it vertical-lines} and
continuous {\it slanted-lines} define regions occupied by O-rich and
C-rich candidate AGB stars, respectively. These appear statistically
distinct from foreground stars and super giant stars of M33. The
position of the tip of the RGB is also indicated.}
\label{cmd}
\end{figure}

Intermediate-age AGB stars are brighter than the tip of the RGB. In
the near-infrared colour-magnitude diagram, \JminK$_0$, $K_{{\mathrm
s,}0}$, they occupy two statistically distinct branches depending on
their spectral type. C-rich, or C-type, AGB stars span a wide range of
\JminK$_0$ colours at an approximately constant $K_{\mathrm s}$
magnitude because of the absorption effect of molecules such as CN and
C$_2$. O-rich, or M-type, AGB stars span a narrow range of \JminK$_0$
colours, but a large range of $K_{\mathrm s}$ magnitudes. They define
a bright extension of the RGB branch. The atmospheres of O-rich AGB
stars is also dominated by molecular absorption but in this case these
are mostly of H$_2$O, TiO and VO.  Foreground stars, mostly dwarfs,
observed towards M33 are distributed in a vertical sequence at
$(J-K_{\mathrm s})_0\sim0.91$ while super giant stars, belonging to
M33, define a slanted branch between foreground stars and O-rich AGB
stars. Figure \ref{cmd} shows the location of the different types of
stars; this diagram probably includes also upper main-sequence stars
as well as Cepheid variable stars which have \JminK$_0$ colours bluer
than AGB stars and cannot be disentangled from foreground stars using
single epoch near-infrared photometry.

Two different statistical criteria can be adopted to disentangle
between C-rich and O-rich AGB stars brighter than the tip of the RGB.
The first criterion (using {\it vertical-lines}) assumes that C-rich
AGB stars are redder than \JminK$_0=1.36$ which is a visually
estimated colour, chosen specifically for this galaxy, marking
the departure of the reddest stellar branch from the vertical one
(Fig.~\ref{cmd}). O-rich stars, including AGB stars and super giant
stars, are bluer than this colour but redder than \JminK$_0=0.91$
which disentangles them from foreground stars.  This criterium,
  although being subjective, relies on spectroscopic identifications
  of C-rich and O-rich AGB stars (Cioni et al. \cite{ci01}). It has
  the advantage of isolating the bulk of the C star population
  producing a reliable but not complete sample of C-rich AGB stars. In
  fact, faint C stars together with super giant stars contaminate the
  region populated by O-rich AGB stars producing a sample, of O-rich
  AGB stars, which is complete but less reliable at bright and faint
  magnitudes. The second criterion adopts {\it slanted-lines},
instead of vertical lines, that better represent the inclination of
the O-rich AGB branch as well as that of the RGB branch. The equations
of the lines are: $K_{\mathrm s} = -13.333(J-K_{\mathrm s})_0 +
30.666$ and $K_{\mathrm s} = -13.333(J-K_{\mathrm s})_0 + 26.933$.
O-rich AGB stars are included within these lines while C-rich AGB
stars are redder (Fig.~\ref{cmd}).  The width of the region occupied
by AGB stars ($0.28$ mag) excludes super giant stars and accounts for
a possible high metallicity spread within the galaxy. This
  criterium was established as a result of simulating the distribution
  of AGB stars in the near-infrared colour-magnitude diagram (Cioni et
  al.  \cite{lf}). It has the advantage of isolating the bulk of the
  O-rich AGB population by removing the contamination by super giant
  and faint C stars. The isolated sample of C stars approaches
  completeness although contamination around the dividing lines is not
  excluded at the expenses of reliability.  Both criteria will be
used throughout the paper as a comparison and a validation of the
procedures as well as to assess uncertainties in the resulting
distribution of mean age and metallicity across the galaxy.

\begin{figure}
\resizebox{\hsize}{!}{\includegraphics{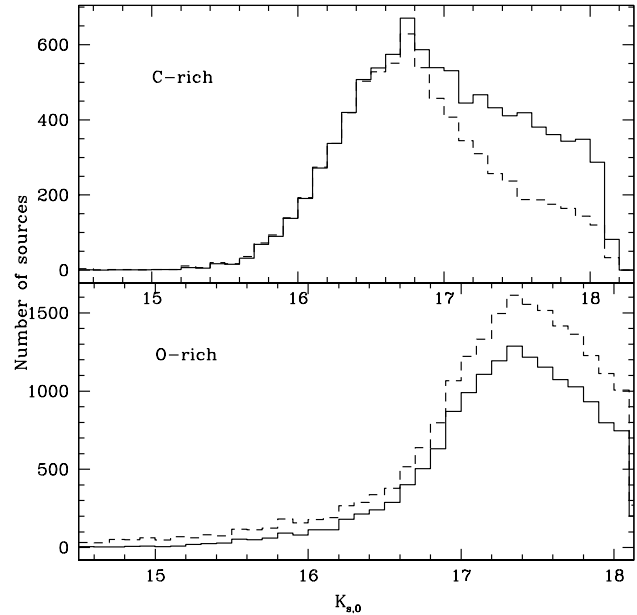}}
\caption{Distribution of the number of C-rich and O-rich AGB stars
versus $K{_{\mathrm s,}}_0$ magnitude. Both the result of selecting
these stars using vertical lines ({\it dashed-line histograms}) and
using slanted lines ({\it continuous-line histograms}) are shown; bins
are of $0.1$ mag.}
\label{hist}
\end{figure}

The histograms of AGB stars selected from the near-infrared
colour-magnitude diagram are shown in Fig.~\ref{hist}. C-rich AGB
stars brighter than $K_{{\mathrm s,}0}=16.6$ are equally recovered by
both selection criteria, while using slanted lines more fainter C-rich
AGB star candidates are obtained than by using vertical lines. The C/M
ratio derived from this selection is $0.60$. In low-metallicity
environments C-rich AGB stars are more numerous and extend to faint
magnitudes. Therefore, it is possible that in M33 there are more faint
C-rich AGB stars than those occupying only the red branch. The
vertical-lines selection criterion selects more O-rich AGB stars,
however, their distribution as function of $K_{\mathrm s}$ magnitude
is very similar to the one obtained using the slanted-lines
criterion. The latter includes super giant stars and gives a C/M ratio
of $0.35$.

\subsection{Surface distributions}

The selection of a pure sample of C and M stars is a delicate
task. In the previous section two criteria, commonly used in the
literature, are presented and a comparison of the respective samples is
briefly discussed but see also Sect.~\ref{areadivision}.

Battinelli, Demers \& Mannucci (\cite{bade07}) have recently assessed
the selection of C and M stars at infrared wavelengths. Their main
conclusion is that the constant ($J-K_{\mathrm s}$) colour
discriminating between the two types of AGB stars depends on the
observed stellar population. This results confirms previous evidences
based on the study of the Magellanic Clouds and NGC 6822 galaxies
(Cioni \& Habing \cite{cm}, Cioni et al. \cite{ngc0}). In these cases
the dividing colour is a visual estimate of the departing of the C
stars branch to red colours from the almost vertical branch of M
stars. The different colour used for each galaxy is attributed to the
different metal content among them, although a direct calibration is
not yet available. However, a subsequent study of the Magellanic
Clouds has shown very good agreement between the metallicity estimated
from the C/M ratio and from the $K_{\mathrm s}$ band magnitude
distribution. There, the selection of C and M stars was made using
{\it slanted} instead of {\it vertical} lines that combine the visual
departure of the C stars branch to red colours and the location of C
and M stars from stellar evolution models. The agreement between the
metallicity distributions across each galaxy supports the evidence
that the selection of C and M stars depends on environment.

\begin{figure*}
\hspace{-1.15cm}
\epsfxsize=0.32\hsize \epsfbox{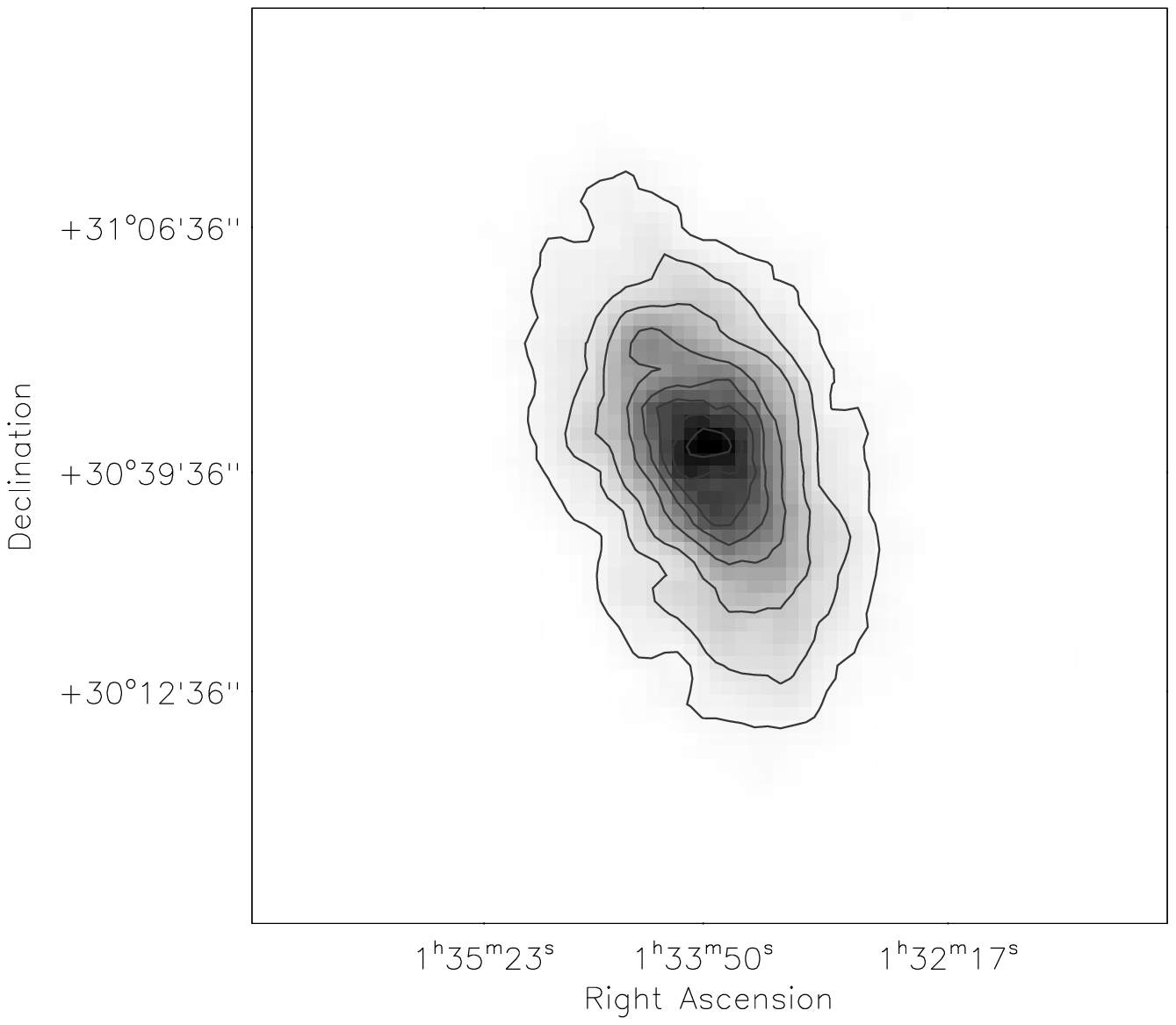}
\hspace{-1.45cm}
\epsfxsize=0.32\hsize \epsfbox{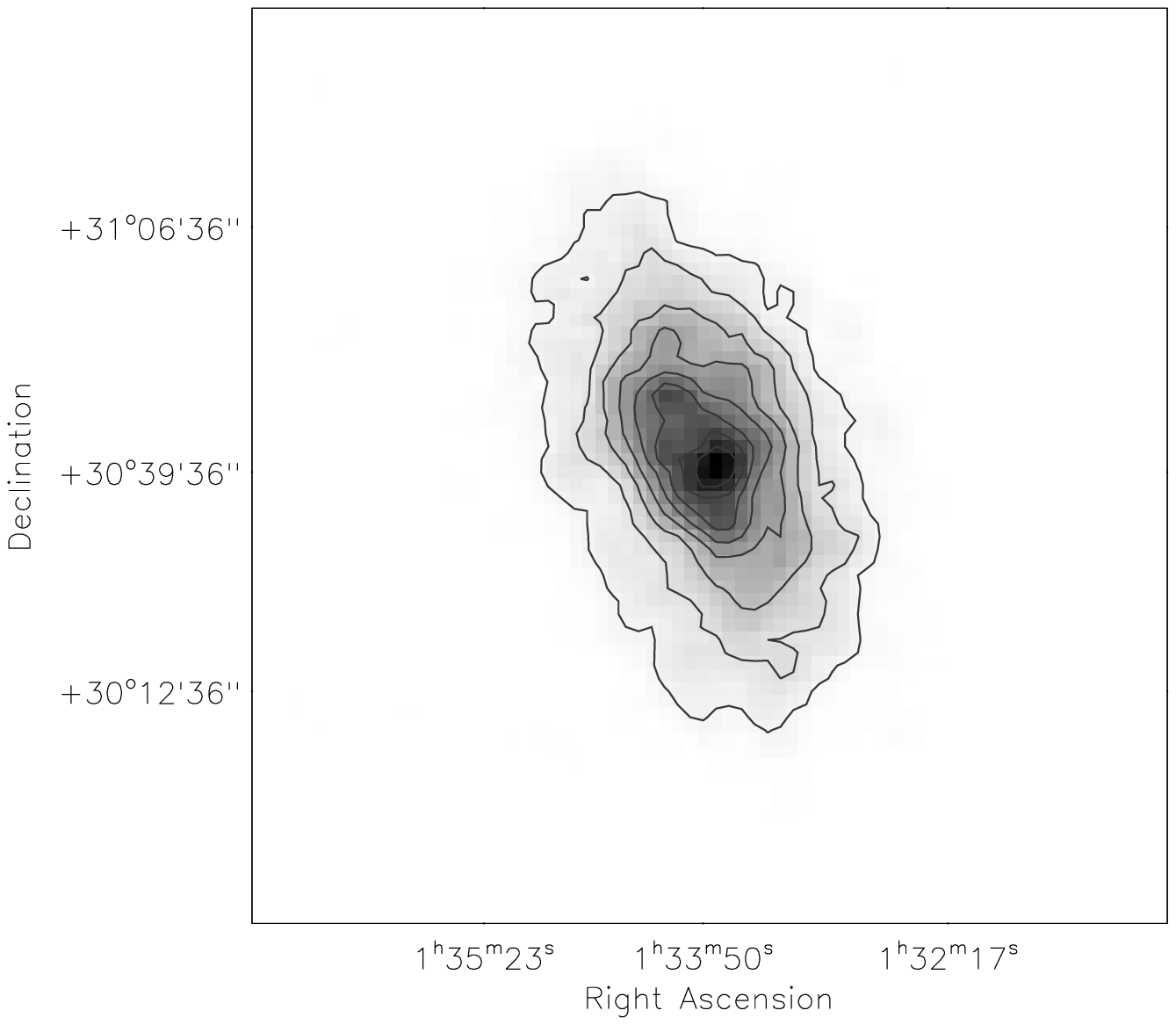}
\hspace{-1.45cm}
\epsfxsize=0.32\hsize \epsfbox{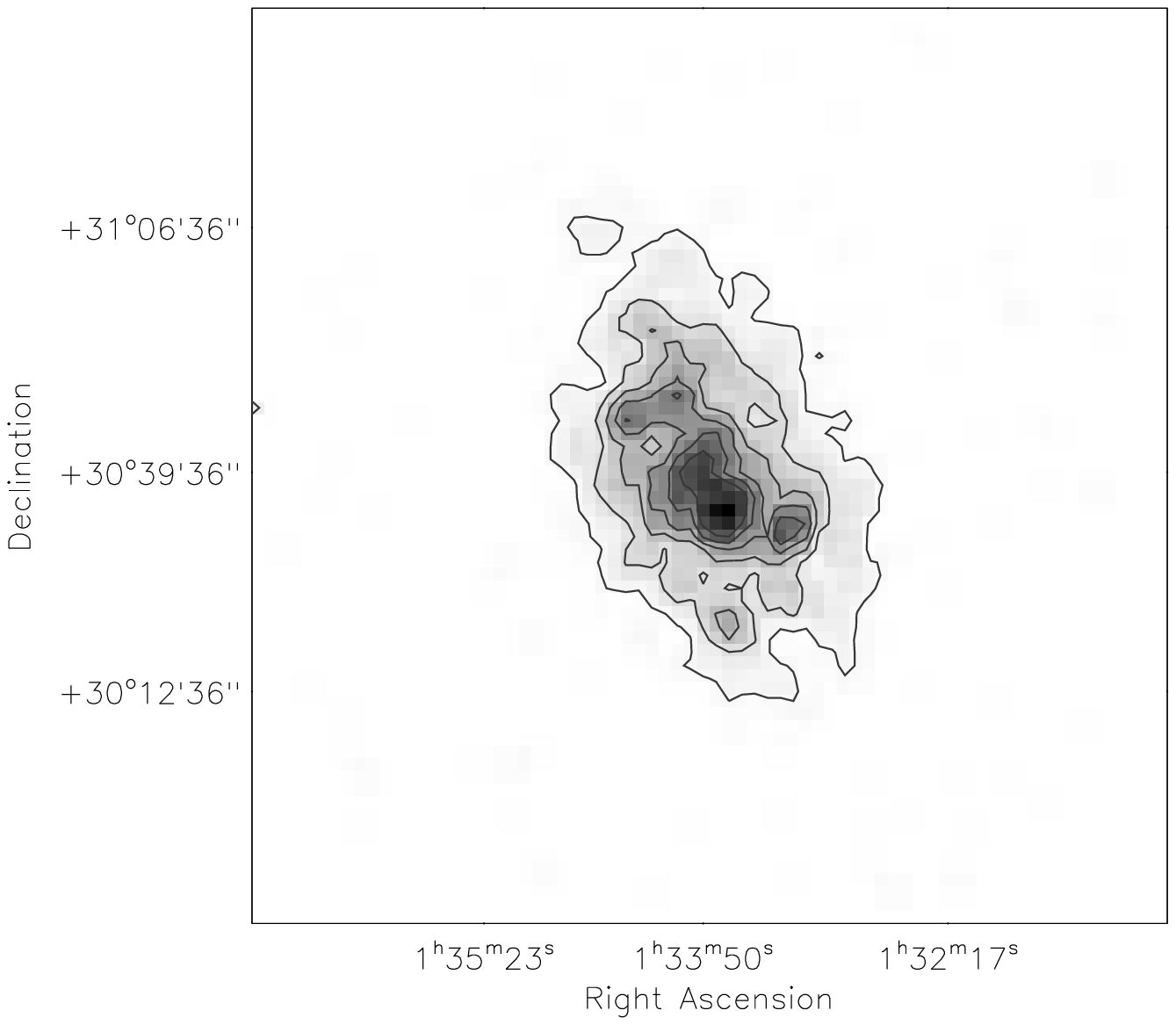}
\hspace{-1.45cm}
\epsfxsize=0.32\hsize \epsfbox{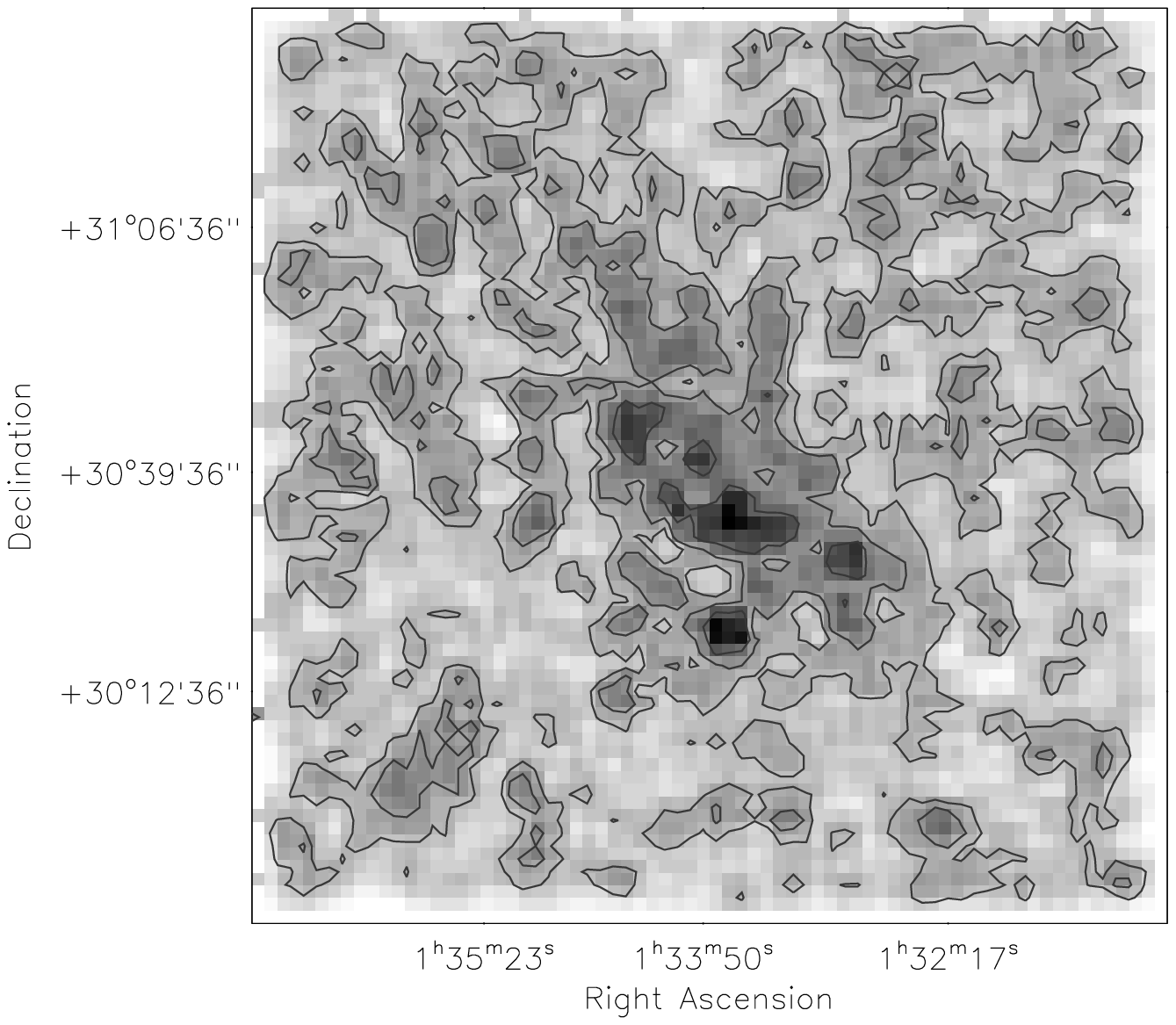}
\caption{{\it (from left to right)} Distribution of the number density
  of C stars, M stars, supergiant stars and mostly foreground
  stars. See the text for the selection of each sample. Contours are
  at: $3$, $9$, $18$, $33$, $45$, $55$, $65$ and $75$ for C stars,
  $2$, $5$, $10$, $15$, $20$, $25$, $30$ and $35$ for M stars, $0.5$,
  $2$, $4$, $6$, $8$ and $10$ for super giant stars, and $1.4$, $2$
  and $3$ for foreground stars. Darker regions correspond to higher
  numbers. North is to the top and East is to the left.}
\label{surf}
\end{figure*}

What is the distribution of the bulk of C, M, super giant and
foreground stars that may affect the selection criteria discussed so
far? The distribution of these stars isolated from the
colour-magnitude diagram is shown in Fig.~\ref{surf}. C stars have
been selected using the {\it vertical lines} criterion; M stars using
the {\it slanted line} criterion; super giant stars have $J-K_{\mathrm
s}>0.91$, $K_{\mathrm s}<-13.333*(J-K_{\mathrm s})+30.666$ and
$K_{\mathrm s}<17$ and foreground stars have $0.61<J-K_{\mathrm
s}<0.91$ and $K_{\mathrm s}<17$. The sample of both super giant and
foreground stars have been limited to stars much brighter than the tip
of the RGB to avoid as much as possible the contamination with other
stars. Maps in Fig.~\ref{surf} have been obtained applying a box car
smoothing to the number density calculated using bins of
$1.2^{\prime}$.

C and M stars are both broadly and  asymmetrically distributed
across the galaxy and show hints of the galaxy spiral arms. Towards
the centre, C stars increase in number following a broad circular
structure while M stars show a well defined concentration at the
location of the galaxy nucleus surrounded by a smoother elongated
lower density structure.  Super giant stars describe a clumpy
distribution which is predominantly confined in the central region of
the galaxy. Their number is considerably lower than that of C or M
stars at these distance from the centre. Therefore, the C/M ratio and
the analysis of the $K_{\mathrm s}$ magnitude distribution that
follows will not be affected by the presence of super giant
stars. Foreground stars are, as expected, more or less homogeneously
distributed within the entire area surveyed in the direction of the
M33 galaxy. It is possible that the darkest concentrations do also
contain some genuine M33 stars that cannot be disentangled using the
near-infrared colour-magnitude diagram alone. However, these stars are
too blue to be AGB stars and their exclusion is appropriate for the
analysis presented in this paper.  A more detailed investigation
of the galaxy structure, combining optical and near-infrared
photometry, will be presented elsewhere.

\subsection{Subdivision of the M33 area}
\label{areadivision}

\begin{figure}
\resizebox{\hsize}{!}{\includegraphics{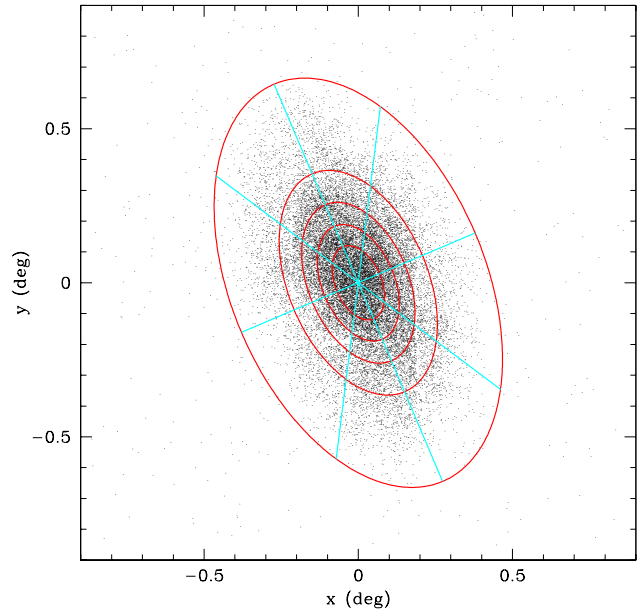}}
\caption{Distribution of C-rich AGB stars across M33.  Coordinates are
in degrees and are centred at $(\alpha, \delta)=(23.46^{\circ},
30.66^{\circ})$ according to van den Bergh (\cite{vdb}).  Sectors of
elliptical rings defining the regions used in this study are clearly
indicated. Each ring contains an approximately equal number of stars.
North is to the top and East to the left.}
\label{fig1}
\end{figure}

To study the spatially resolved SFH, the area occupied by M33 stars
was divided into five elliptical rings and each ring has been divided
into eight sectors. Figure \ref{fig1} shows these regions after
converting to polar coordinates using as a reference centre $(\alpha,
\delta)=(23.46^{\circ}, 30.66^{\circ})$ (van den Bergh \cite{vdb}).
The position angle of the major axis of each ellipse is $23^{\circ}$
(van den Berg \cite{vdb}). Each ellipse is such that the semi-minor
axis $b=a/3.4$ where $a$ is the semi-major axis and corresponds to
$0.127^{\circ}$, $0.199^{\circ}$, $0.275^{\circ}$, $0.384^{\circ}$ and
$0.7^{\circ}$ from the inner to outer ellipse respectively. This
relation has been obtained from the galaxy parameters listed by NED:
$2a=70.8^{\prime}$ and $2b=41.7^{\prime}$.  The size of the semi-major
axis of each ellipse has been determined such as each elliptical ring
contains approximately the same number of C-rich AGB stars ($\approx
1470$ and $\approx 1900$ using the {\it vertical-} and the {\it
slanted-lines} selection criteria, respectively). Each ring comprises
$\sim 3500$ O-rich candidate AGB stars depending on the selection
criteria.  These numbers give a statistical significant sample of AGB
stars per sector to use for the determination of the mean age and
metallicity across the galaxy.

The distribution of C-rich and O-rich AGB stars as a function of
$K_{{\mathrm s,}0}$ magnitude within each ring are shown in
Fig.~\ref{cring} and Fig.~\ref{oring}. Sources selected using both
criteria are compared. The shape of the histograms is likely to affect
the determination of the SFH parameters more than the overall number
of sources, because theoretically produced $K_{\mathrm s}$ magnitude
distributions are scaled to the absolute number of C-rich AGB stars
observed. A separation of C-rich AGB stars at \JminK$_0=1.36$ clearly
selects less faint stars, but also less stars of mid-range brightness,
see in particular the distribution in the outermost ring
(Fig.~\ref{cring}). This criterion includes super giant stars in the
distribution of O-rich AGB stars as well as faint candidate AGBs which
are excluded by the other criterion. However, the shape of the overall
distribution of O-rich AGB stars in the different rings is rather
similar (Fig.~\ref{oring}).

\subsubsection{Completeness and confusion}
By selecting stars with an uncertainty in magnitude better than $10$\%
within each ellipse it is possible to evaluate the effect of crowding
on the faintest detectable source. In the $K_s$ band the mean
magnitude of stars with photometric errors of $0.10\pm0.01$ is $18.15$
in the central ellipse, which corresponds to the discontinuity caused
by the tip of the RGB, and increases with a step of $\sim0.03$ mag
outwards for each ellipse attaining a value of $18.30$ in the
outermost ellipse. In the $J$ band it is $18.89$ in the central
ellipse and $19.17$ in the outermost ellipse with a step of $\sim0.07$
mag. All these values, and especially those in the three innermost
ellipses, are well above the expected position of the tip of the RGB
in the $J$ band. On the other hand, it is common practice to accept
sources with a S/N$=5$ as good detections. This corresponds to
increasing the tolerance to photometric errors as large as $\sim0.2$
mag. Repeating the same calculation in each ellipse we obtain:
$K_s=18.98$ (inner ellipse) and $K_s=19.75$ (outer ellipse) as well as
$J=19.12$ and $J=20.05$, respectively. These values indicate that any
AGB star brighter then the tip of the RGB was sufficiently well
detected to be included in the following analysis.

However, it remains to be checked how these values compare with the
level of confusion. This can be done by extrapolating the cumulative
distribution of the extracted sources and calculating the source
density within each ellipse. A source density of $1$ per $30$ beams
represents the confusion limit but for very dense fields $1/50$ is a
better rule (Hogg \cite{hogg}). Considering that the average seeing of
$1.07^{\prime\prime}$ (Sect.~$2$) is much larger than the instrumental
point spread function ($0.4^{\prime\prime}$), it is the seeing that
defines the beam and the confusion limit corresponds to
$50\times\pi(1.07^{\prime\prime}/2.35)^2=32$ arcsec$^2$.
assuming a gaussian point spread function. This occurs at $K_s=18.55$
and $J=18.48$ in the central ellipse suggesting that AGB stars might
be confusion limited in the $J$ band. For any other ellipse the
magnitudes of AGB stars are well above the confusion limit of one
source per $32$ arcsec$^2$.

\begin{figure}
\resizebox{\hsize}{!}{\includegraphics{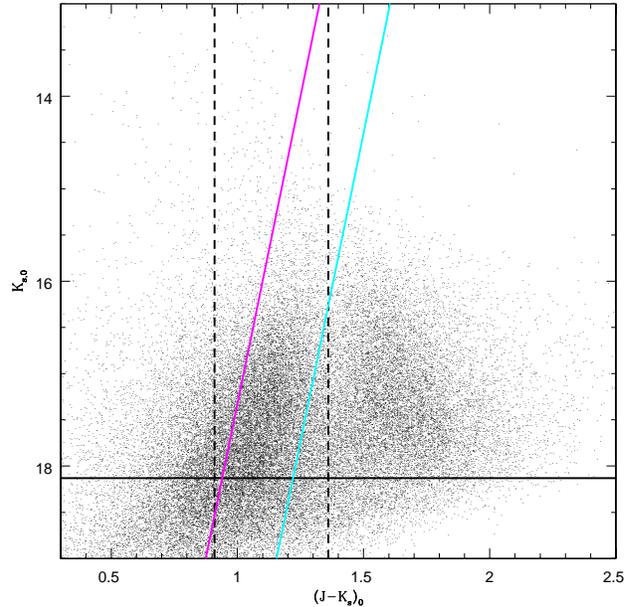}}
\caption{Colour-magnitude diagram of the non-stellar sources detected
  in $J$ and $K_s$ across M33. Many of these sources are blends but
  there are also HII regions, associations and background
  galaxies. Lines are like in Fig.~\ref{cmd}.}
\label{nonstars}
\end{figure}

The number density has been estimated utilising all extracted
sources. In particular, most of these are stellar-like or non-stellar
objects in roughly equal numbers. The spatial distribution of the
latter is mostly confined within the inner ellipses of the galaxy
suggesting that many of these sources are the result of merging and
crowding instead of genuine HII regions or associations. Others,
distributed throughout the field observed, are likely background
galaxies. The distribution of non-stellar objects in the ($J-K_s$,
$K_s$) colour-magnitude diagram (Fig.~\ref{nonstars}) outlines two
broad branches: the most populous overlapping the distribution of
O-rich giants and the other, almost symmetric, at $0.5$ redder
colours.

\begin{figure}
\resizebox{\hsize}{!}{\includegraphics{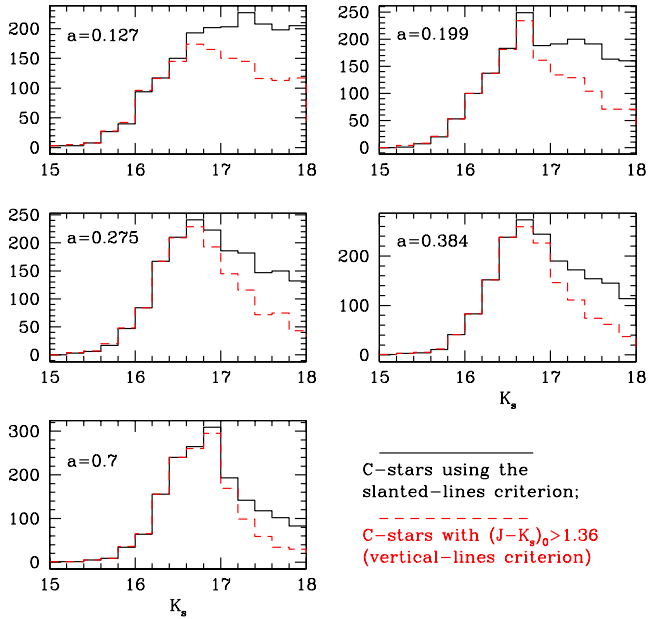}}
\caption{Distribution of the number of C-rich AGB stars within each
elliptical ring. Rings are marked with the value of the semi-major
axis (degrees). The selection of sources performed using {\it slanted
lines} is shown by continuous-line histograms while the selection
using {\it vertical lines} is shown by dashed-line histograms. Bins
are of $0.2$ mag.}
\label{cring}
\end{figure}

\begin{figure}
\resizebox{\hsize}{!}{\includegraphics{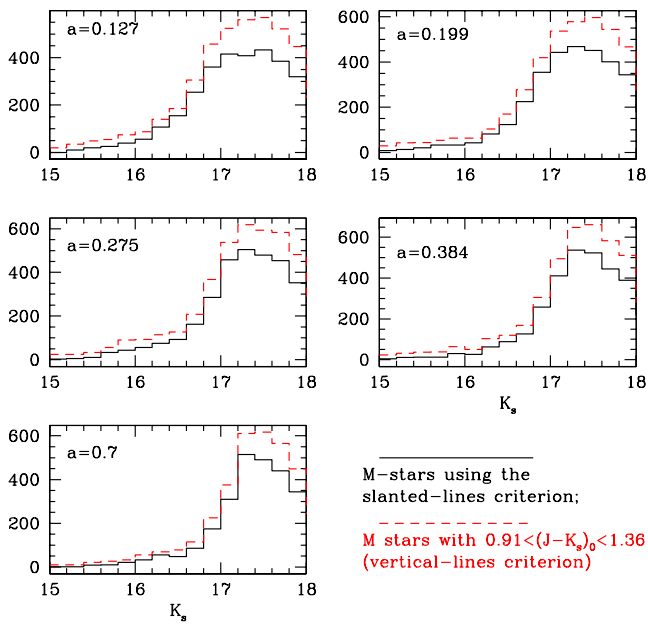}}
\caption{The same as for Fig.~\ref{cring} but for O-rich AGB stars.}
\label{oring}
\end{figure}

\subsubsection{Structure and extinction}

In order to interpret the distribution of age and metallicity across
M33 we need to account for its structure and in particular for
the parameters that characterise the AGB disc (position angle and
inclination). The plane of the galaxy is inclined with respect to the
line of sight by $i=56^{\circ}\pm 1^{\circ}$ (Zaritsky et
al. \cite{za89}). According to van der Marel \& Cioni (\cite{vdm})
this would produce a sinusoidal variation of the magnitude of objects
with identical properties at a given distance from the centre with
position angle ($\Phi$). Based on geometrical considerations the
amplitude of this variation can be described as $A=0.038\rho$tan$i$
where $\rho=\sqrt(x^2+y^2)$ is the angular distance from the centre
expressed in degrees. AGB stars in M33 can be traced to a maximum
distance of $0.7^{\circ}$ which corresponds to detecting a maximum
amplitude of $0.04$ mag.

The presence of differential extinction may also significantly alter
the determination of the mean metallicity and age using the $K_s$
magnitude distribution of AGB stars. In order to evaluate the effect
of differential extinction across M33 we need to study the variation
of features tracing extinction across the galaxy. In the near-infrared
the colour $J-K_s$ as well as the mode of the distribution of
individual magnitudes represent these features. Note that any, albeit
small, variation due to the orientation of the galaxy in the sky (see
above) will also be included. In particular, magnitudes are degenerate
in both parameters (structure and extinction) while colours better
represent reddening.

By subdiving the M33 area into the same eight sectors as shown in
Sect.~3.2 three elliptical ring areas were defined. Their outer
semi-major axis corresponds to $0.2^{\circ}$, $0.4^{\circ}$ and
$0.7^{\circ}$, respectively, and comprises a sufficiently large
statistical number of AGB stars to measure the effect of orienation
and extinction. Histograms of the number of AGB stars versus each
near-infrared magnitude and $J-K_s$ colour were constructed adopting a
bin size of $0.1$ in mag and $0.05$ in colour within each elliptical
ring. In particular, C and M stars, brighter than the tip of the RGB,
were selected within narrow ranges of colours: $1.5<(J-K_s)_0<2.0$ and
$1.0<(J-K_s)<1.3$, respectively, to avoid contamination from one
spectral type to the other. The bin defining the mode of each
histogram contained $20-30$ stars each with a photometric error of
$\le0.1$ in magnitude and $\le0.25$ in colour which correspond to an
uncertainty of at most $0.02$ mag in the determination of the peak
position.

\begin{figure}
\resizebox{\hsize}{!}{\includegraphics{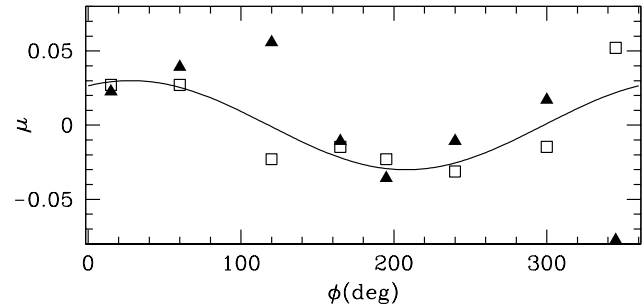}}
\caption{Distribution of the difference between the peak of magnitudes
  (squares) and colours (triangles) of C stars within $8$ sectors and
  their average of a unique ellipse with a semi-major axis of
  $0.7^{\circ}$ versus PA (see text for details). The best fit
  sinusoid has an half amplitude of $0.03\pm0.01$ mag and corresponds
  to $\phi=28^{\circ}\pm6^{\circ}$. This angle is measured on the
  plane of the galaxy where $0^{\circ}$ corresponds to the NE point of
  the major axis and increasing to the East. Errors on each point are
  at most $0.02$ mag and have been omitted for clarity.}
\label{sinu}
\end{figure}

The average distribution of the peak position of magnitudes and
colours of C stars for all ellipses traces a well defined sinusoid
with a half amplitude of $0.03\pm0.01$ mag and a position angle
$\phi=28^{\circ}\pm6^{\circ}$ (measured on the galaxy disc where
$0^{\circ}$ corresponds to the NE point of the major axis and
increasing to the East) with $\chi^2=0.001$ (Fig.~\ref{sinu}). Each
point in this figure corresponds to equal sectorial areas of a single
ellipse with semi-major axis of $0.7^{\circ}$.  In the usual
astronomical convention where sky angles are measured from the North
increasing to the East this is equivalent to
$\Phi=51^{\circ}\pm6^{\circ}$. Despite the moderately small error
associated to the position angle, the sinusoid has a rather flat
maximum as well as minimum which suggest a larger uncertainty in the
accuracy of these features. M stars are consistent with a similar
pattern but present a larger scatter resulting in a larger $\chi^2$ while a 
variation in the bin size has a small effect on these numbers. 

 It is, however, surprising that both magnitudes and colours describe
 a very similar sinusoidal pattern. If all near-infrared magnitudes
 would strictly describe the same variation then the variation in the
 colours should be close to zero. By inspecting the variation in the
 $J$, $H$ and $K_{\mathrm s}$ bands separately we conclude that while
 $H$ and $K_{\mathrm s}$ behave similarly, and produce a best fit
 sinusoid with $\phi=353^{\circ}$ and amplitude $0.04$ mag, the
 variation in the $J$ band is at odds ($\phi=80^{\circ}$ and amplitude
 $0.04$ mag). This has two effects: decreasing the amplitude of a
 sinusoid that aims to fit all three near-infrared magnitude
 variations and inducing a sinusoidal variation in the $J-K_{\mathrm
   s}$ colour (Fig. \ref{sinu}). If the detected variation in colour
 would only be due to extinction the magnitude variation could be
 corrected accordingly. Interstellar absorption in a given wave band
 is usually stronger at $J$ than at $K_{\mathrm s}$, this might
 explain why the variation in $J$ disagrees with the variation in the
 other redder wave bands.  Because this study focuses on the
 $K_{\mathrm s}$ magnitude we have used the average sinusoid shown in
 Fig. \ref{sinu} to correct for structural and reddening effects.
   This is a conservative approach in view of the uncertainties in the
   previous considerations.

The mean magnitude within each ellipse varies as a function of radius
increasing by about $0.08$ mag from the inner to the next outer
ellipse. This variation is much larger than the variation derived
above attributed to the orientation and extinction of the galaxy and
suggests a non-negligible variation in the age and/or metallicity of
the stellar population. These variations are usually of radial
type. The size of the three ellipses considered in this study is such
that the middle one corresponds to the metal-poor ring-like structure
suggested from the distribution of the C/M ratio (Fig.~\ref{cm72}). 

 Dust in the M33 disc might follow a patchy rather than regular
  distribution, as assumed above. If the size of a dusty patch is
  comparable to the size of one or a few bins, thus much smaller than
  individual sectorial regions used to build histograms, then the
  method adopted in this paper does not account for it. A detailed
  reddening map that will allow us to correct the photometry of each
  source prior the construction of histograms of their distribution is
  not yet available.

\subsection{Distribution of the C/M ratio}

The ratio between C-type and M-type stars is a simple indicator of
metallicity. A high number of C stars, and therefore a high C/M ratio,
is typical of metal-poor environments. Stars are of C-type when their
atmosphere contains more carbon than oxygen atoms apart from those
that are coupled into CO molecules. C atoms are dredged-up from the
stellar interior to the atmosphere during stellar evolution.  If the
metal content at the time when stars formed was low it is easier to
form C-stars than vice-versa. The correlation between the C/M ratio
and [Fe/H] has been empirically determined and was recently calibrated
by Battinelli \& Demers (\cite{bade05}).  Cioni \& Habing (\cite{cm})
have used this ratio to clearly show the presence of a metallicity
gradient within the Large Magellanic Cloud (LMC). However, this ratio
depends on the age of the underlying stellar population as was shown
by Cioni et al. (\cite{lf}). The age parameter is further discussed in
Sect.~\ref{s:sfh}.

\begin{figure}
\begin{minipage}{0.45\textwidth}
\resizebox{\hsize}{!}{\includegraphics{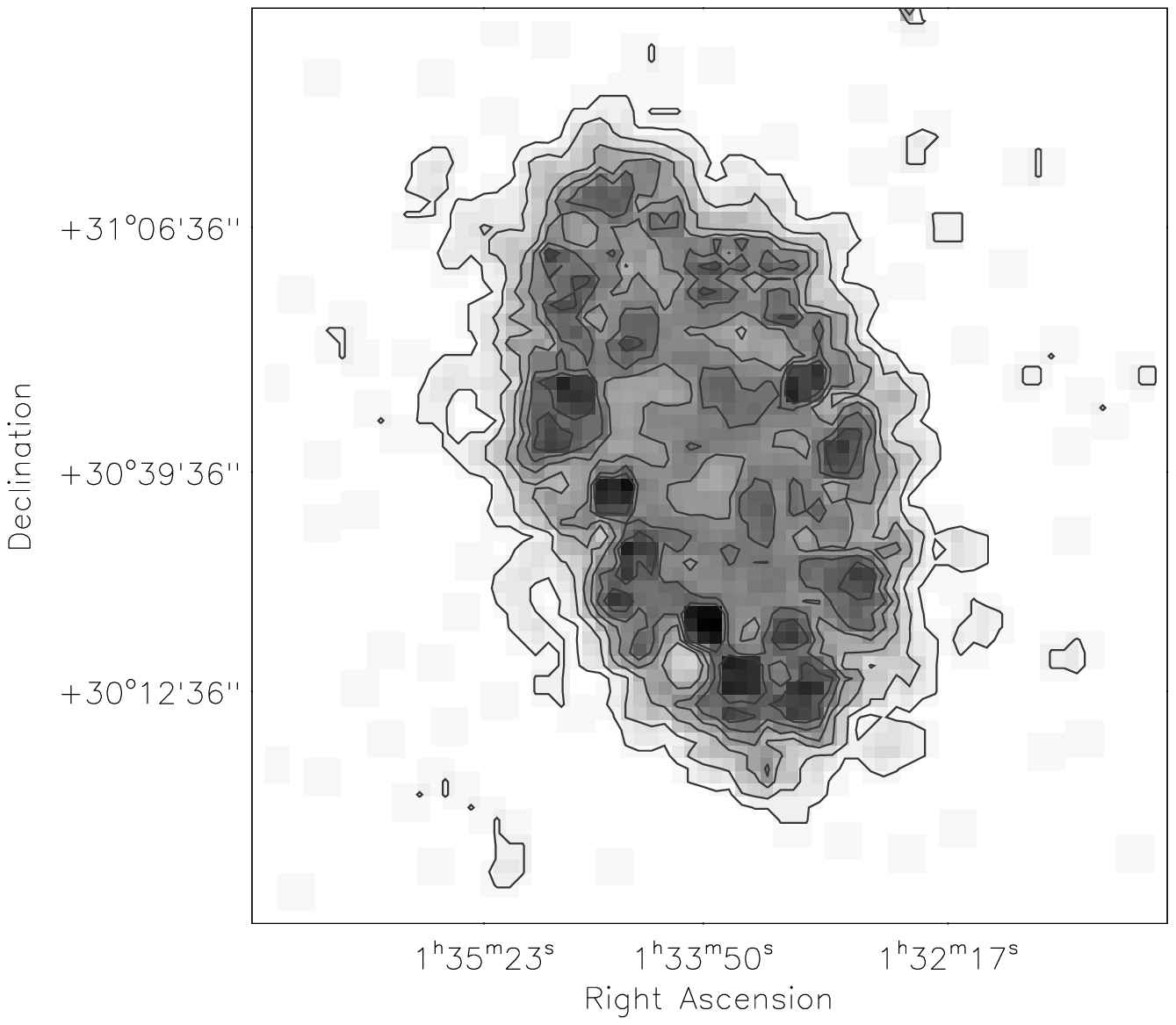}}
\end{minipage}
\hfill
\begin{minipage}{0.45\textwidth}
\resizebox{\hsize}{!}{\includegraphics{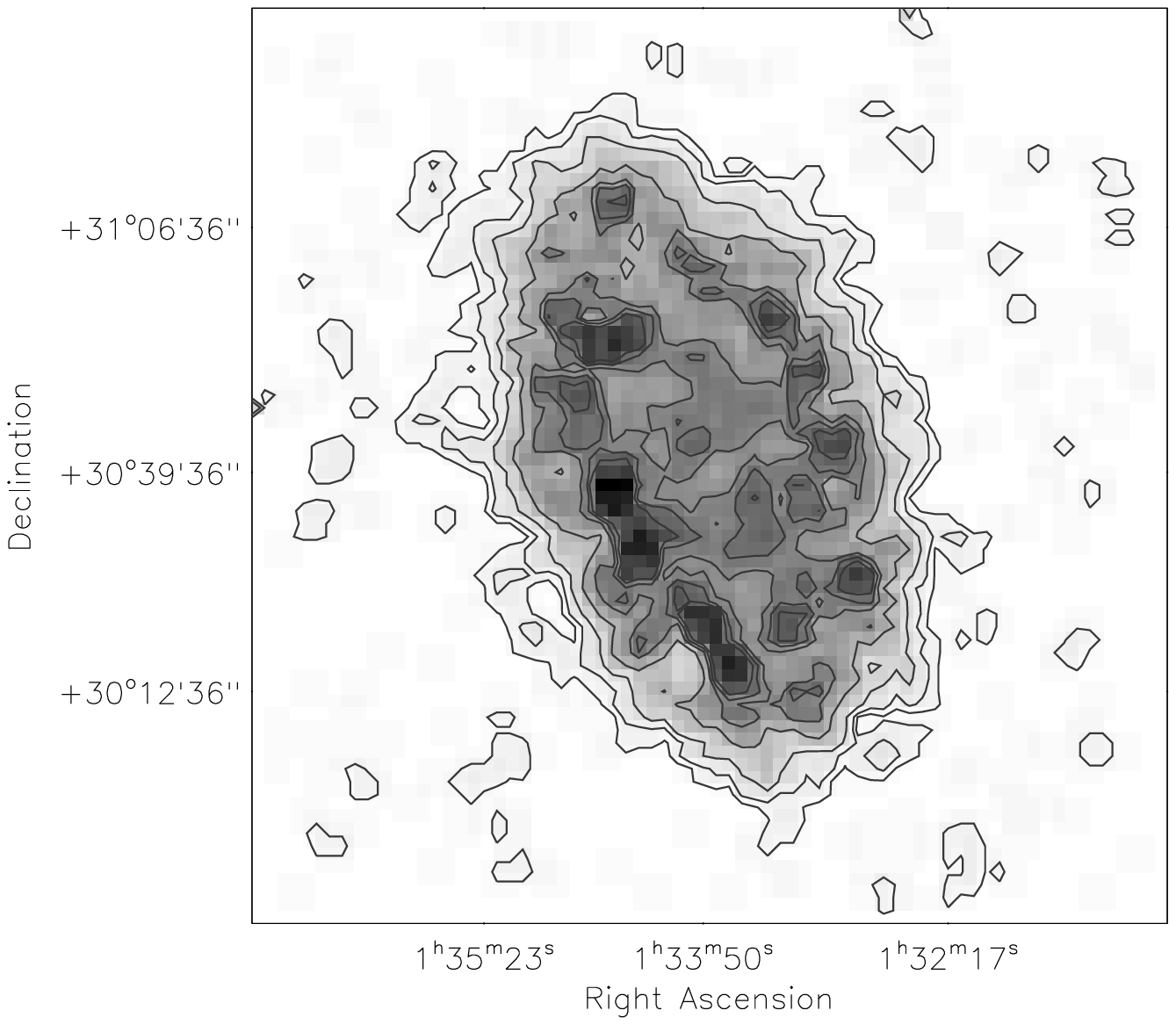}}
\end{minipage}
\caption{Distribution of the C/M ratio across M33. C-rich and O-rich
AGB stars above the tip of the RGB have been selected using slanted
lines ({\it top}) or using vertical lines ({\it bottom}). Darker
regions correspond to higher ratios. Contours are at: $0.2$, $0.6$,
$1.2$, $1.6$, $2.0$ and $2.4$ in the {\it top} panel and at $0.25$,
$0.5$, $1.0$, $2.0$, $3.0$, $3.5$ and $4.0$ in the {\it bottom}
panel.}
\label{cm72}
\end{figure}

\begin{figure}
\resizebox{\hsize}{!}{\includegraphics{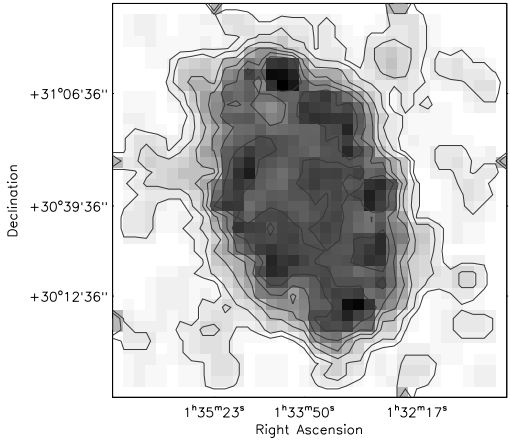}}
\caption{The same as the {\it bottom} panel of Fig.~\ref{cm72} but
  using bins of $2.4^{\prime}$. Contours are at: $0.25$, $0.5$, $1.0$,
  $1.5$, $2.0$ and $2.5$.}
\label{cm36}
\end{figure}

The surface distribution of the C/M ratio across M33 is shown in
Fig.~\ref{cm72}. The top panel shows the distribution obtained
selecting stars using the {\it slanted-lines} criterion while the
bottom panel shows the distribution obtained using the {\it
vertical-lines} criterion. Both distributions have been obtained
applying a box car smoothing to the ratio calculated using bins of
$1.2^{\prime}$. Grey-scaled distributions appear remarkably similar:
the outermost parts of the galaxy show a low ratio most probably
affected by a low number statistics, immediately inwards dark regions
of a high ratio are distributed in a ring-like structure surrounding
an inner region with a low ratio but also rather patchy. Maps with
twice a lower resolution emphasise the ring-like distribution traced
by regions of a high ratio compared to the inner part of the galaxy
(Fig.~\ref{cm36}). However, using larger bins obviously smoothes fine
details inducing the NW region of enhanced ratios to become more
prominent than the corresponding SE region.  According to Battinelli
\& Demers (\cite{bade05}) the range spanned by the C/M ratio in
Fig.~\ref{cm72} corresponds to a spread in [Fe/H] of at least $0.6$
dex. In particular, the most metal-poor regions of the galaxy have
[Fe/H]$=-1.54$ dex while the most metal rich have [Fe/H]$=-0.91$ dex.

% This figure was obtained using slanted lines but shifted of 0.05
% at each extreme. 

\subsection{Determination of metallicity and mean age}
\label{s:sfh}

The distribution of AGB stars as a function of $K_{\mathrm s}$
magnitude in each sector have been compared with theoretical
distributions as in Cioni et al. (\cite{lf}) where a detailed
description of the stellar evolutionary models adopted is also
given. Briefly:

-- $JK_{\mathrm s}$--band photometry has been simulated using the
TRILEGAL population synthesis code (Girardi et al. \cite{gi05}), that
randomly generates a population of stars following a given SFR,
age-metallicity relation and initial mass function;

-- bolometric magnitudes, used to derive $K_{\mathrm s}$ magnitudes,
   were obtained by applying the extended tables of bolometric
   corrections (BCs) from Girardi et al. (\cite{gi02}) for O-rich
   stars and by empirical relations for C-rich stars: the BC in
   the K band is taken from the relation by Costa \& Frogel
   (\cite{cf96}) and the $(J-K)$ colour is derived from the
   T$_{\mathrm {eff}}-(J-K)-$C/O relation from Marigo et
   al. (\cite{ma03}), which itself is based on to the empirical data
   by Bergeat et al. (\cite{berge});

-- the stellar properties are interpolated over a large grid of
stellar evolutionary tracks, based on Bertelli et al. (\cite{be94})
and Girardi et al. (\cite{gi03}) for massive stars, Girardi et
al. (\cite{gi00}) for low- and intermediate-mass stars and
complemented with grids of thermally pulsing AGB tracks calculated by
means of Marigo et al.'s (\cite{ma99}, \cite{ma03}) synthetic code.
We refer the reader to the paper published very recently by Marigo et
al.~(\cite{ma07}) where details about different stellar isochrones,
released at different moments in time, are discussed and new
computations presented. The main difference between the isochrones
adopted in this study and those just released is in the inclusion of
the treatment of dust in the thermally pulsing AGB phase. This affects
more AGB stars with thick circumstellar envelopes, which have
$J-K_{\mathrm s}>2$, than AGB stars with thin circumstellar envelopes
which represent the bulk of the AGB population analysed in this study
of M33. In particular, Marigo et al.~(\cite{ma07}) show that very few
O-rich AGB stars, with high mass-loss rates, occupy the region of
($J-K_{\mathrm s}$) colours where most C-rich AGB stars are located.
Note also that the M33 data are photometrically calibrated using 2MASS
which is suitable for comparison with the Cioni et al.~(\cite{lf})
isochrones.

Theoretical distributions were created for $5$ different metallicities
(Z $=0.0005$, $0.001$, $0.004$, $0.008$ and $0.016$) and for $5$
different star formation rates (SFRs). The latter adopts a simple
family of exponentially increasing/decreasing SFRs, $\psi(t)\propto
exp(t/\alpha)$, where $t$ is the stellar age in Gyr, and $\alpha$ is a
free parameter that was taken to have values of: $-5$, $-2$, $2$ and
$5$; the special case of a SFR constant in time corresponds to
$\alpha=1000$. The mean age of all stars formed in a model with a
given value of $\alpha$ is given in Table $1$ of Cioni et
al. (\cite{lf}). The youngest population explored corresponds to a
mean age of $2$ Gyr and the oldest one to $10.6$ Gyr. The step in mean
age between adjacent models is at least $2$ Gyr. Models were
constructed at the distance of the LMC ($(m-M)_0=18.4$) and have been
shifted by $5.9$ mag to fit the population of M33. Both {\it
vertical-} and {\it slanted-lines} criteria to distinguish M-type from
C-type AGB stars have been applied to the simulated colour-magnitude
diagrams and were adjusted accordingly by the metallicity effect on
the \JminK~colour (see Cioni et al.  \cite{lf} for details).  Note
that, similarly to Cioni et al. \cite{lf}, only the $K_{\mathrm s}$
magnitude distribution of AGB stars is used. This is the wavelength
that best approximates the bolometric luminosity of AGB stars (with
thin circumstellar envelopes), in particular carbon stars for which
these theoretical models are well calibrated (i.e. using the C stars
luminosity function of Magellanic Cloud clusters; Marigo, Girardi \&
Bressan~\cite{ma99}). Before comparing observed with theoretical
distributions we corrected for the sinusoidal pattern derived in
Sect. 3.3.2. This variation represents an average effect between
extinction and orientation of the AGB disc. In practice, $K_{\mathrm
s}$ histograms have been shifted by $0.001-0.029$ mag depending on
their mean $\phi$ coordinates, regardless of their distance from the
centre, in a direction ($+/-$) that compensates for the sinusoidal
variation shown in Fig.~\ref{sinu}.

\begin{figure}
\resizebox{\hsize}{!}{\includegraphics{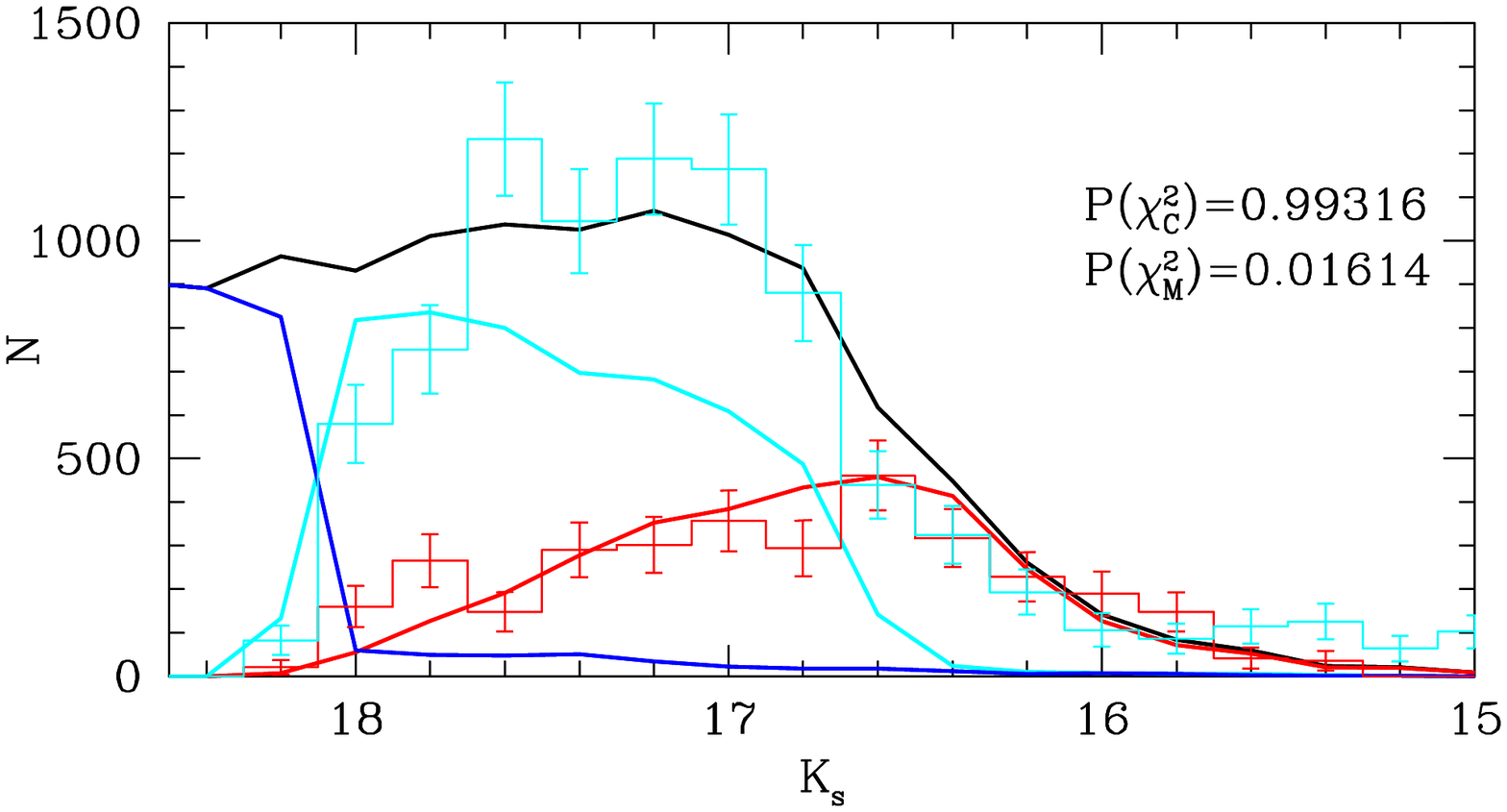}}
\resizebox{\hsize}{!}{\includegraphics{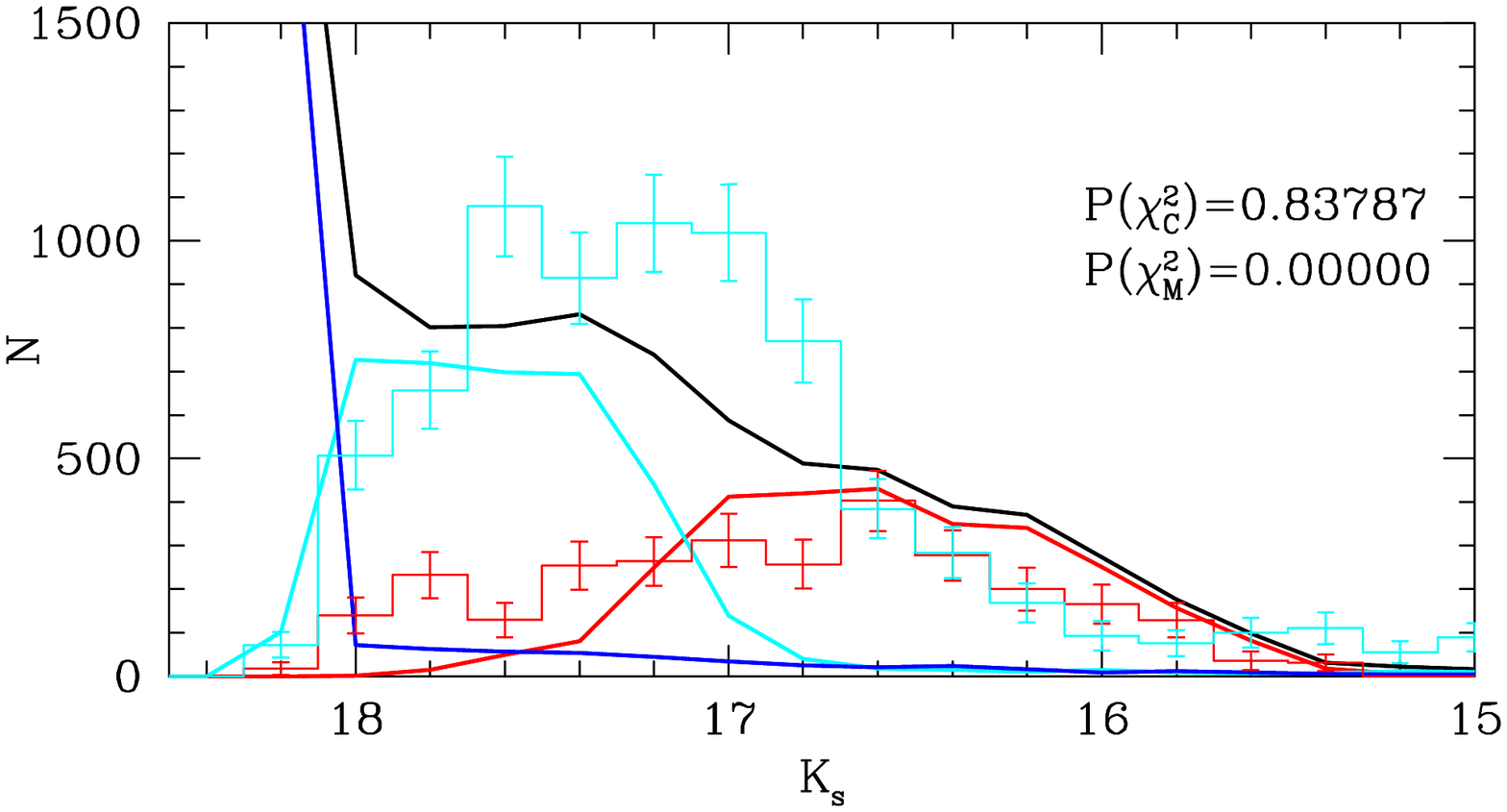}}
\caption{Comparison between the observed $K_{\mathrm s}$ magnitude
  distribution of C and M stars (histograms) and theoretical
  distributions. The probability that a given model represents the
  observed distributions, associated to the $\chi ^2$ value, is
  indicated. These histograms are fitted with models corresponding to
  a SFR with $\alpha=5$ (or a mean age of $8.7$ Gyr) and a metallicity
  Z $=0.0005$ ({\it top panel}) or Z $=0.001$ ({\it bottom panel}). In
  the electronic version of this figure C stars are in red, M stars in
  light blue, the sum of C and M stars is in black while RGB stars are
  in dark blue.}
\label{lumfex}
\end{figure}

\begin{figure}
\resizebox{\hsize}{!}{\includegraphics{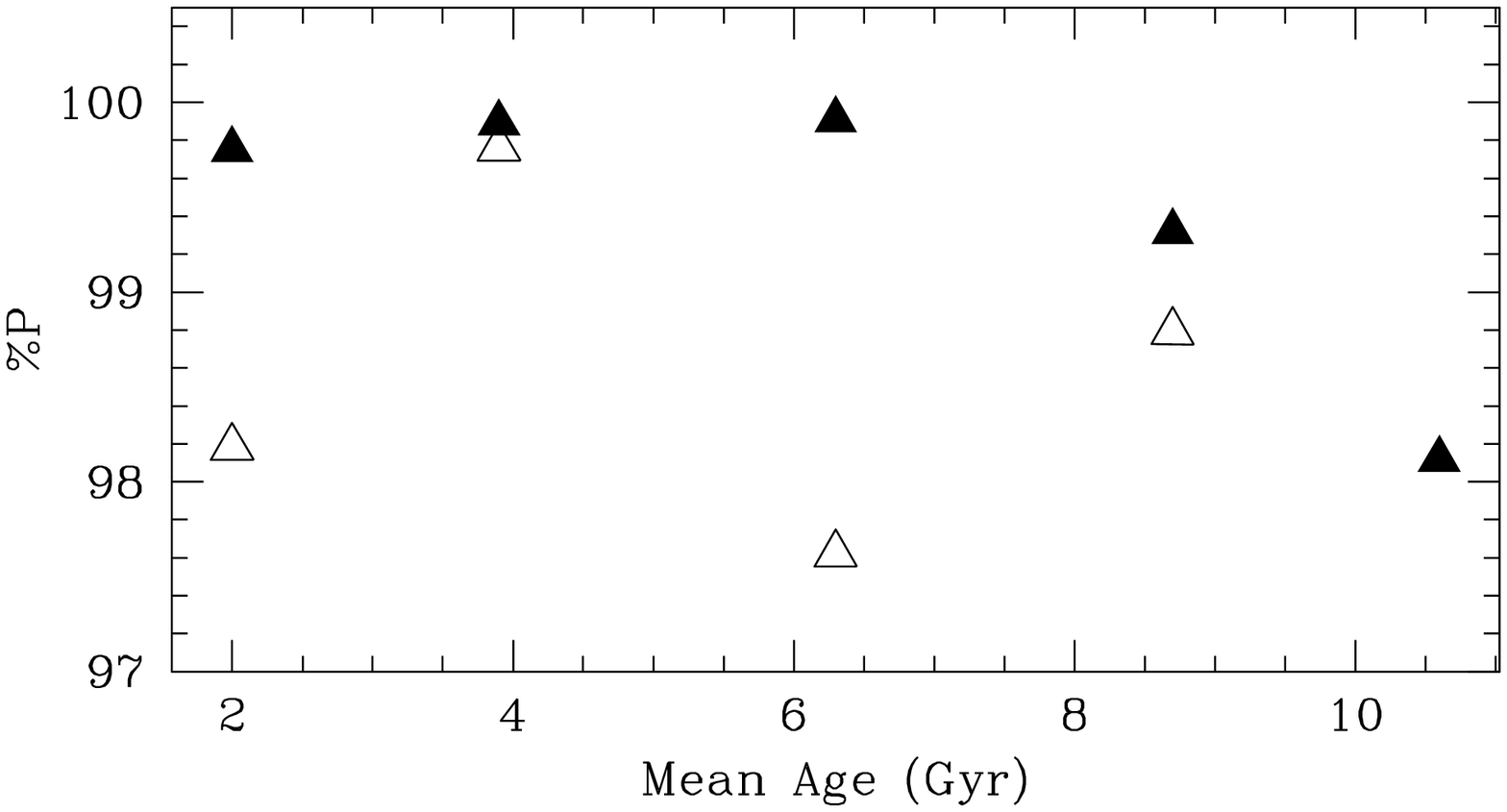}}
\caption{Probability as a function of mean age for models of a given
metallicity which represent the stellar population of M33 in a sector
of a ring. Different symbols refer to a different metallicity as
follows: Z=0.0005 (filled triangles), Z=0.001 (empty triangles). Other
models explored in this study give a probability much lower than the
values plotted here and are not shown.}
\label{fitunique}
\end{figure}

\begin{figure*}

\hspace{-0.6cm}
\vspace{-0.1cm}
\epsfxsize=0.24\hsize \epsfbox{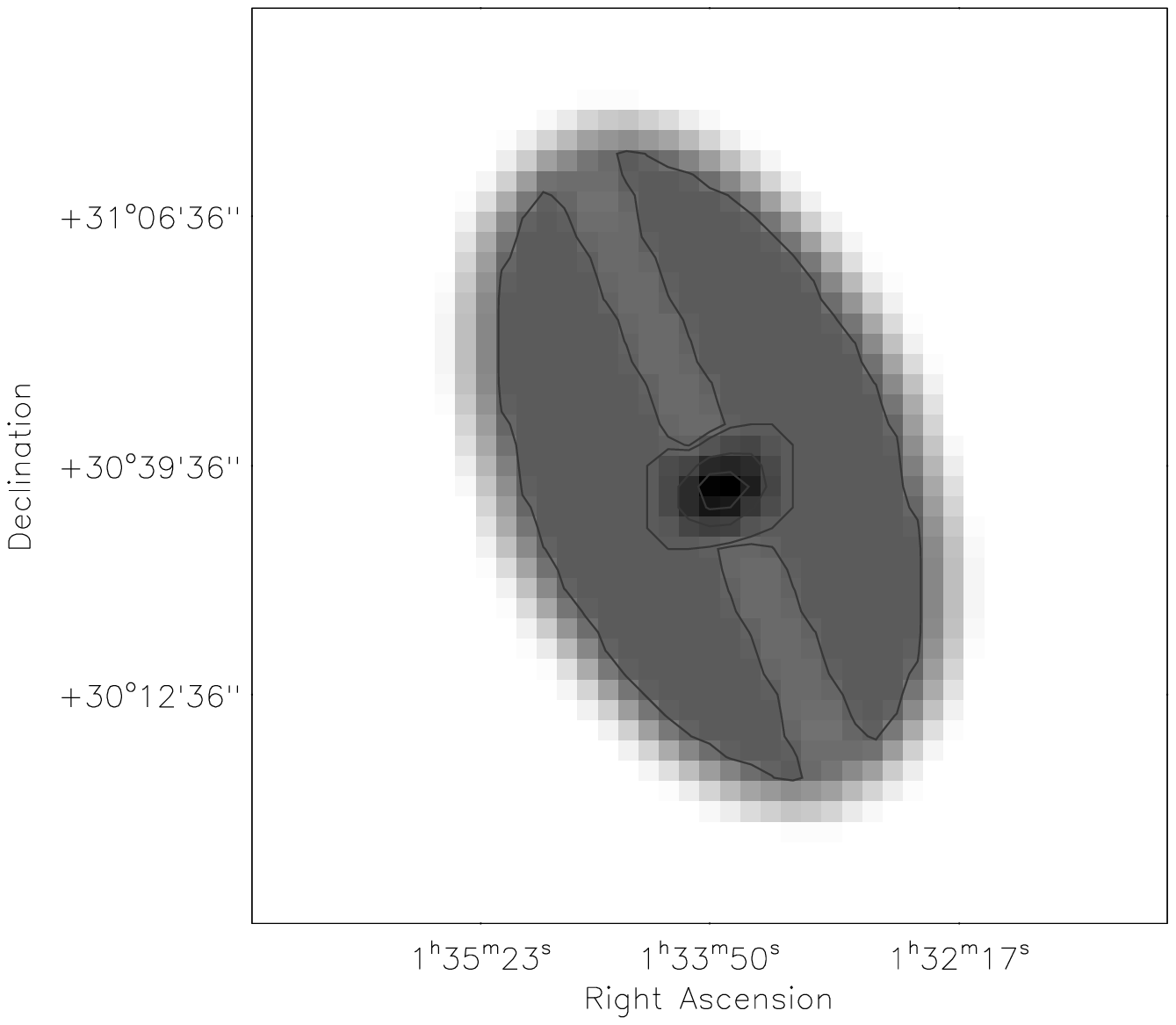}
\hspace{-1.0cm}
\epsfxsize=0.24\hsize \epsfbox{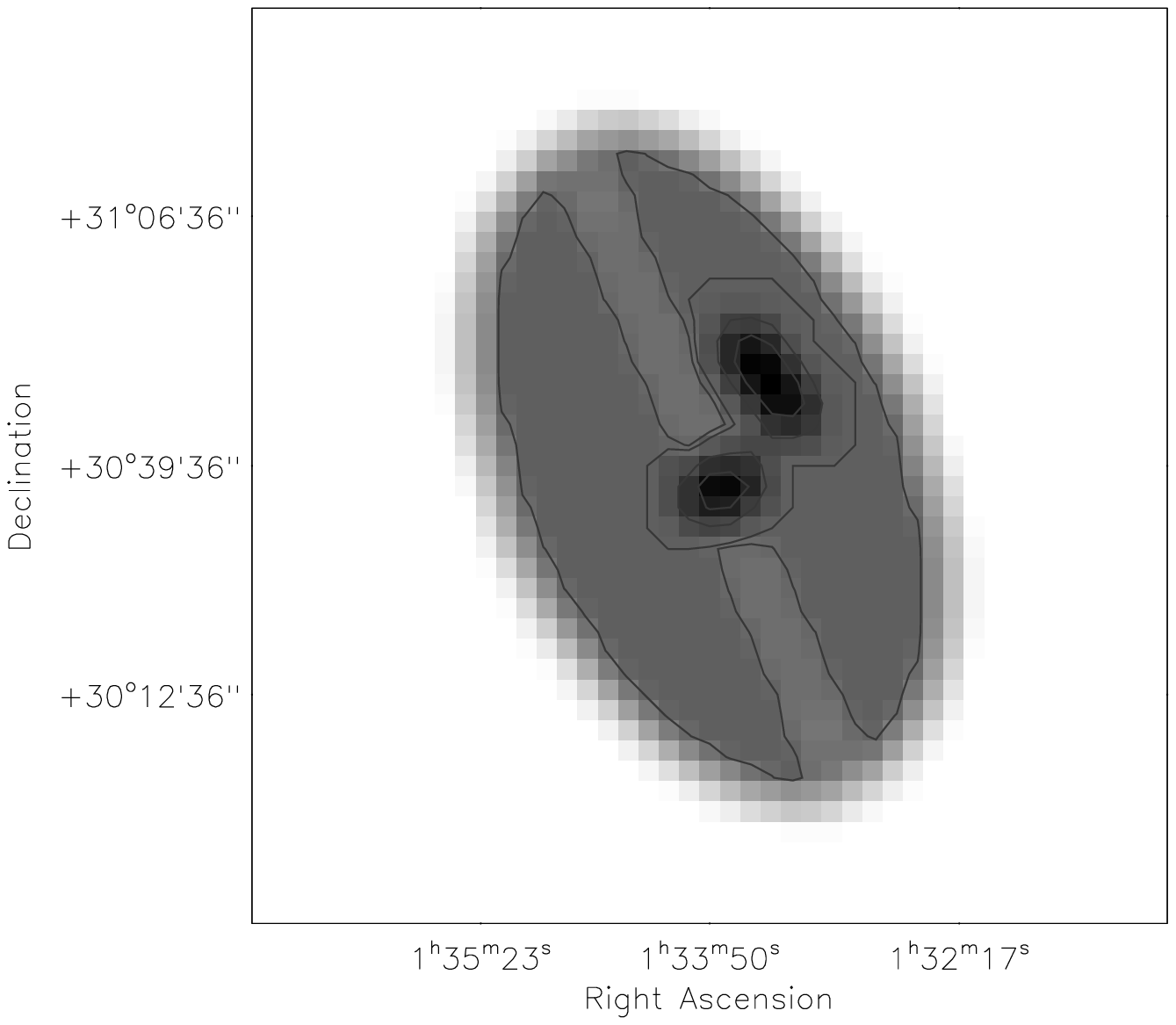}
\hspace{-1.0cm}
\epsfxsize=0.24\hsize \epsfbox{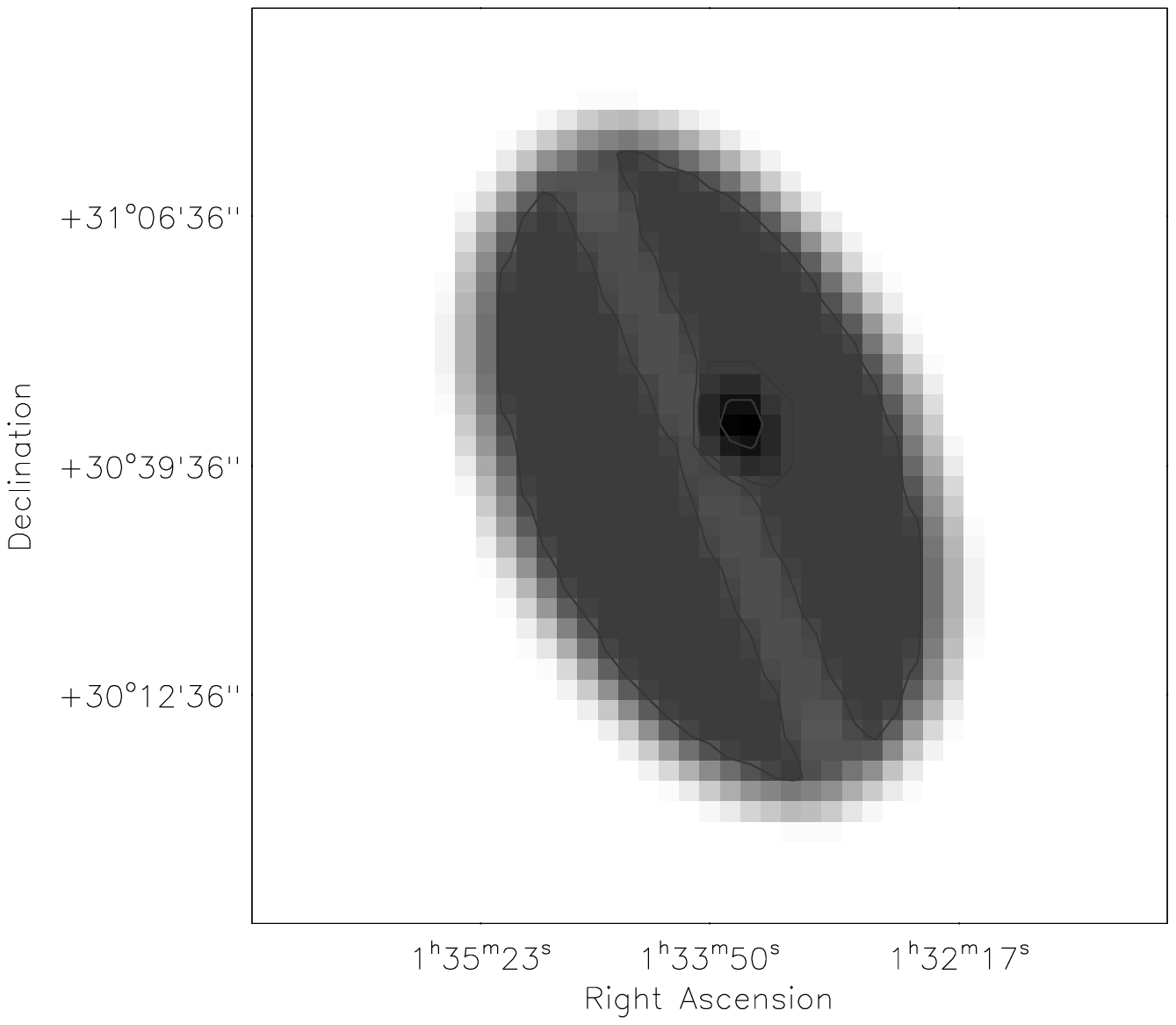}
\hspace{-1.0cm}
\epsfxsize=0.24\hsize \epsfbox{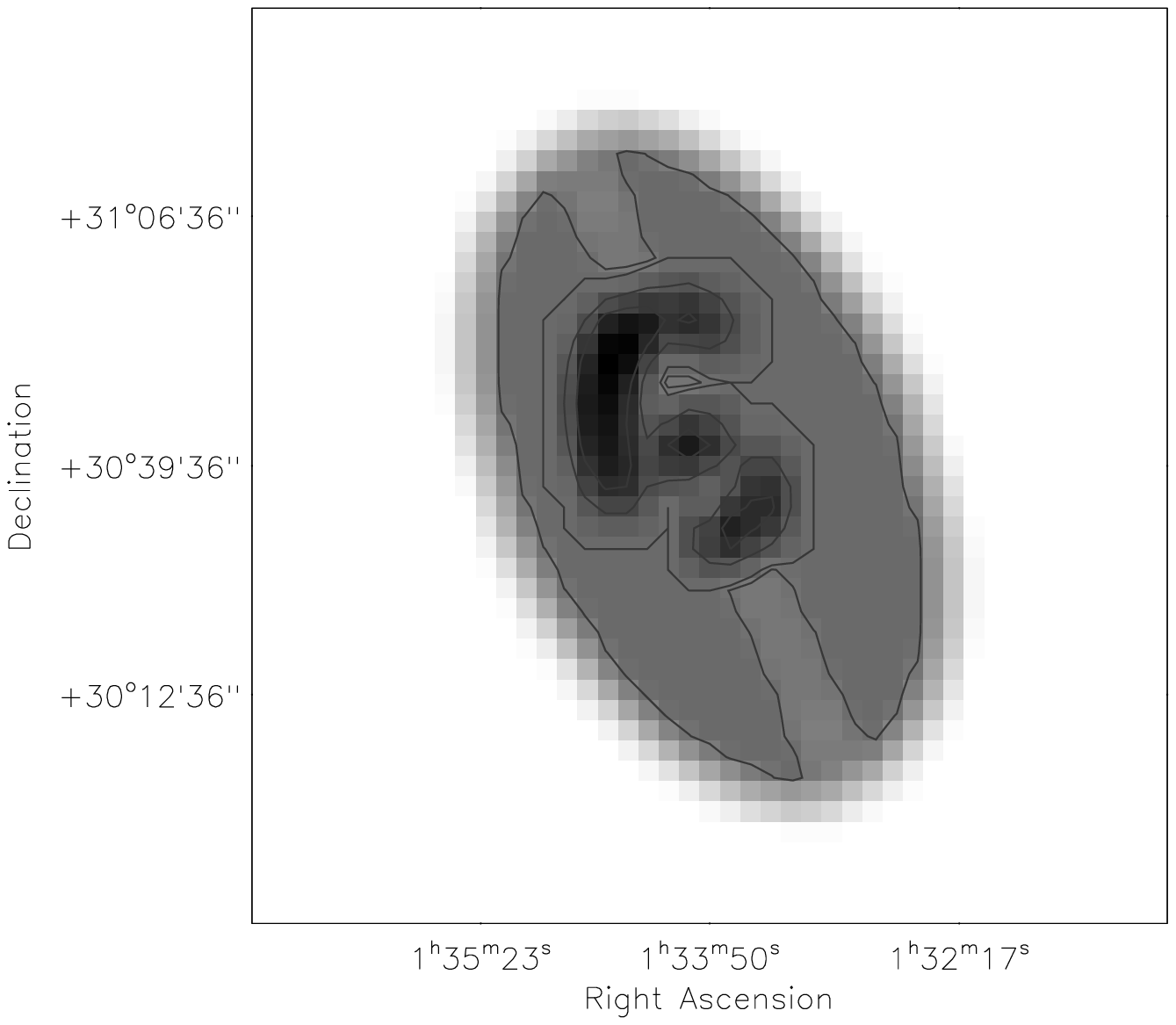}
\hspace{-1.0cm}
\epsfxsize=0.24\hsize \epsfbox{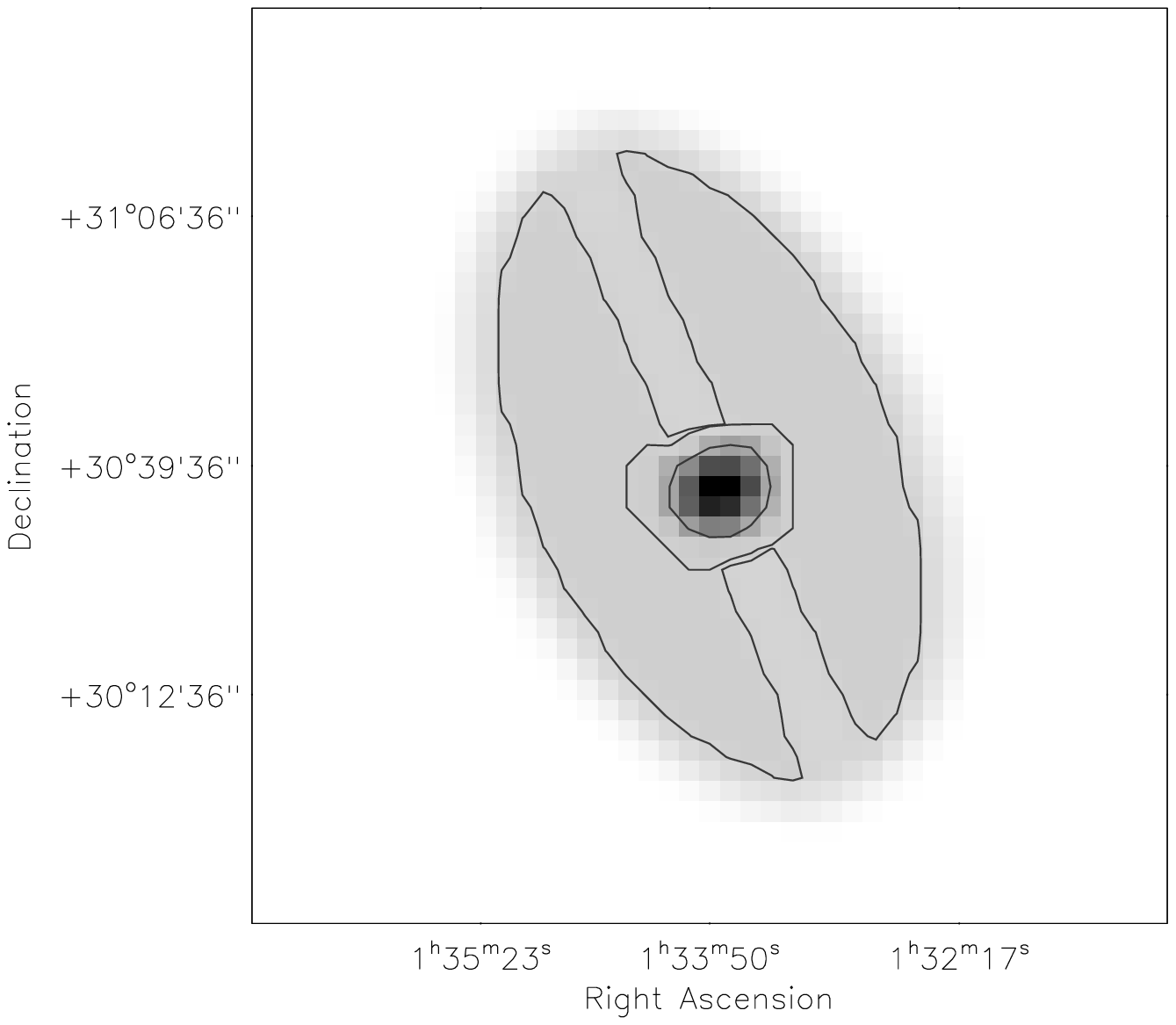}

\hspace{-0.6cm}
\epsfxsize=0.24\hsize \epsfbox{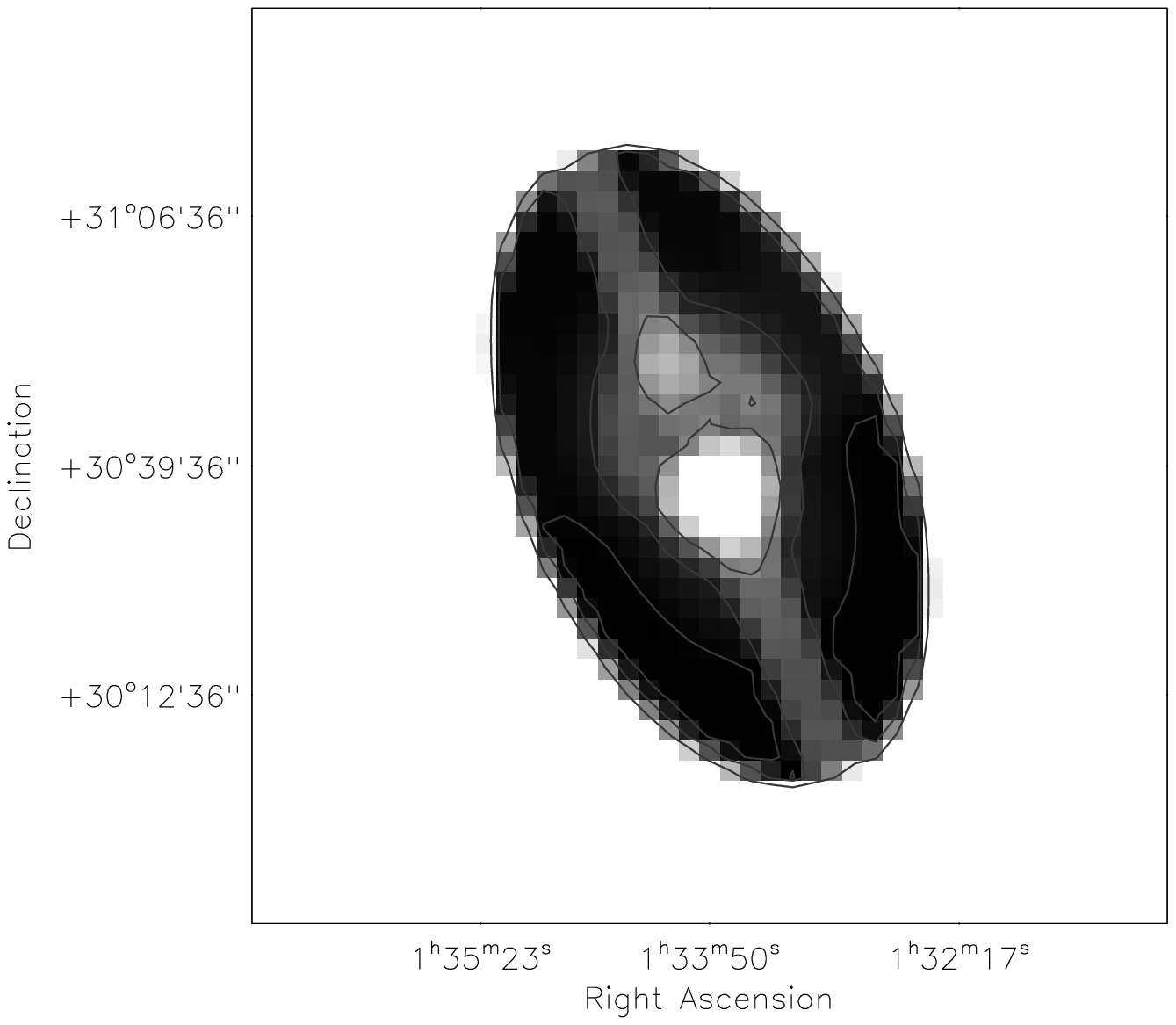}
\hspace{-1.0cm}
\epsfxsize=0.24\hsize \epsfbox{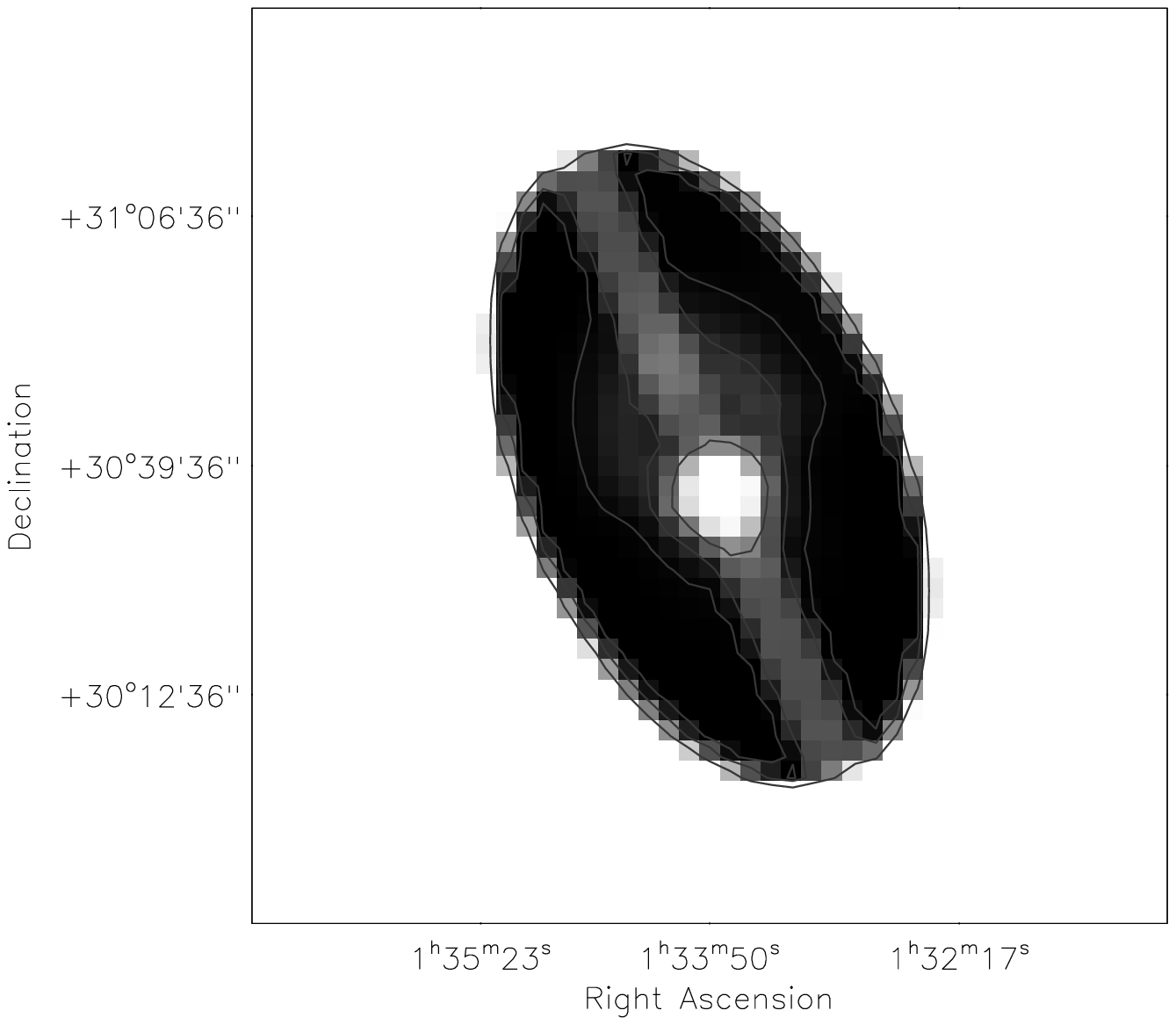}
\hspace{-1.0cm}
\epsfxsize=0.24\hsize \epsfbox{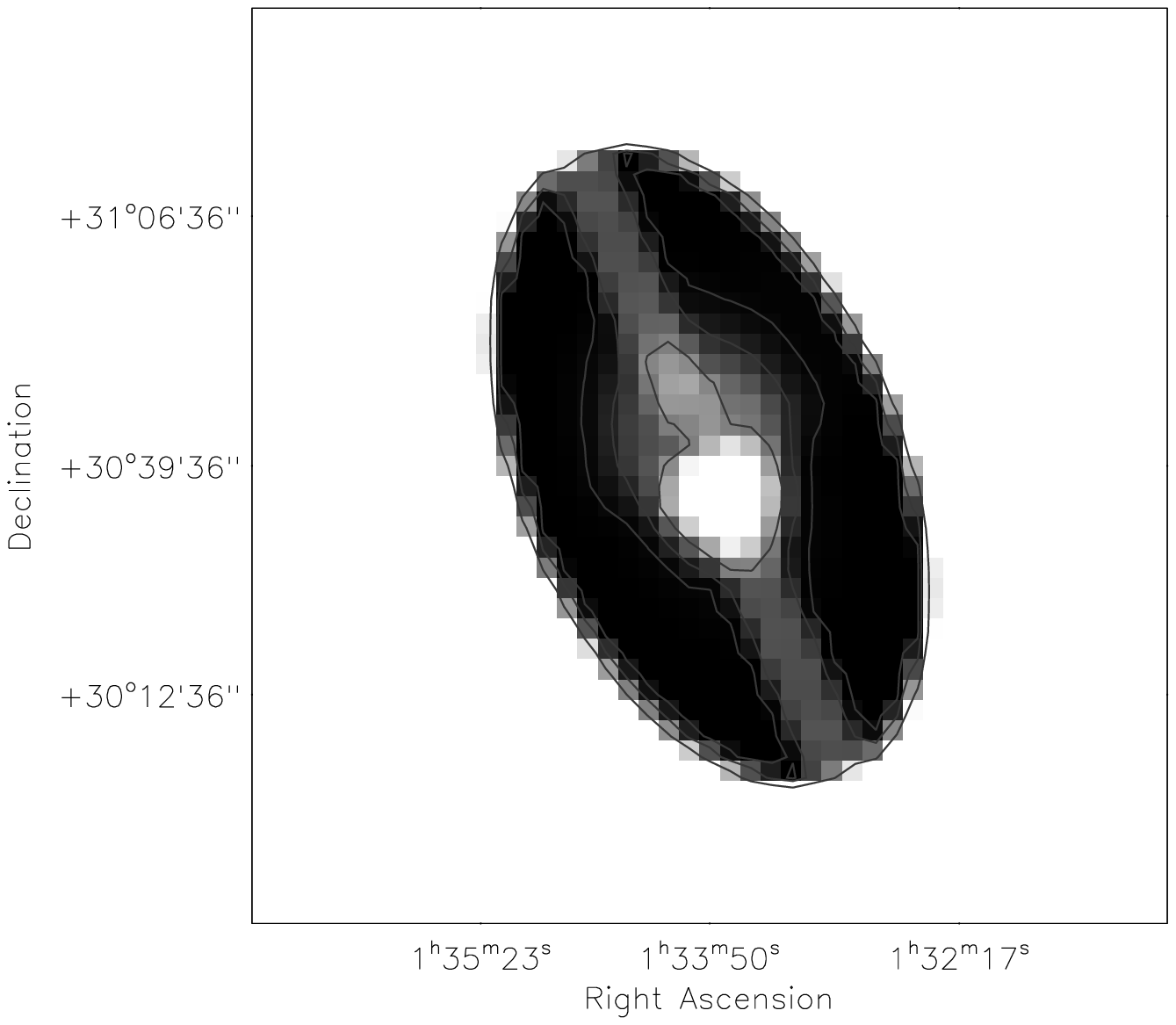}
\hspace{-1.0cm}
\epsfxsize=0.24\hsize \epsfbox{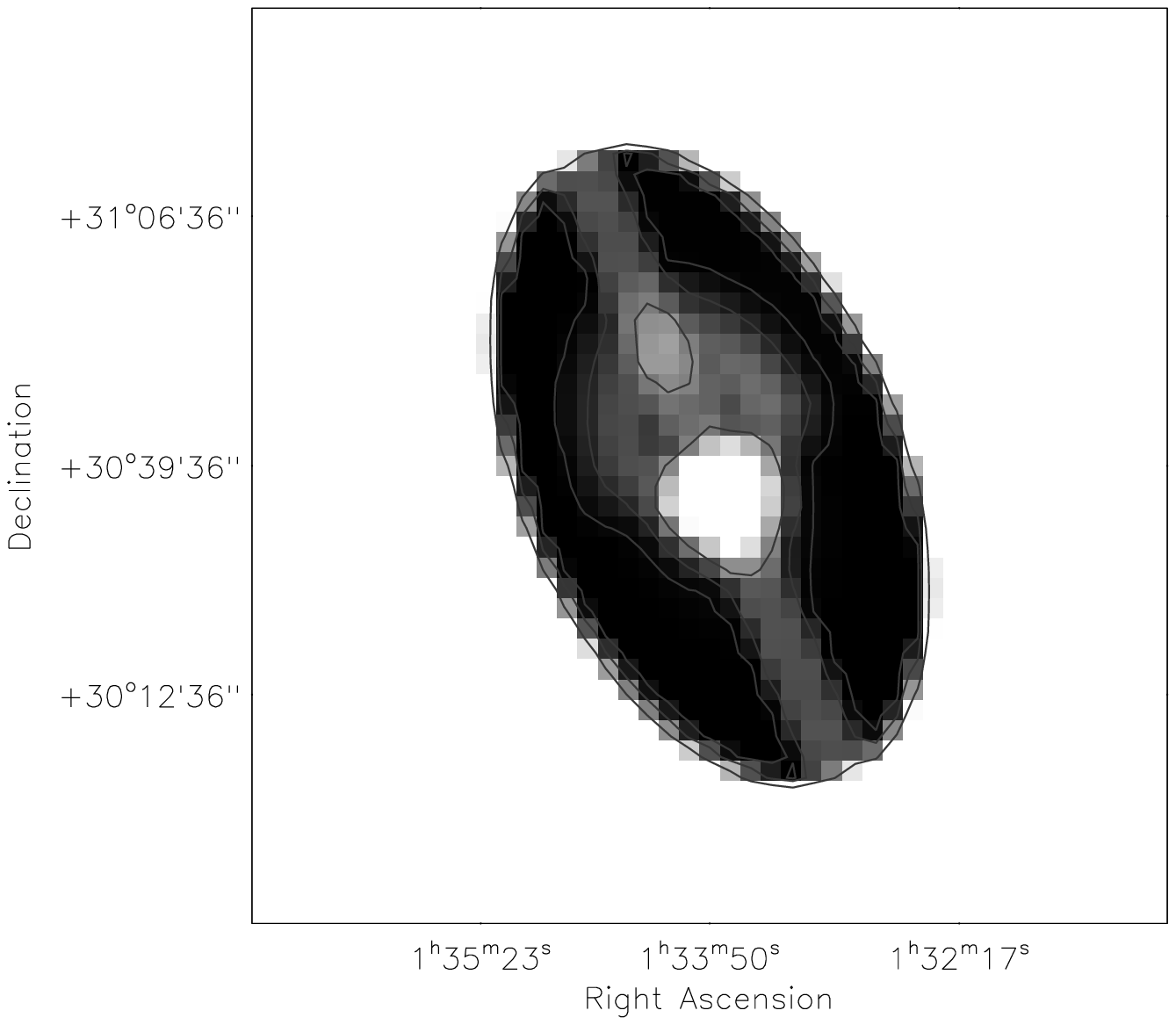}
\hspace{-1.0cm}
\epsfxsize=0.24\hsize \epsfbox{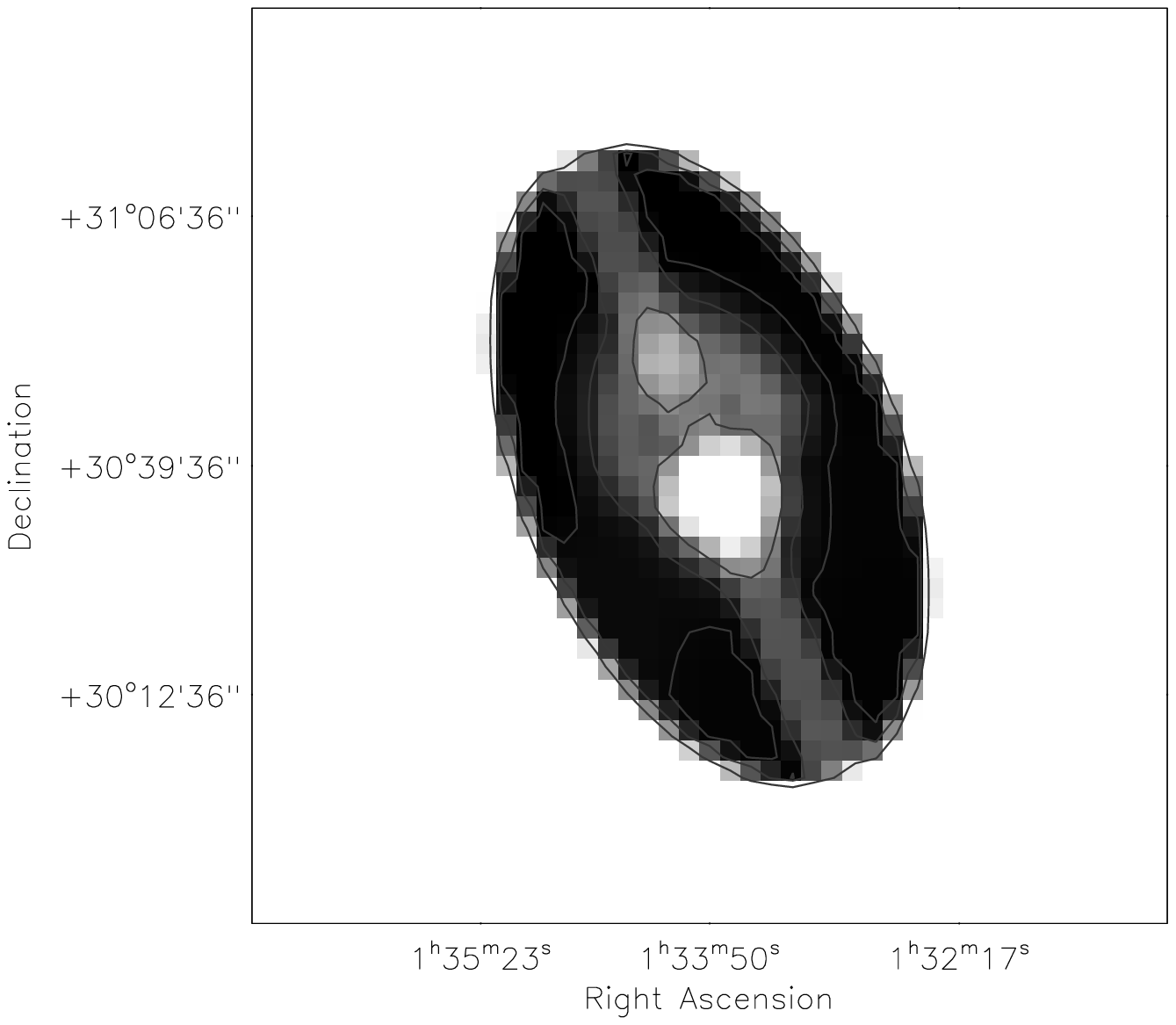}

\hspace{-0.6cm}
\vspace{-0.1cm}
\epsfxsize=0.24\hsize \epsfbox{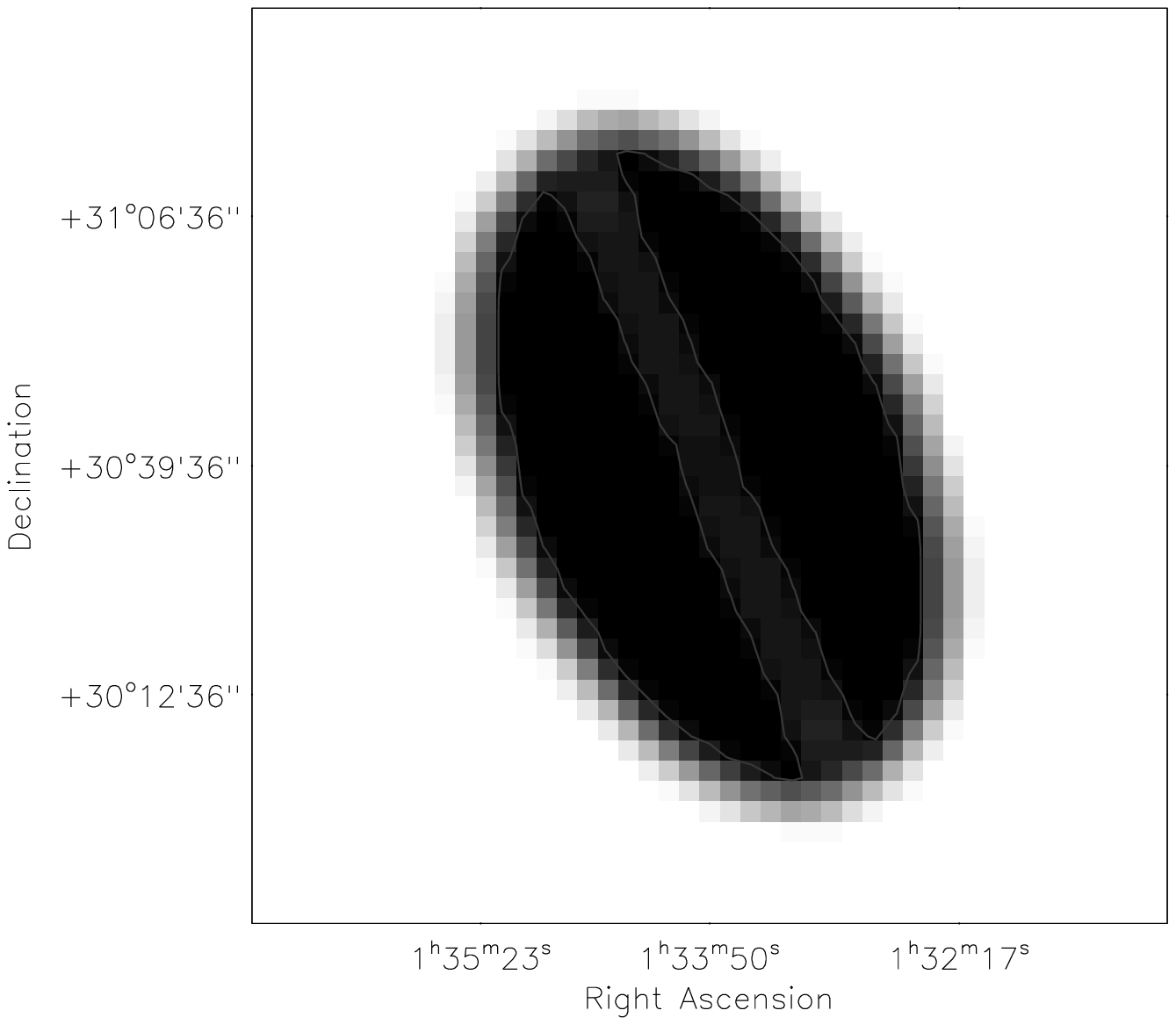}
\hspace{-1.0cm}
\epsfxsize=0.24\hsize \epsfbox{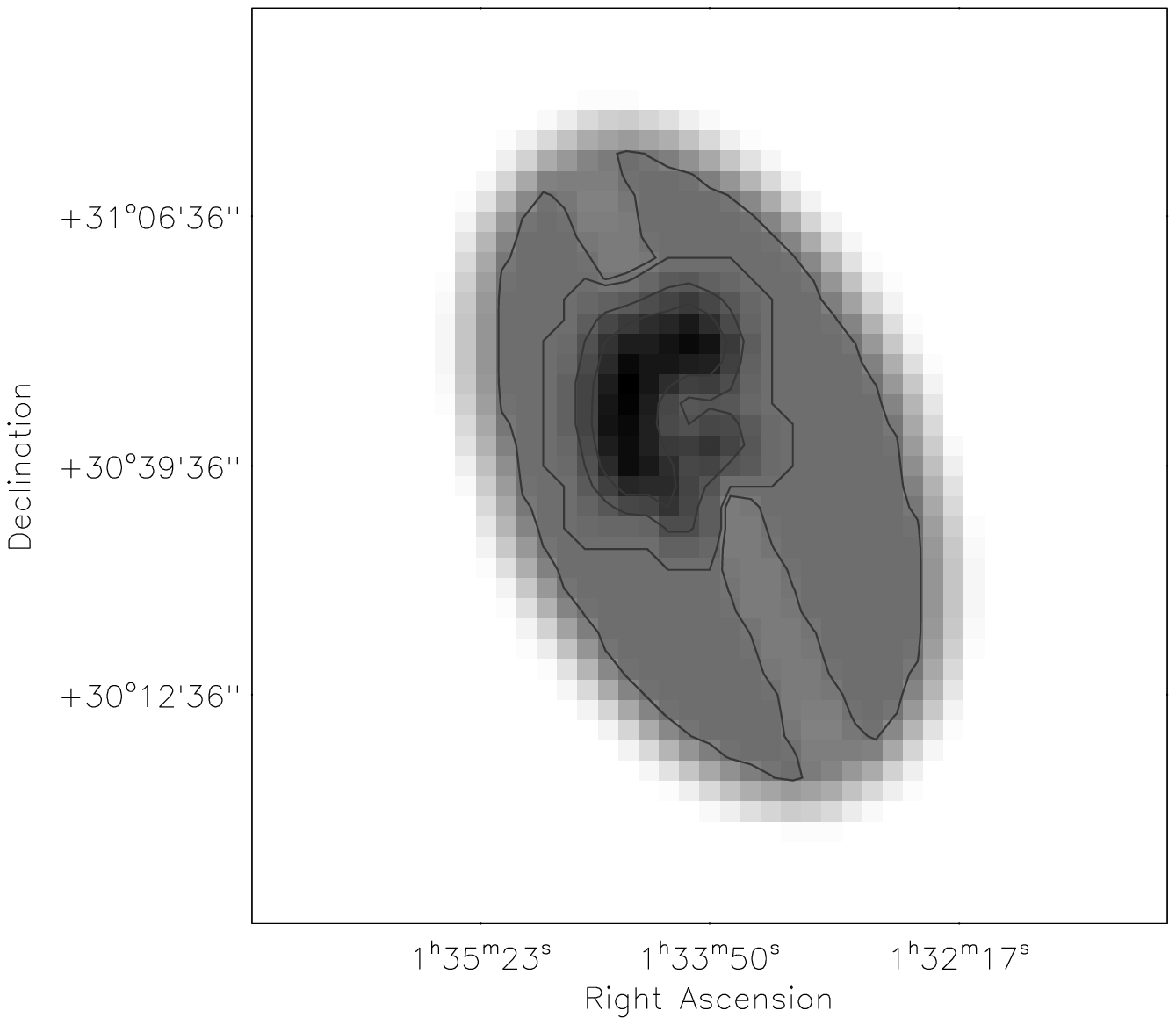}
\hspace{-1.0cm}
\epsfxsize=0.24\hsize \epsfbox{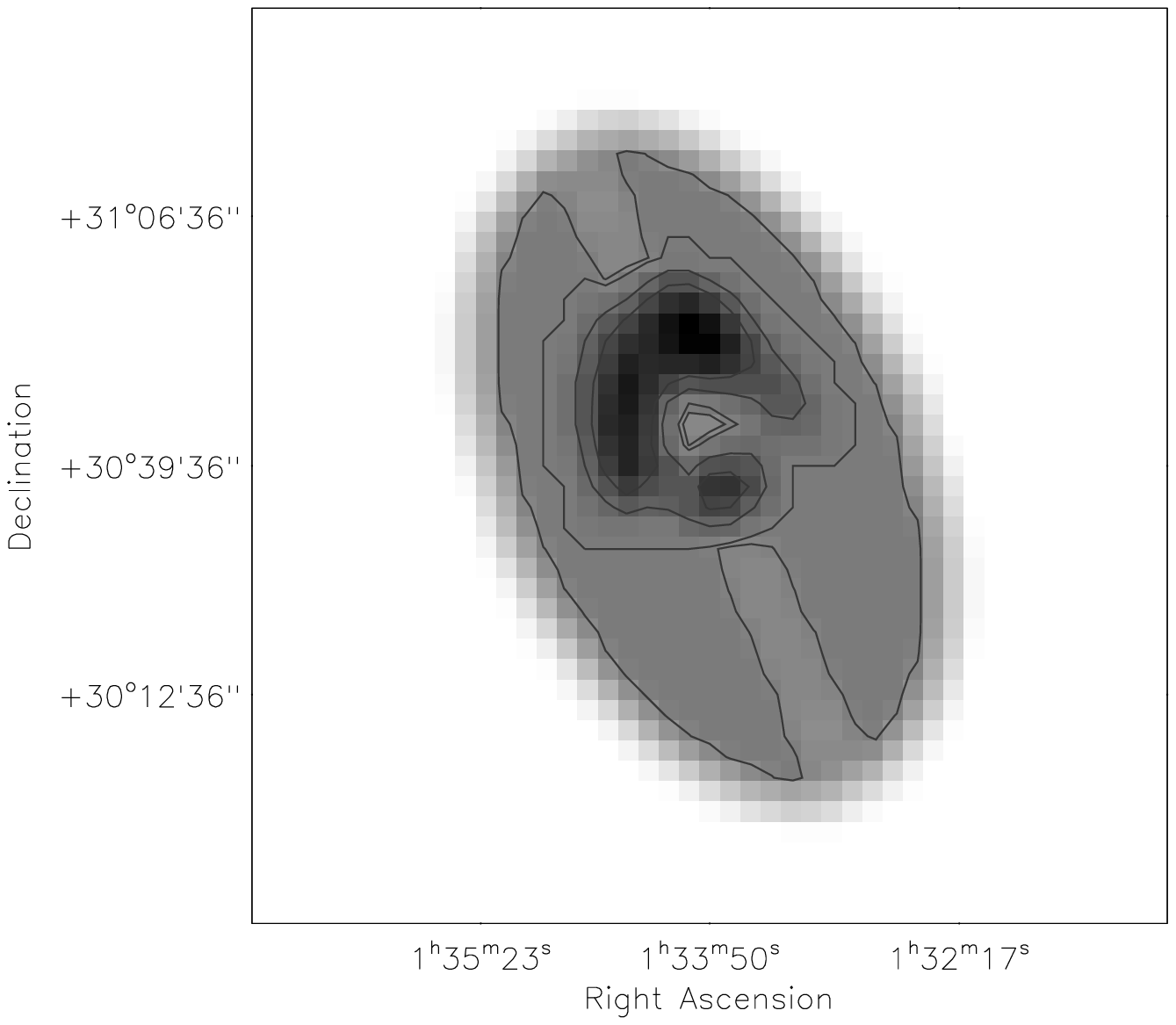}
\hspace{-1.0cm}
\epsfxsize=0.24\hsize \epsfbox{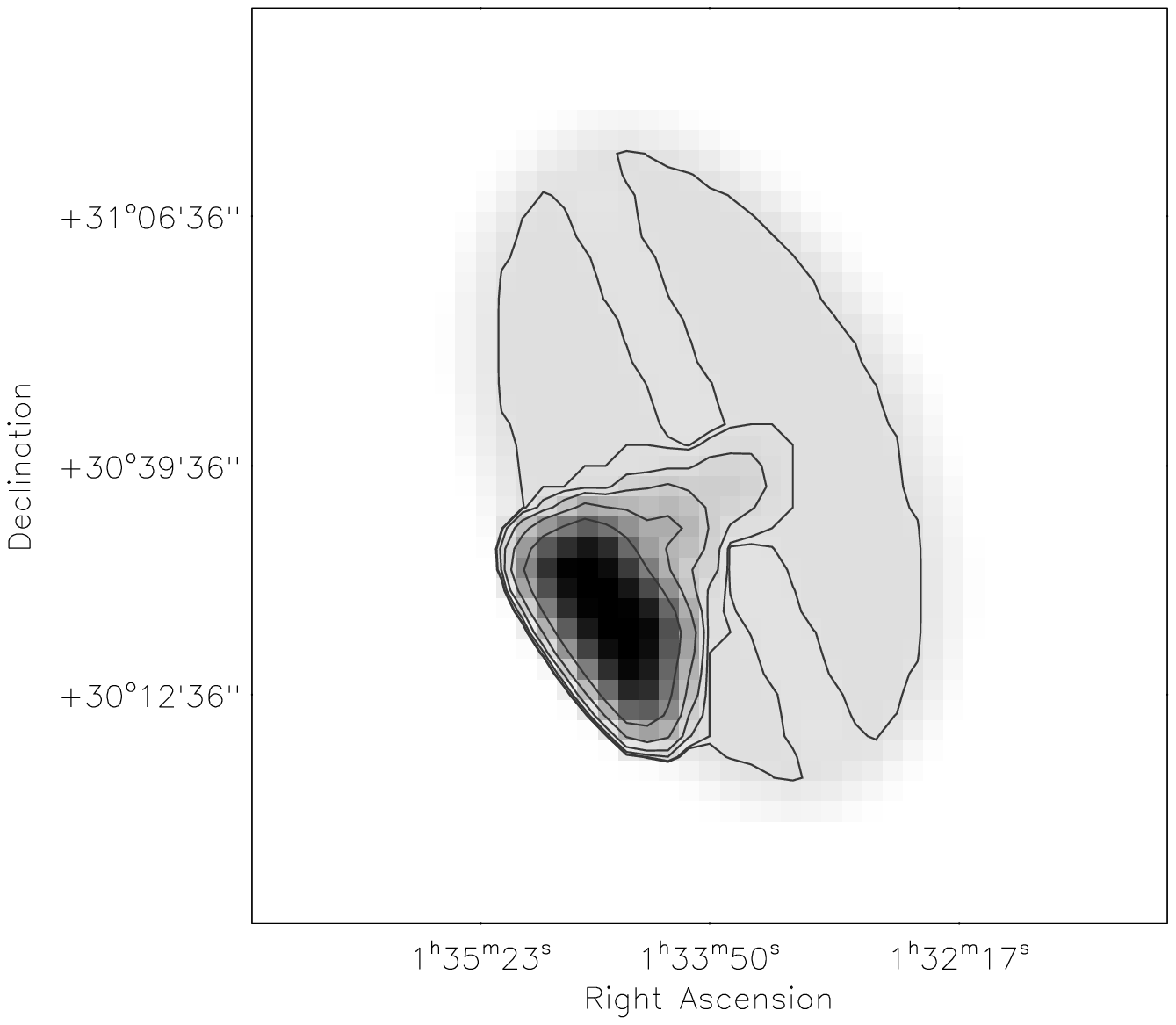}
\hspace{-1.0cm}
\epsfxsize=0.24\hsize \epsfbox{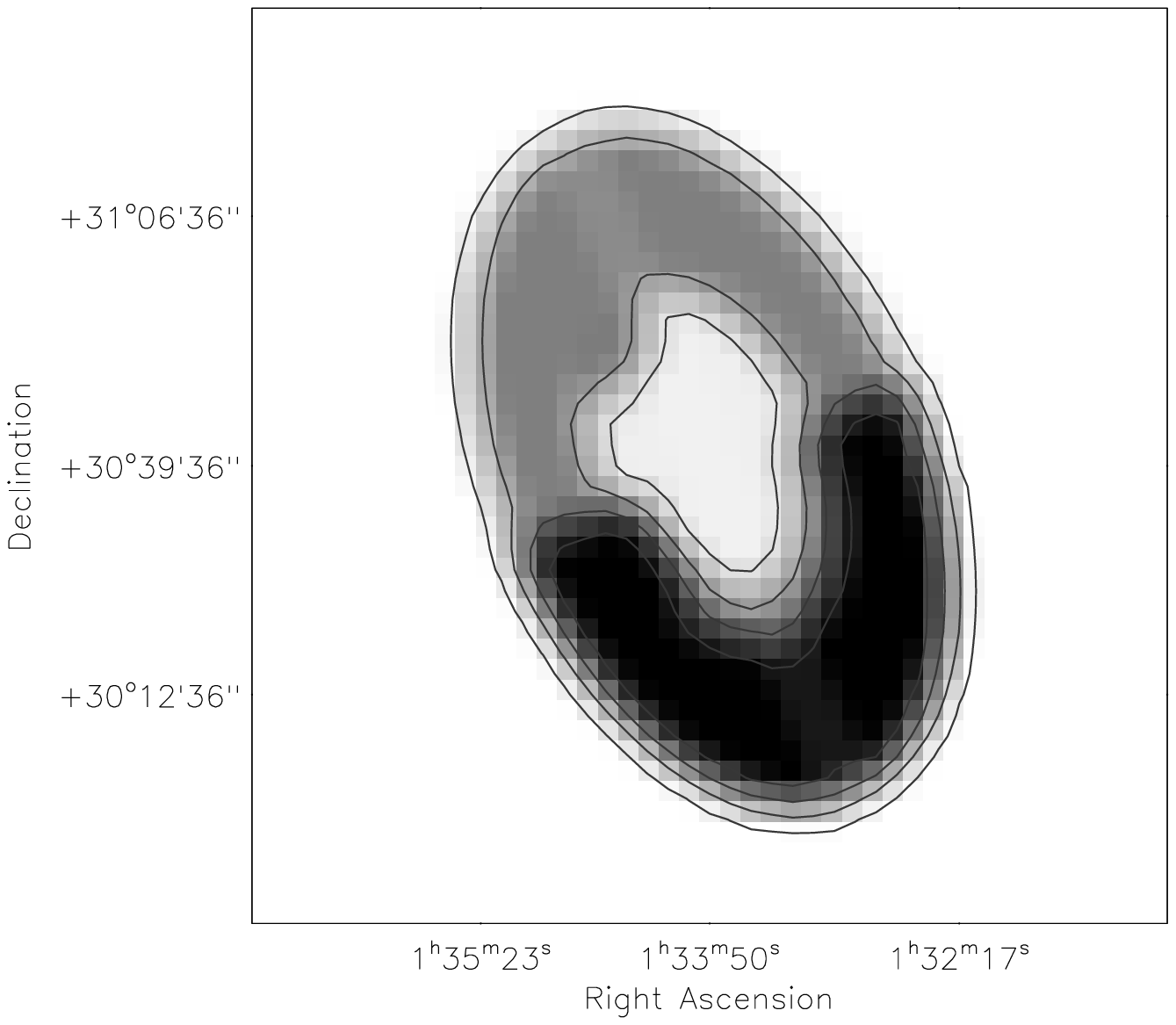}

\hspace{-0.6cm}
\vspace{-0.1cm}
\epsfxsize=0.24\hsize \epsfbox{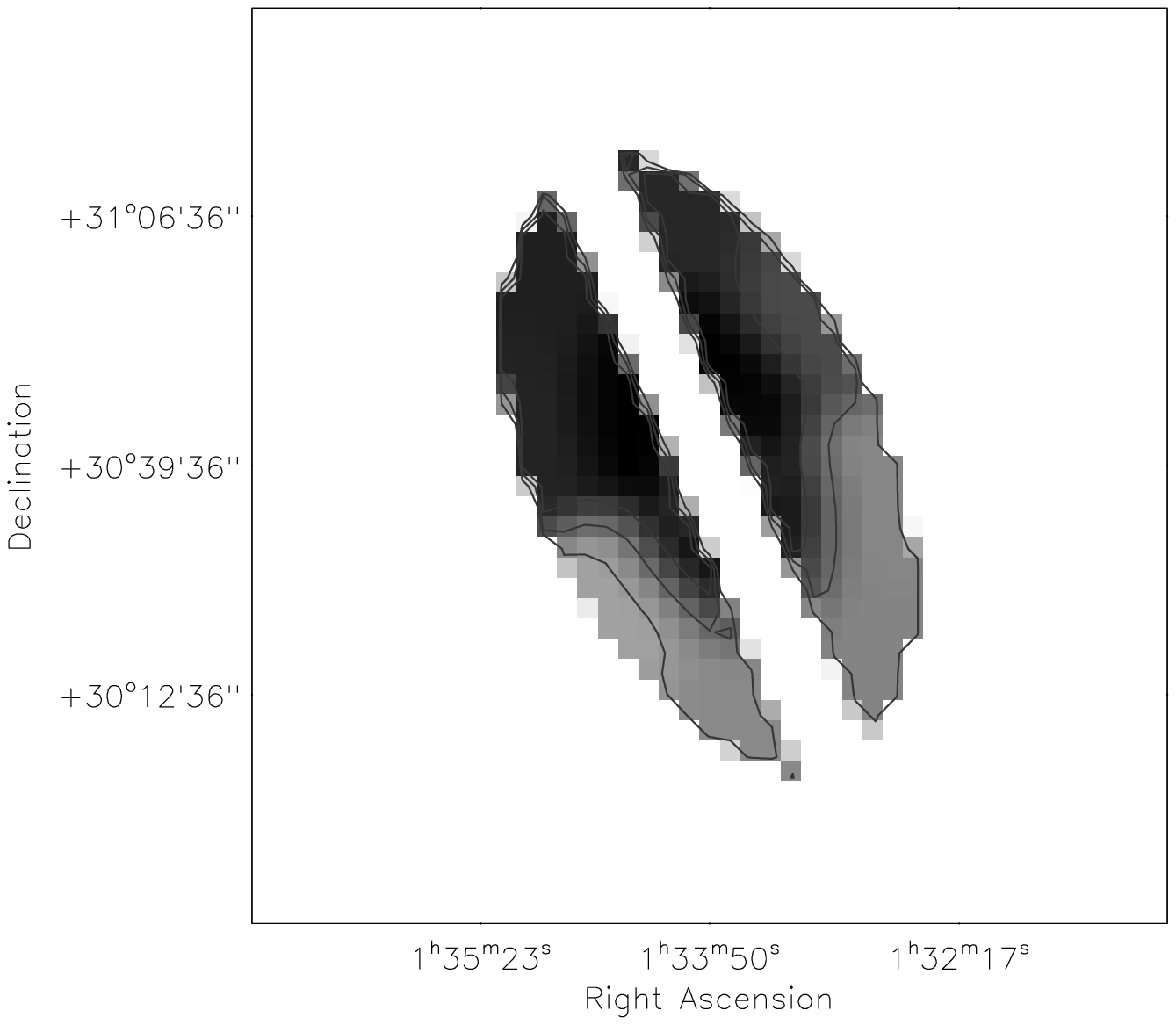}
\hspace{-1.0cm}
\epsfxsize=0.24\hsize \epsfbox{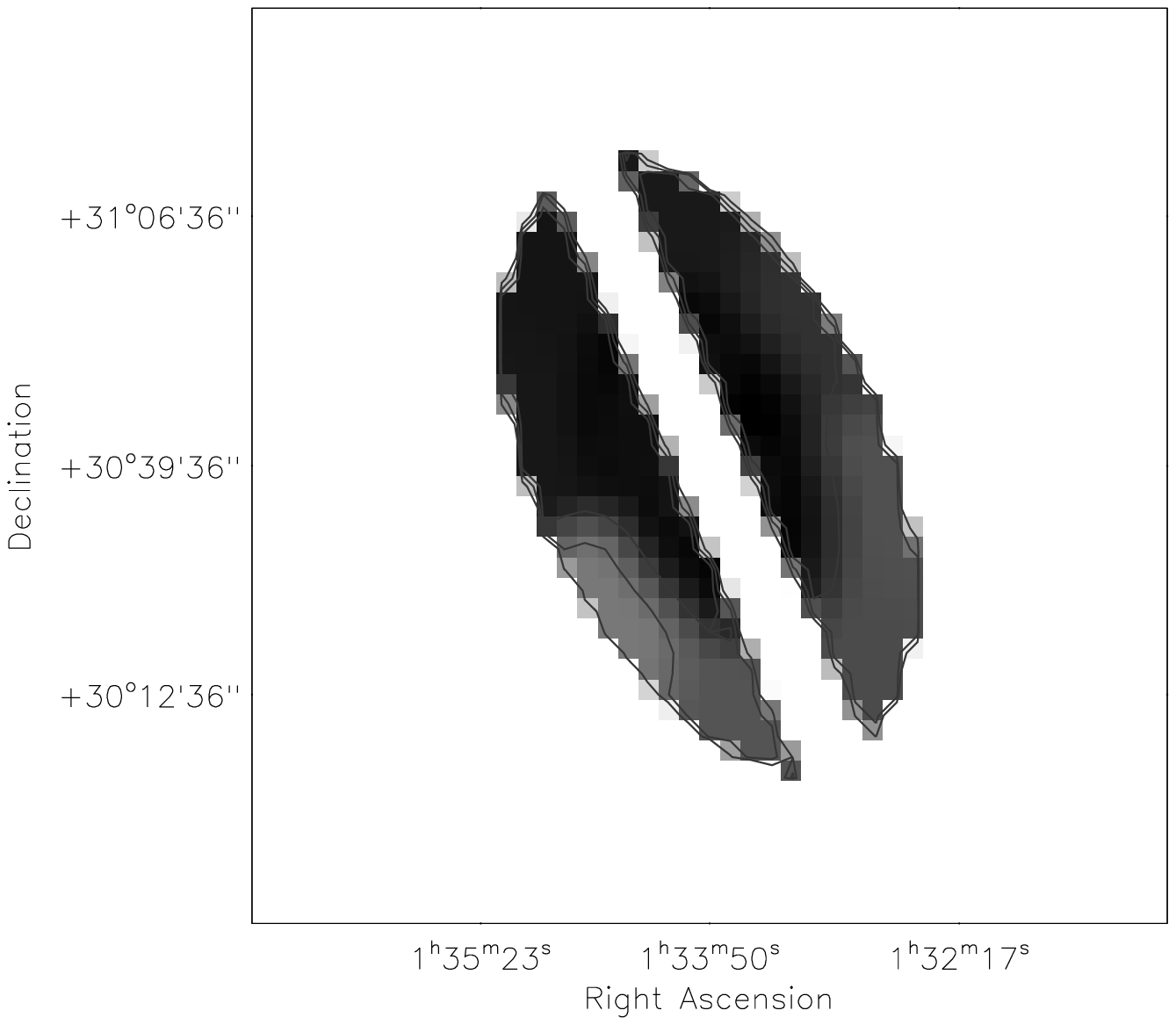}
\hspace{-1.0cm}
\epsfxsize=0.24\hsize \epsfbox{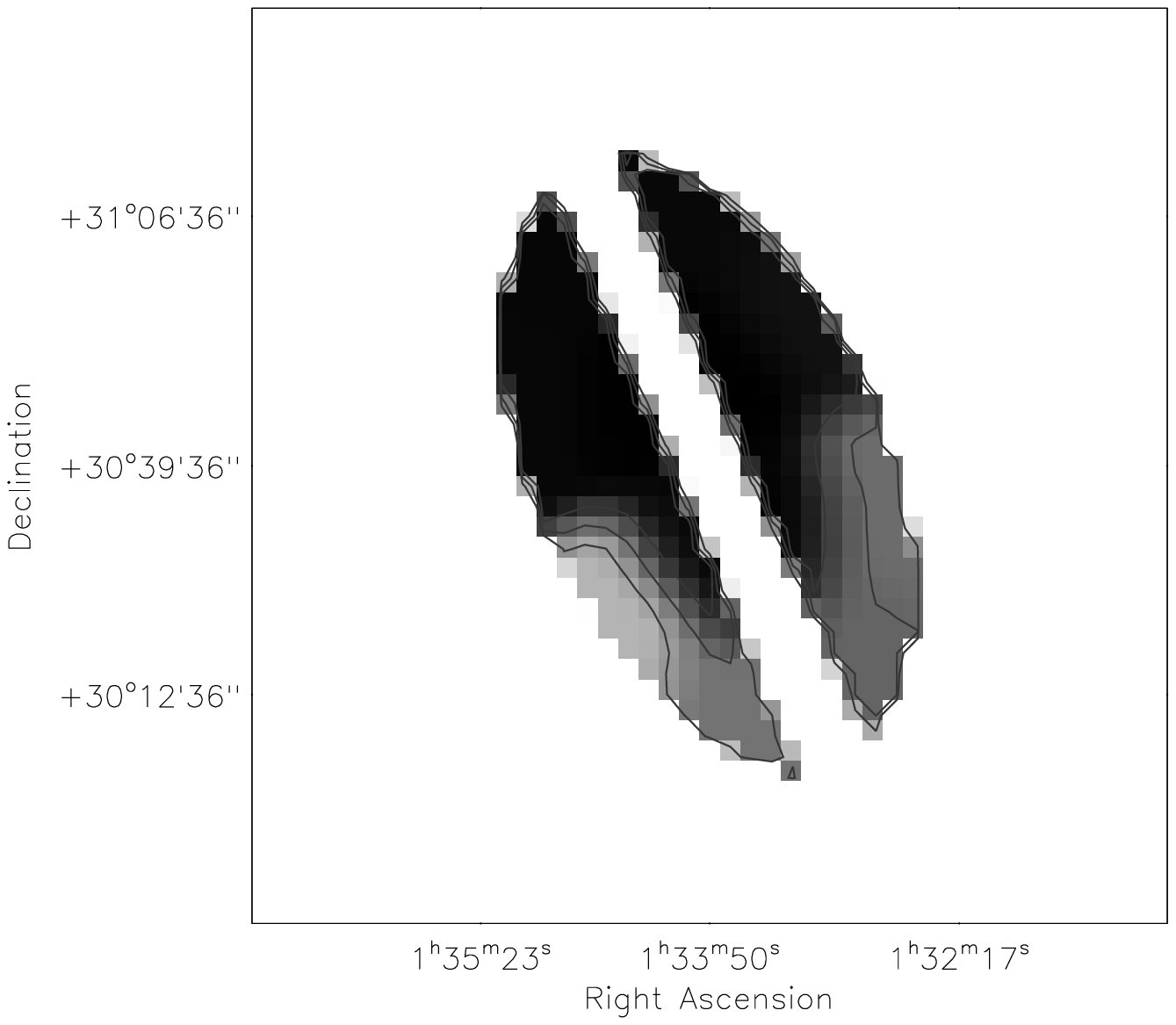}
\hspace{-1.0cm}
\epsfxsize=0.24\hsize \epsfbox{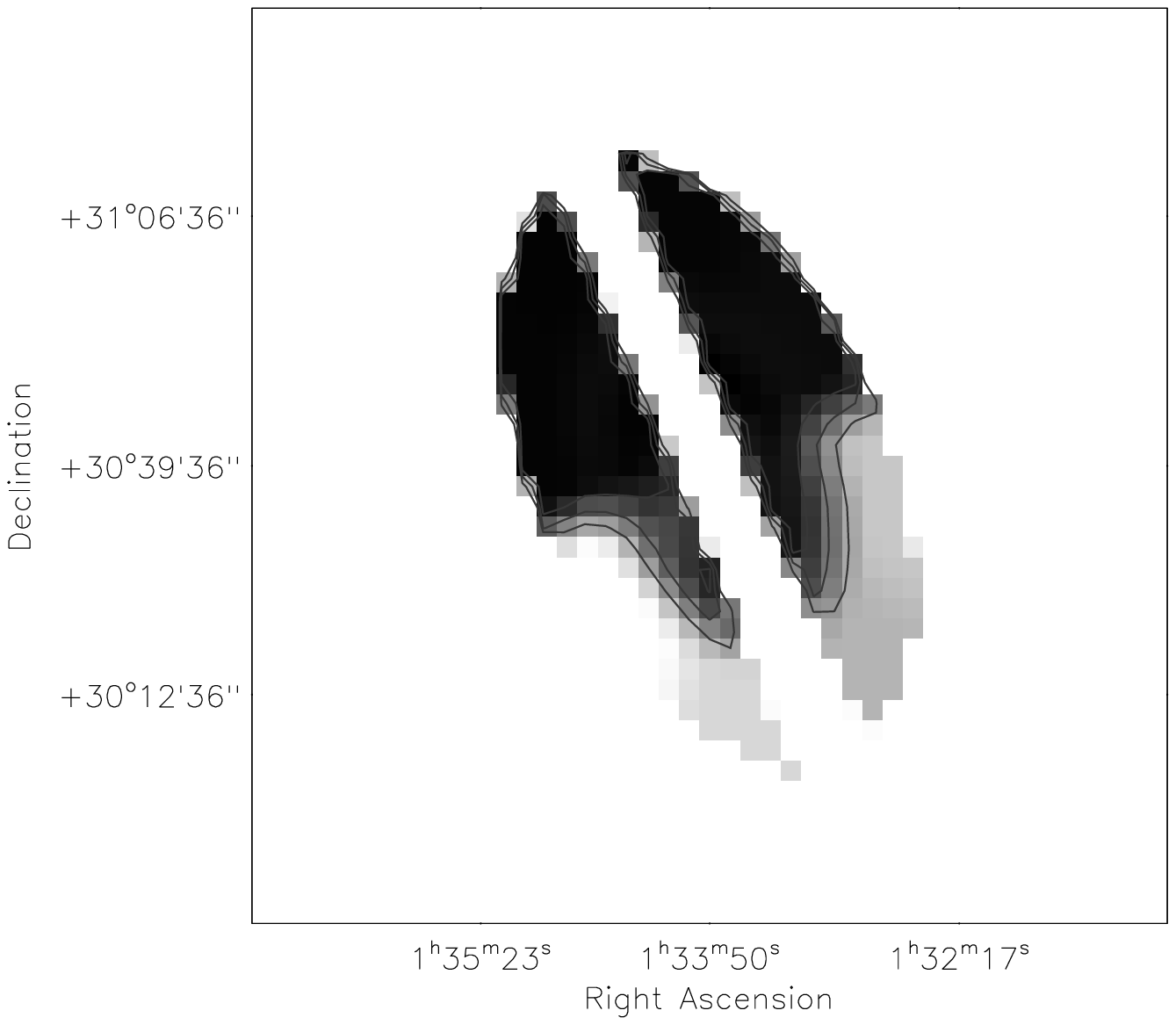}
\hspace{-1.0cm}
\epsfxsize=0.24\hsize \epsfbox{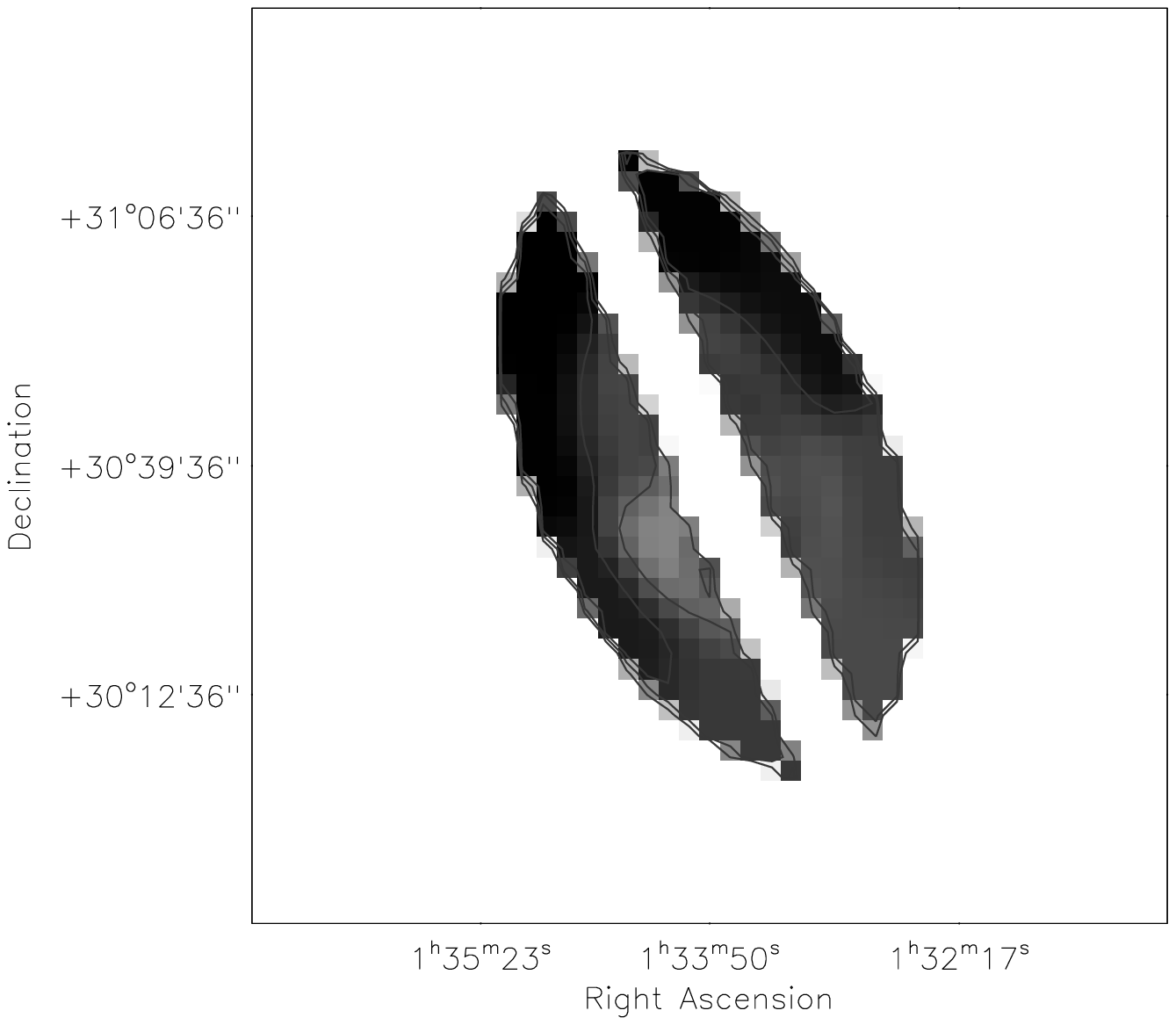}

\hspace{-0.6cm}

\caption{Surface distribution of the most probable metallicity ({\it
  first and third rows)} at a given SFR (or mean age) and the
  associated probability distribution ({\it second and fourth rows})
  obtained by fitting the $K_{\mathrm s}$ magnitude distribution of C
  stars. C stars have been selected from the near-infrared
  colour-magnitude diagram (Fig.~\ref{cmd}) using vertical lines ({\it
  top two rows}) and using slanted lines ({\it bottom two rows}).
  {\it From left to right each column} refers to a mean age of $2$,
  $3.9$, $6.3$, $8.7$ and $10.6$ Gyr, respectively.  Contours for each
  panel in the {\it top row} are at Z $= 0.00048$ (panel 1); $0.00048$,
  $0.0005$, $0.0006$ and $0.0007$ (panels 2 and 3); $0.00048$,
  $0.0005$, $0.0006$, $0.0008$, $0.0012$ and $0.002$ (panel 4);
  $0.001$, $0.003$, $0.005$ and $0.007$ (panel 5). Contours for each
  panel in the {\it third row} are at Z $= 0.00048$, $0.0005$, $0.0006$ and
  $0.0007$ (panels 1, 2, 3, 4); $0.00048$, $0.0005$ and $0.001$ (panel
  5). The grey scale for the probability distributions shows only
  values above $0.7$ (contours are at $0.85$, $0.95$ and $0.99$) and
  $0.95$ (contours are at $0.97$, $0.98$ and $0.99$) for the {\it
  second} and {\it fourth} rows respectively. Darker regions
  correspond to higher values.}
\label{sep}
\end{figure*}

Figure \ref{lumfex} shows examples of the fit of the observed
distributions of C and M stars with theoretical distributions obtained
from a given model. In this case the observed number counts refer
to a sector of a ring while models corresponds to a SFR with $\alpha =
5$ (or a mean age of $8.7$ Gyr) and metallicity Z $=0.0005$ or Z
$=0.001$. In the top panel the interpretation of C stars is good at
the $99$\% level and it reduces to $84$\% for an increasing
metallicity, for M stars it is rather poor in both cases
($0-1$\%). The probability of fitting the same observed distribution
of C stars with the whole range of model distributions examined in
this study is shown in Fig.~\ref{fitunique}.  The point corresponding
to the highest probability indicates that the overall population,
within this sector, is metal poor (Z $=0.0005$) and with a mean age of
$\sim 6$ Gyr. This figure also shows the level of uncertainty
associated to the metallicity and age quantities. For example: Z
$=0.0005$ gives systematically better fits at any age while the
difference between a $2$, $4$ or $6$ Gyr mean age does not seem
sufficiently strong.  Note that this is just an example for one
sector of one ring, and the uniqueness of a model fit across the whole
galaxy can therefore be estimated from the probability maps
(Fig.~\ref{sep}).   The similarity of some of the maps shown
suggests that the difference in age is not robustly determined or on
the contrary that there is a large spread in mean age.

\begin{figure*}
\epsfxsize=0.38\hsize \epsfbox{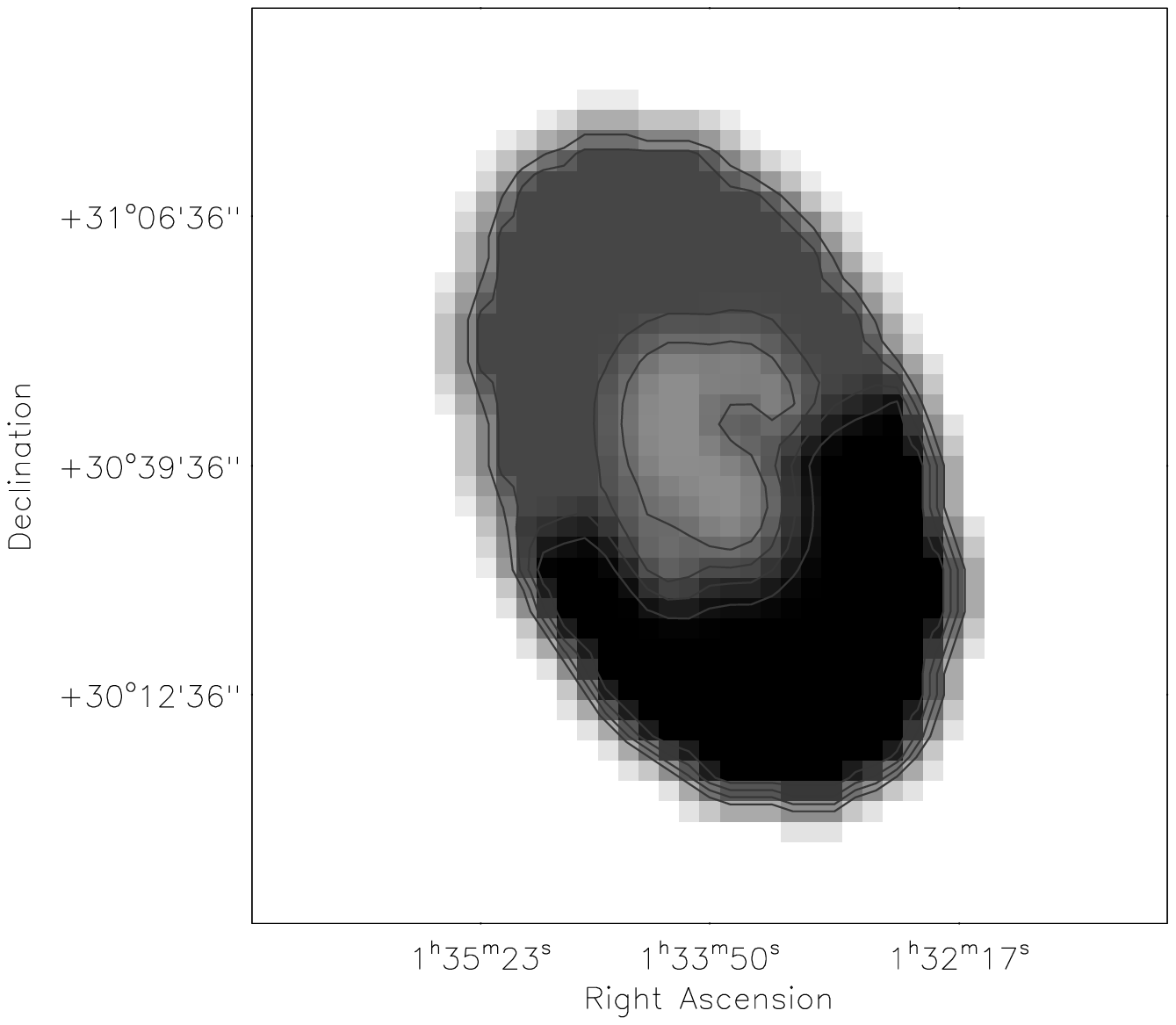}
\hspace{-1cm}
\epsfxsize=0.38\hsize \epsfbox{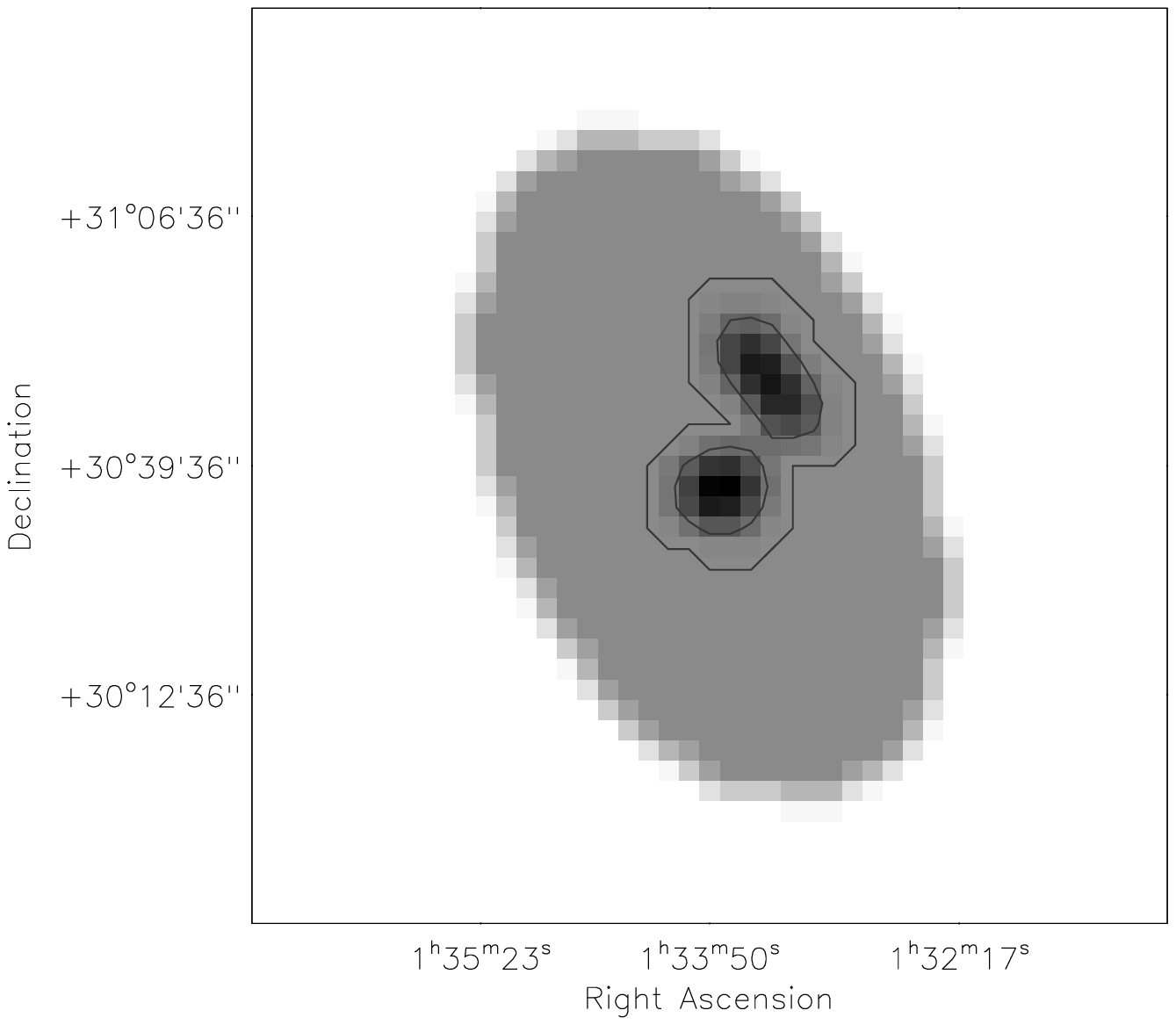}
\hspace{-1cm}
\epsfxsize=0.38\hsize \epsfbox{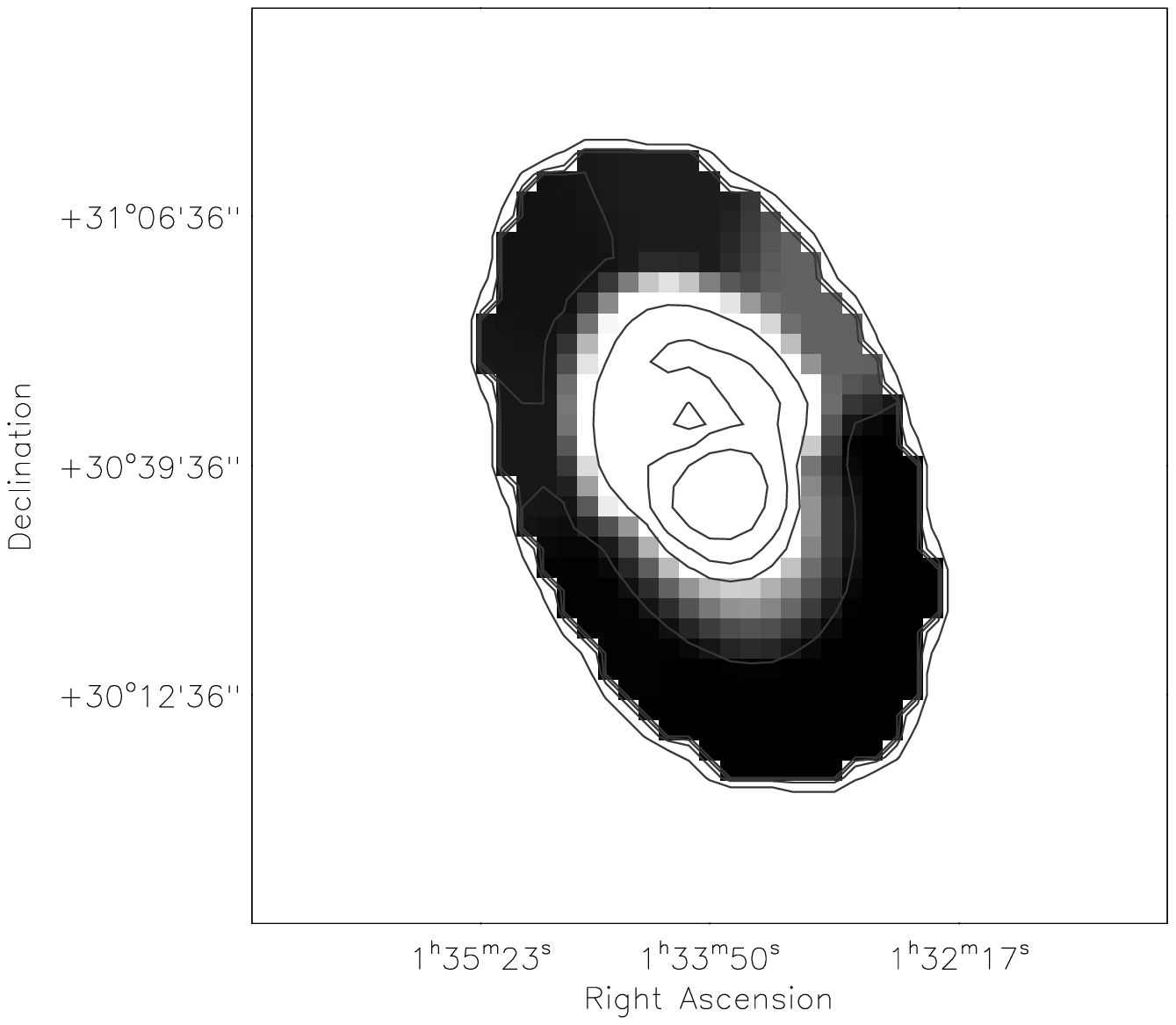}

\epsfxsize=0.38\hsize \epsfbox{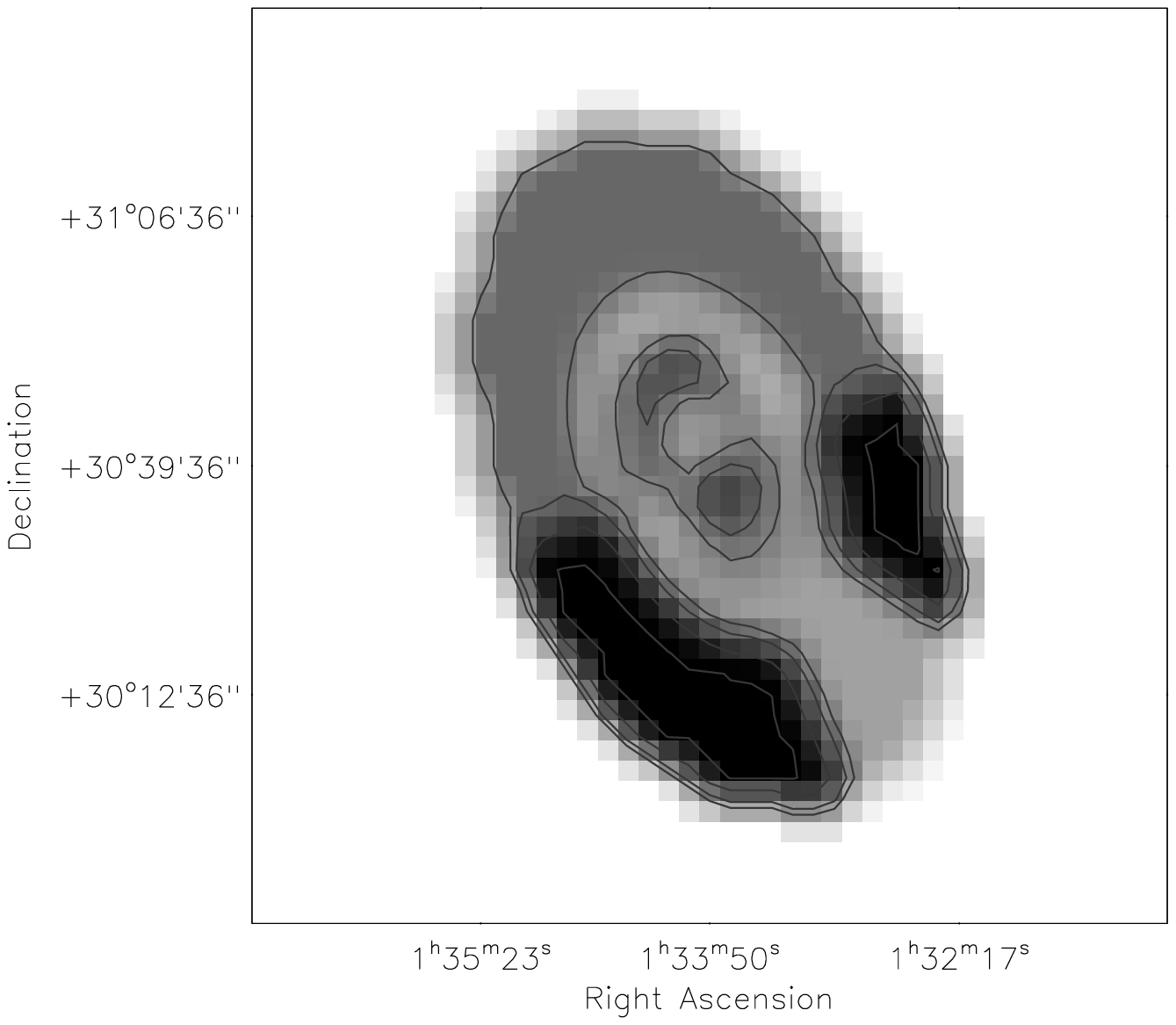}
\hspace{-1cm}
\epsfxsize=0.38\hsize \epsfbox{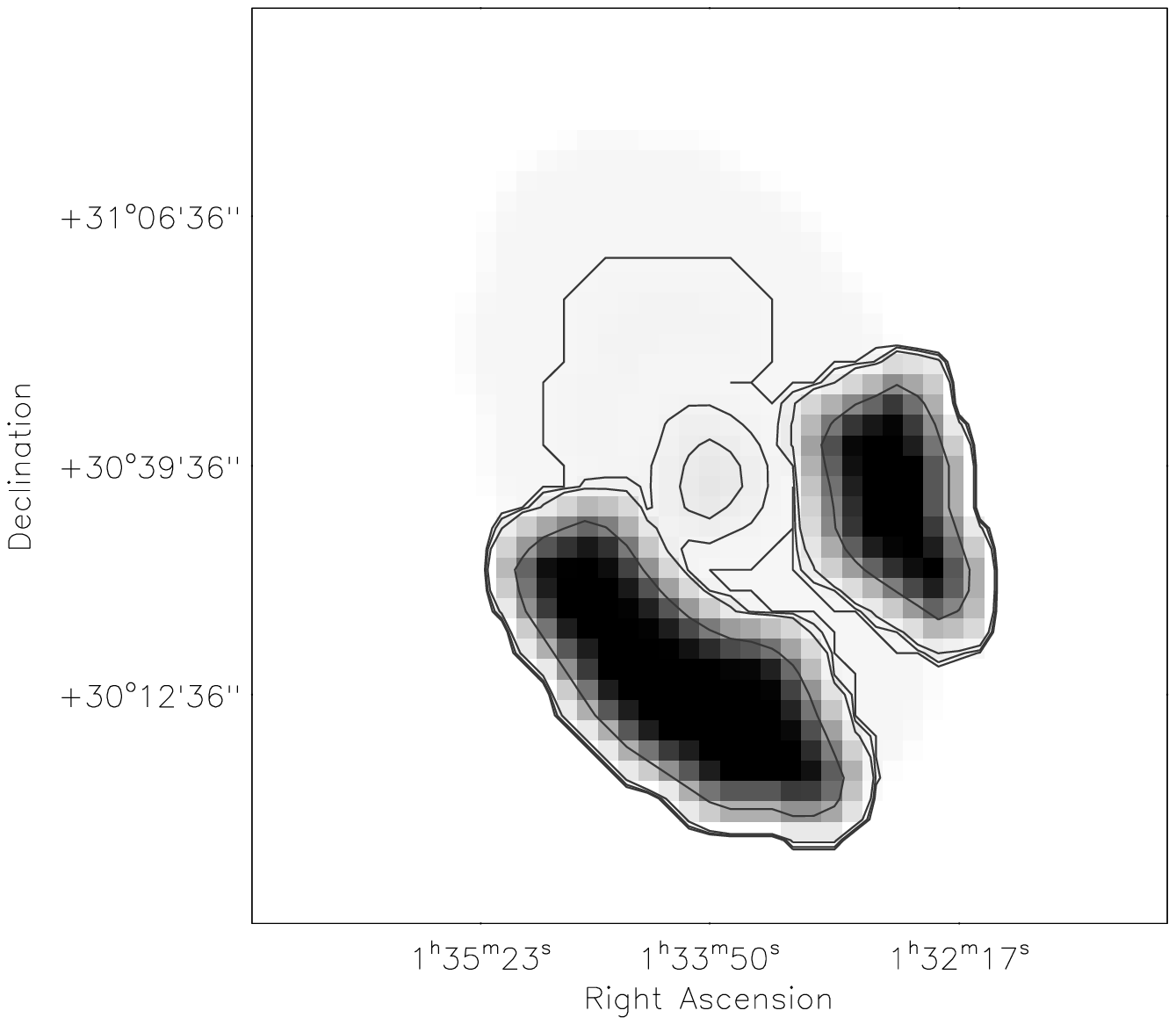}
\hspace{-1cm}
\epsfxsize=0.38\hsize \epsfbox{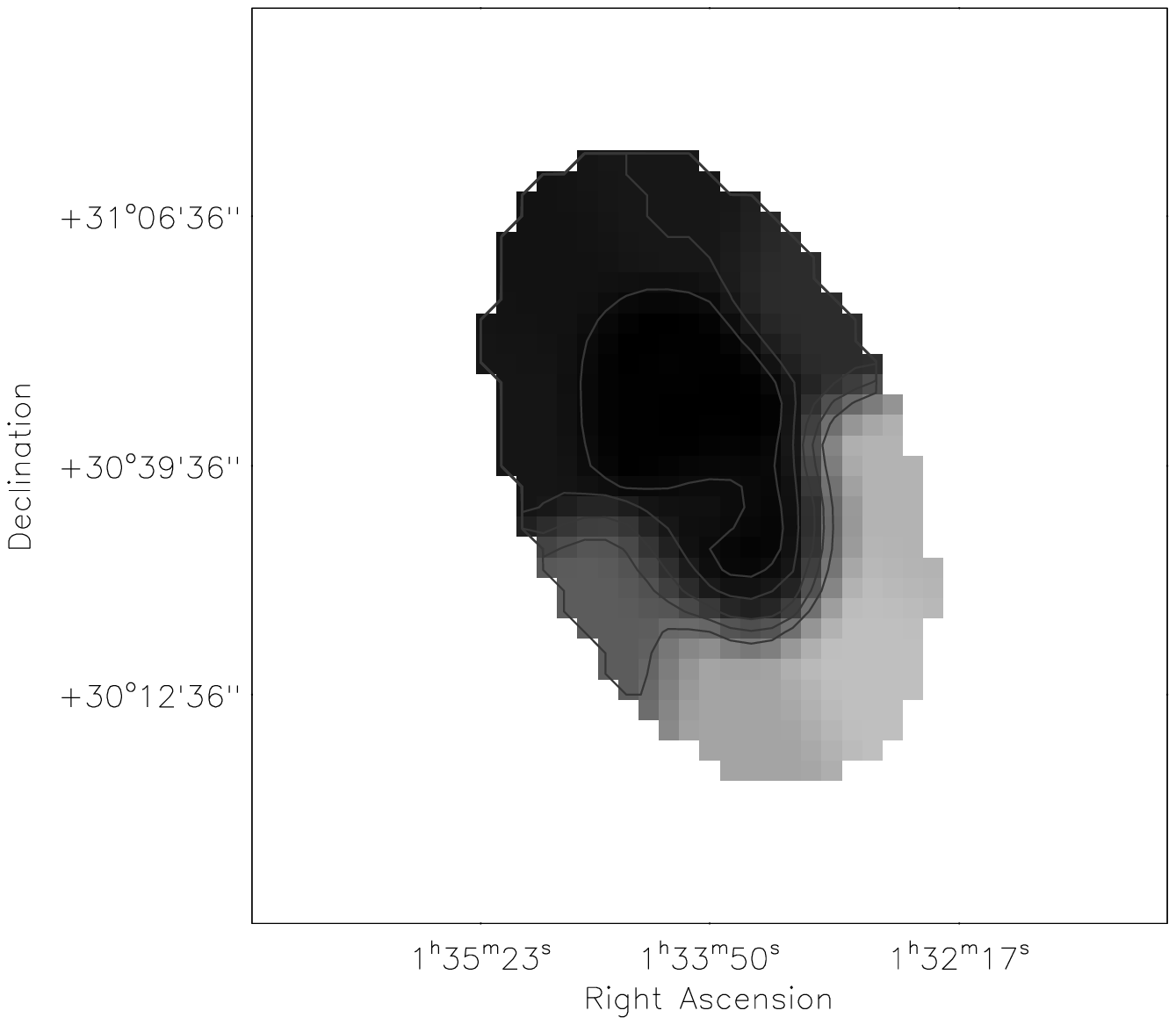} 
\caption{Spatial distributions of the mean-age of the stellar
population of M33 (left), of the metallicity (middle) and of the
statistical probability that expresses the confidence level of the
previous distribution. These distributions have been constructed from
the comparison between the observed $K_{\mathrm s}$ magnitude
distribution of C-type AGB stars with theoretical distributions. C
stars have been selected from the colour-magnitude diagram
(Fig.~\ref{cmd}) using vertical lines ({\it top row}) or slanted lines
({\it bottom row}).  Bins are of $2.4^{\prime}$ and dark regions
correspond to high numbers.  {\it From left to right} contours are at: 
$5$, $6$, $7$, $8$ (top) and $5.5$, $6.5$, $8.5$, $10.5$ (bottom) Gyr
for age; $0.0005$, $0.0006$ (top) and $0.0005$, $0.0006$, $0.0008$,
$0.004$ (bottom) for metallicity where only values above $0.0002$ are
displayed; $0.992$, $0.994$, $0.996$, $0.998$, $0.999$ where only
values above $0.98$ are displayed (top) and $0.85$, $0.95$, $0.98$,
$0.999$ where only values above $0.98$ are displayed (bottom) for
probability.}
\label{combc}
\end{figure*}

\begin{figure*}
\epsfxsize=0.38\hsize \epsfbox{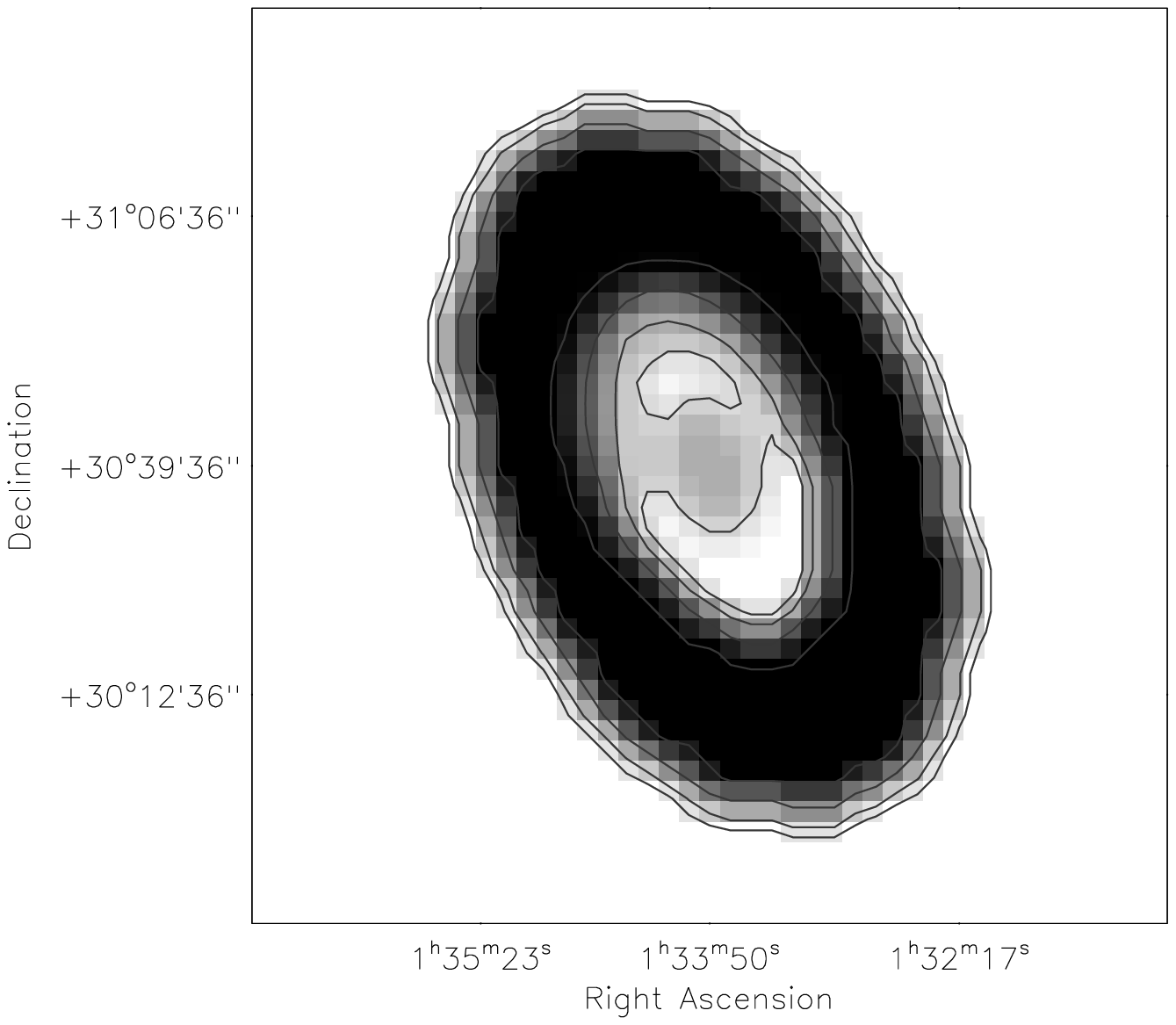}
\hspace{-1cm}
\epsfxsize=0.38\hsize \epsfbox{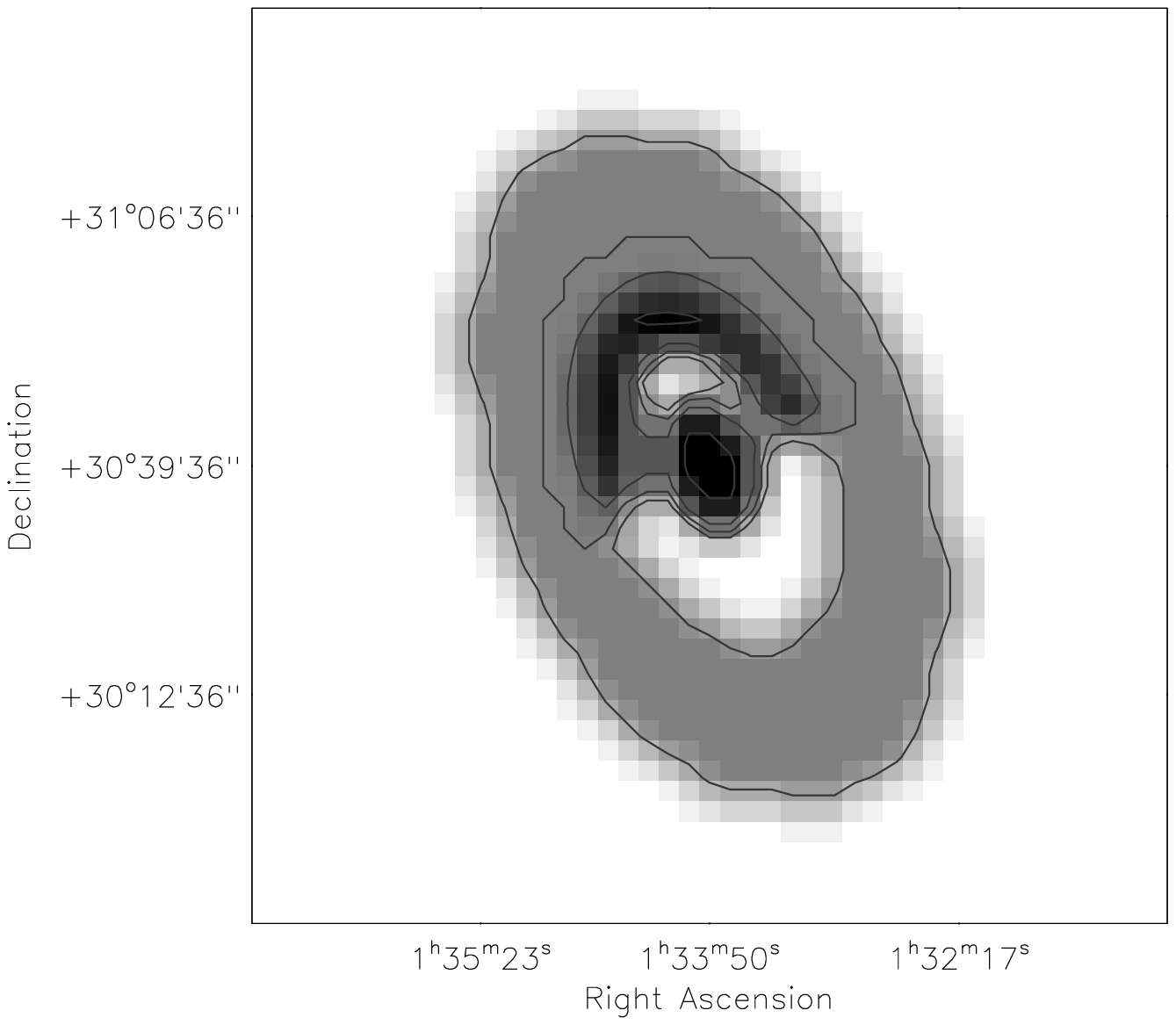}
\hspace{-1cm}
\epsfxsize=0.38\hsize \epsfbox{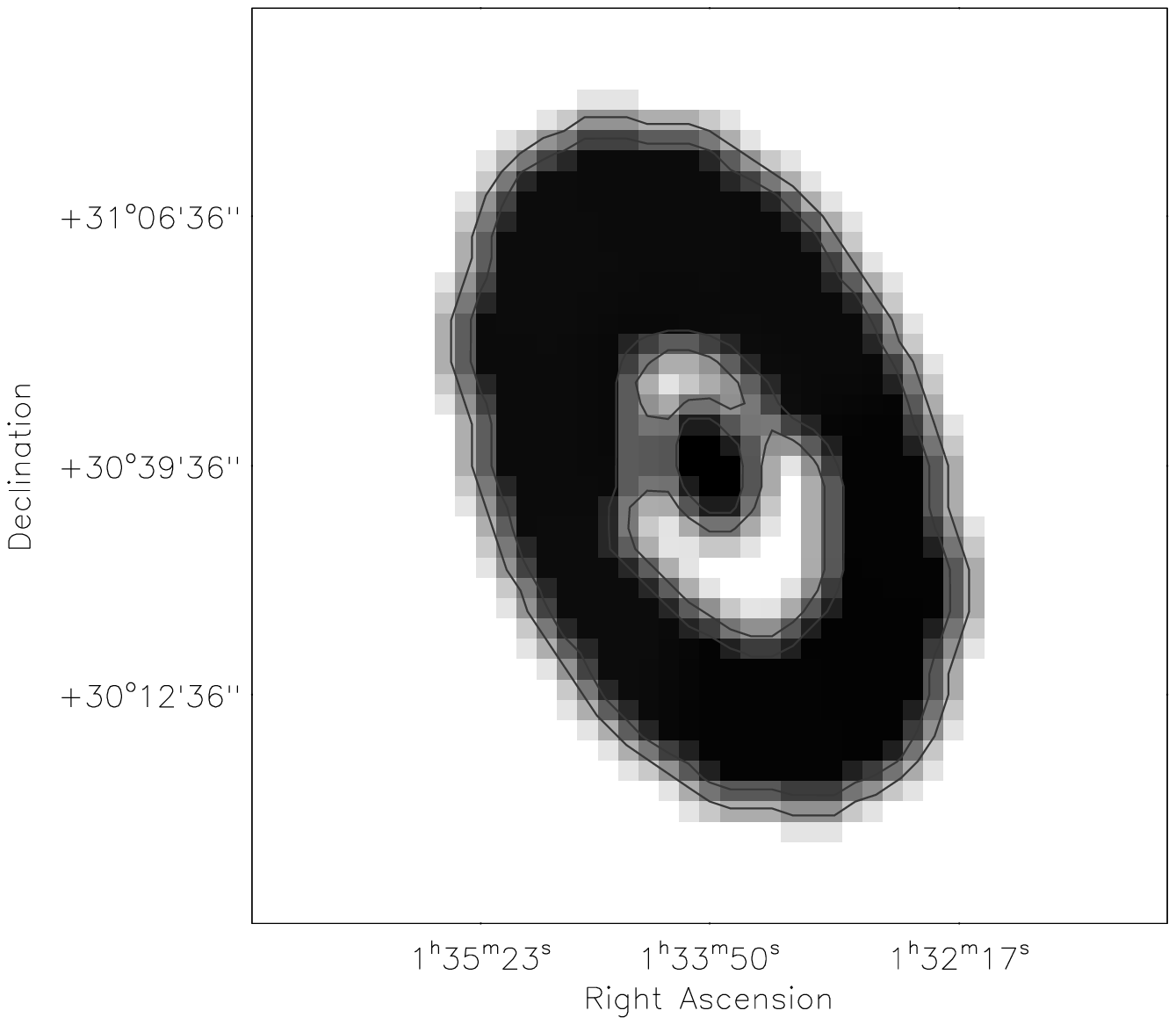}

\epsfxsize=0.38\hsize \epsfbox{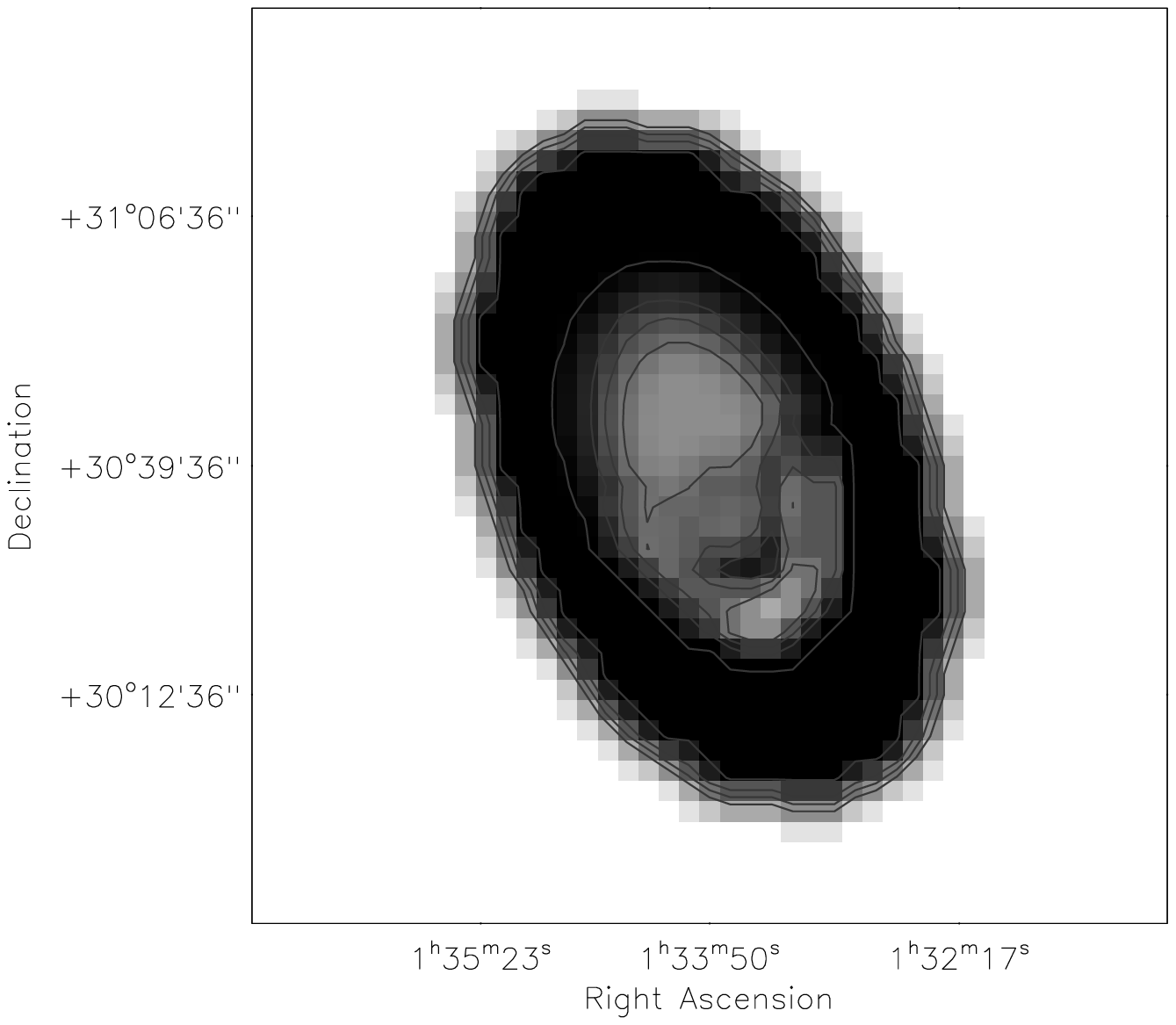}
\hspace{-1cm}
\epsfxsize=0.38\hsize \epsfbox{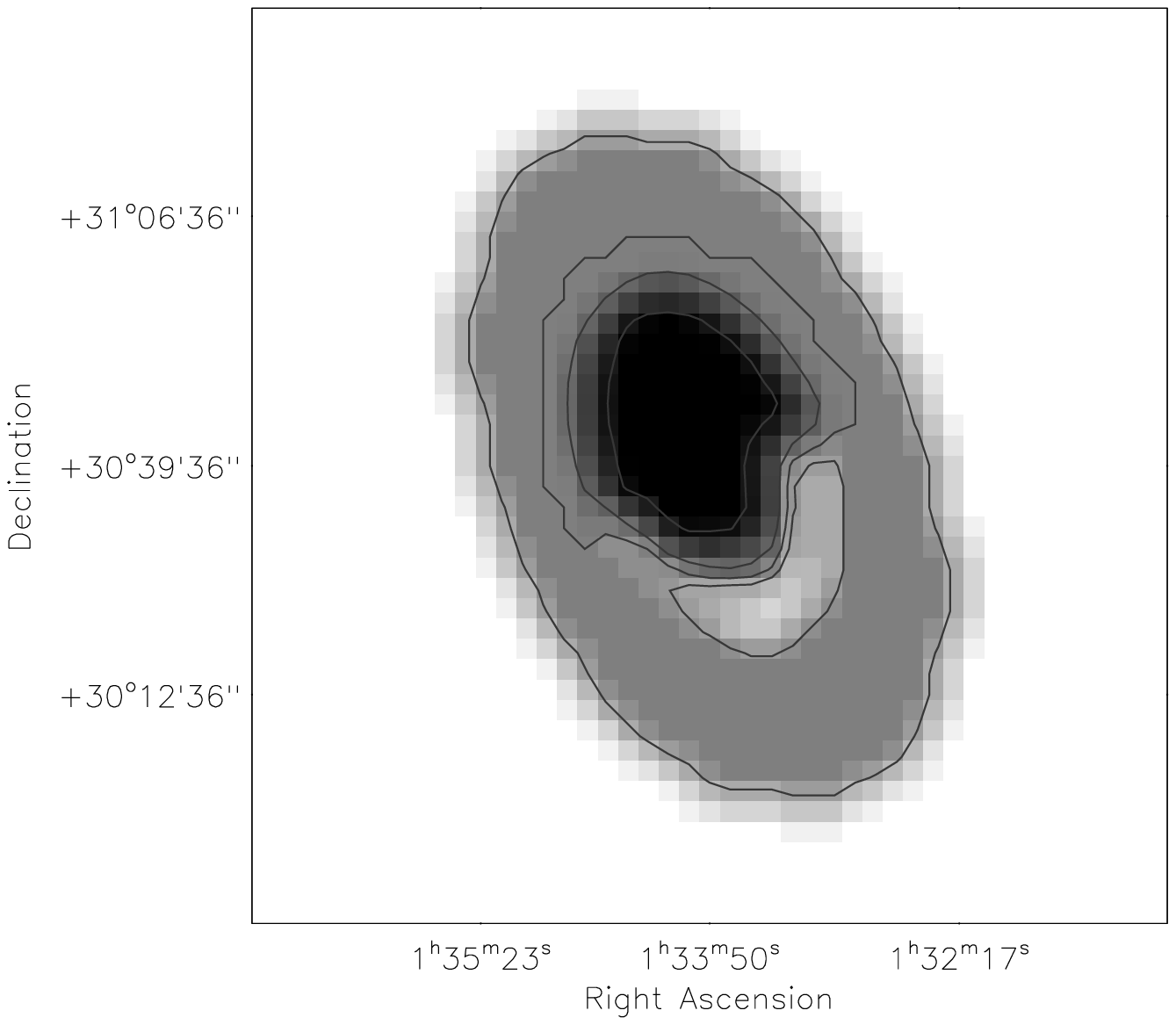}
\hspace{-1cm}
\epsfxsize=0.38\hsize \epsfbox{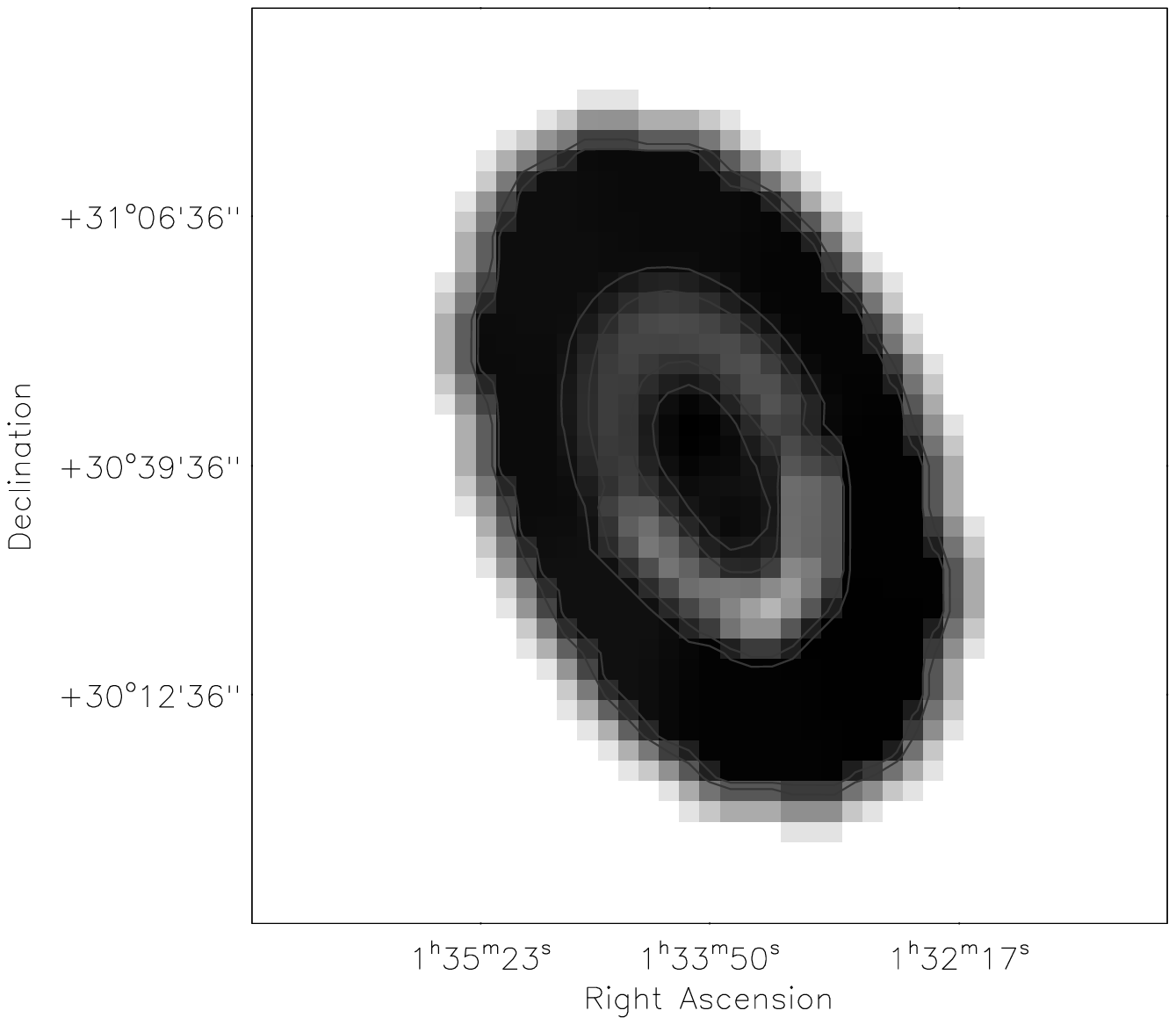} 
\caption{Spatial distributions of the mean-age of the stellar
population of M33 (left), of the metallicity (middle) and of the
statistical probability that expresses the confidence level of the
previous distributions (right). These maps have been constructed from
the comparison between the observed $K_{\mathrm s}$ magnitude
distribution of M-type AGB stars with theoretical distributions. M
stars have been selected from the colour-magnitude diagram
(Fig.~\ref{cmd}) using vertical lines ({\it top row}) or slanted lines
({\it bottom row}).  Bins are of $2.4^{\prime}$ and dark regions
correspond to high numbers. {\it From left to right} contours are at:
$1$, $2$, $4$, $6$ (top) and $5$, $6$, $7$, $8.5$ (bottom) for age;
$0.0065$, $0.008$, $0.01$, $0.015$ (top \& bottom) for metallicity;
$0.5$, $0.8$ (top) and $0.8$, $0.9$ (bottom) for probability.}
\label{combm}
\end{figure*}

The distribution of metallicity separately for different SFRs and the
probability that a given model represents the observed distribution of
C stars, selected with both criteria, are shown in Fig.~\ref{sep}. In
order to create each of the maps shown, first, we made a grid of
$13,395$ points with coordinates $-0.47^{\circ}\le x\le 0.47^{\circ}$
and $-0.7^{\circ}\le y\le 0.7^{\circ}$, in the plane of the galaxy,
equally spaced with a step of $0.01^{\circ}$. Then, we assigned to
each point the quantity (age, metallicity and likelihood -- the
probability of getting a $\chi^2$ value) accordingly to which sector a
point belongs. We re-binned the distribution of values in bins equal
to $0.04$ (this corresponds to a resolution of $2.4^{\prime}$),
smoothed the intensity with a $2\times2$ box car function and restored
the orientation of the galaxy in the sky. Finally, we constructed
greyscale maps where darker regions correspond to higher numbers.
Similar maps were created also for the distribution of O-rich AGB
stars, however, these individual maps are not shown here while
Fig.~\ref{combc} and \ref{combm} show combined maps for both AGB
spectral types. These maps were obtained by choosing the theoretical
distributions of AGB stars that correspond to the smallest $\chi^2$
value among those generated from each combination of SFR and
metallicity, for a given sector of a given ring.  Note that due to the
approximations involved in building-up the theoretical distribution
absolute values of mean age and metallicity should be taken with
care. Much more important are their variations across the galaxy.

\subsubsection{Distribution of metallicity versus mean-age}
The maps resulting from almost all different cases of SFR and AGB
selection criteria shown in Fig.\ref{sep} suggest that the metallicity
in the centre of the galaxy is different from the metallicity in the
outer regions. In particular, the best metallicity that fits the
overall disc of M33 is the lowest explored in this study which
suggests a stellar population metal poorer than Z$=0.0005$ or at least
as metal poor as [M/H]$=-1.6$ dex assuming Z$_{\odot}=0.02$
(Fig.~\ref{sep}, \ref{combc}) and using the conversion
[M/H]$=log($Z$/$Z$_{\odot})$ where [Fe/H]$\le$[M/H]. Approaching the
centre of the galaxy there are well defined and relatively small
regions rich in metals ([M/H]$=-1.2$ dex).  These regions change
location depending on the SFR and AGB selection criterion
considered. The time sequence shown in the top row of Fig.~\ref{sep}
shows that both for a very old and a very young age a metal-rich
nucleus is present and it is slightly displaced from the centre of the
galaxy. However, this is the region where the data are the least
reliable as shown from the probability maps. At other intermediage
ages an additional metal-rich clump is located NW of the nucleus. The
metal-rich spiral pattern corresponding to a mean age of $8.7$ Gyr
does not have to be associated with the NW bump. It is popular to
speculate that a structure has been accreted by M33 in the past and is
now well mixed with the disc population. The variation of mean
metallicity perhaps suggests the propagation of the star formation
with time. The comparison with maps shown in the third row of
Fig.~\ref{sep} should guide the reader on the differences due to the
different way AGB stars can be selected from the near-infrared
CMD. Here, the metal-poor disc is also recovered while the metal-rich
internal structure describes a broken ring in the Northern part of the
galaxy, a minor clump in the centre and for the oldest ages a wide
metal-rich area encompassing the outer Southern region of the galaxy
prior to a clump enriching in metals from the centre to the SE. The
probability that the latter is a real effect is not very high as well
as for the values recovered in a stripe across the major axis of the
galaxy. There is currently no explanation on why the major axis
appears as a critical region in most of the maps.

\subsubsection{Combined distributions of mean-age and metallicity across M33}

Figure \ref{combc} shows the combination of the individual maps of
Fig.~\ref{sep} corresponding to the lowest $\chi^2$ for each sector of
each ring and pair of metallicity and SFR parameters. Taking the
values that correspond to the highest probability (lowest $\chi^2$)
represent a first order approximation to the average parameters of the
stellar population at a given location within the galaxy. In some
cases, e.g. Fig.~\ref{fitunique}, a combination of those values with
high probabilities may be more appropriate. Contrary to
individual maps, combined maps do not show a region of low statistical
significance around the galaxy major axis. This is because at each
point in the map the most reliable metallicity is chosen and
grayscales and countours are adapted accordingly.  The lowest contour
of the metallicity distribution for both selection criteria confirms a
disc population metal poorer than [M/H]$=-1.6$ dex. Regions with an
associated higher metallicity are often those which are less reliable
(have larger values of $\chi^2$); the reliability is higher if C stars
are selected using the {\it vertical line} criterion. The nucleus, as
seen before, is as metal rich as [M/H]$=-1.2$ dex as well as a region
NW of it.

The distribution of mean age is quite similar between both selection
criteria. It shows a broad outer ring which is older ($\sim6$ Gyr)
than the region within it, except perhaps for the nuclear region. If C
stars are selected using the {\it slanted-lines} criterion both the
centre and a small region NE of it are as metal rich as the outer
ring. On the other hand, if the other criterion is used then the broad
old ring is slightly wider while the centre is the metal poorest. It
also appears that the Southern parts of the outer ring are older than
the Northern parts of it. This apparent asymmetry might be a
residual from the correction for the orientation and extinction within
the galaxy (Sect.3.3.2). We have taken a convervative approach of
averaging the variation obtained from colours and magnitudes across
the galaxy. If only the azimutal variation in the $K_{\mathrm s}$ band
were considered then stars in the North would be fainter than stars in
the South -- the peak of the sinusoidal variation would be at $0^{\circ}$
for an amplitude up to $0.05$ mag (slightly larger than the one
adopted in the correction).  These new parameters, however, would
also affect the distribution of metallicity, this might be true for
the {\it slated-lines} criterion but it is not the case for the other
criterion. It is possible that including fainter and bluer C-rich
stars that might be instead O-rich reveals a N-S asymmetry. This
feature has therefore a low statistical significance.

Overall the significance level of both the distribution of mean age
and metallicity across the surface of the galaxy is high ($>98$\%)
which indicates that there is a theoretical model (a combination of
metallicity and SFR) that describes the observed $K_{\mathrm s}$--band
distribution rather well.  Simply by comparing the probability
distributions obtained from the two selection criteria adopted it is
evident that selecting C stars using {\it vertical lines} produces a
sample that is better fit by theoretical distributions across the
galaxy (see also Fig.~\ref{sep}). This criterion is perhaps
conservative because it excludes faint AGB stars, known to exist in
mixed stellar populations, but it does show the distribution of the
bulk of the C star population of this galaxy. Therefore we regard the
top row of Fig. \ref{combc} as trustworthy in the interpretation of
age and metallicity variations across M33 as derived from C stars.

Similar maps derived from the distribution of M-type stars are shown
in Fig.~\ref{combm}. The most reliable results are obtained for the
outer ring and for the centre. It is well confirmed that the outer
ring is old ($6-8.5$ Gyr depending on selection criteria) and metal
poor although the latter corresponds to a much higher metallicity than
that derived from C stars. These maps suggest that the centre of the
galaxy is metal rich and younger which is also in agreement with the
pattern obtained from C stars. It is also possible that North of the
centre another broken ring-like structure supports a higher
metallicity while in the South the same structure will be metal
poor. The latter is the least constrained result. Note that the fit of
the sample of M stars obtained using the {\it slanted-lines} criterion
corresponds to a higher probability than the fit obtained from a
sample selected using the {\it vertical-lines} criterion. Figure
\ref{cmd} shows that the latter isolates the bulk of the M star
population of M33, this is perhaps the reason why, despite the current
uncertainties on the theoretical representation of M stars, the fits
are more reliable although the sample is not complete.  The
significance level of both the distribution of mean age and
metallicity is below the level obtained from C stars. This was also
the case in Cioni et al.~(\cite{lf}) for a similar study of the
Magellanic Clouds. It is possible that the models fail to interpret
correctly the distribution of M stars or that the model parameters
were not sufficiently explored. Very recently Marigo et
al.~(\cite{magi07}) show that improving the lifetime of M-type stars
alleviates the problem of underestimating the stellar populations. In
the subsequent discussion more emphasis is placed on C stars.

Summarizing, the most reliable interpretation of the $K_{\mathrm s}$
magnitude distribution of C and M stars is obtained by selecting
reliable, whilst not complete, sample of stars using the {\it
vertical-lines} criterion for C stars and the {\it slated-lines}
criterion for M stars. A complete sample can only be obtained via
spectroscopic observations of the entire AGB population of M33 (see
Groenewegen et al.~\cite{} for the Magellanic Clouds).

\section{Discussion}
\label{dis}

\subsection{Distance to M33}
Several authors have measured the distance to the M33 galaxy using
different stellar indicators. In particular, Bonanos et
al.~(\cite{bono}) obtains a distance modulus of $24.92\pm0.12$ to a
detached elipsing binary, Sarajedini et al.~(\cite{sara}) obtains
$24.67\pm0.08$ using RR Lyrae stars located in two fields NW and SE of
the galaxy centre. Using the tip of the RGB method: Galletti et
al.~(\cite{gall}) obtains $24.64\pm0.15$ in a field in the outskirts
of the galaxy, Tiede et al.~(\cite{ti04}) obtains $24.69\pm0.07$ in a
different halo field while 
%McConnachie (\cite{thesis}) obtains
%$20.57\pm0.03$ in a diametrically opposite field, and 
Kim et al.~(\cite{km02}) obtains $24.81\pm0.04$ from the average of 10
independent fields distributed throughout the galaxy, mostly in the
outer halo but two in the centre. The latter also provide individual
measurements for each of these fields. There are many other
measurements that have not been cited here and this is because they
were obtained from indicators distributed througout the galaxy and
provided only one measure of the distance to the galaxy without
investigating its variation across it.  The point here is not to
derive a new distance to M33 but to show that inconsistencies among
previous measurements may be explained by accounting for the
orientation and extinction of the disc.

\begin{figure}
\resizebox{\hsize}{!}{\includegraphics{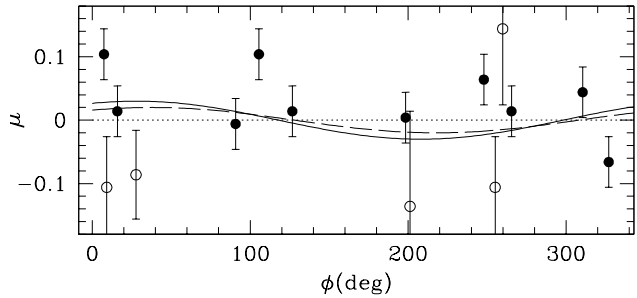}}
\caption{Distribution of the difference between the individual
  distance moduli derived by Kim et al~(\cite{km02}; {\it filled
  circles}) and other authors (see text; {\it empty circles}) and
  their mean versus position angle. Note that the point corresponding
  to McConnachie~(\cite{thesis}) measurement is outside the range
  shown of this figure. The {\it continuous} sinusoid is the same as
  in Fig.~\ref{sinu} while the {\it dashed} sinusoid is the best fit
  to Kim et al.~(\cite{km02}) data only; the zero line is also
  indicated ({\it dotted}). Error bars correspond to the random error
  for Kim et al.~(\cite{km02}) measurements, sytematic errors amount
  to $+0.15$ and $-0.11$, and the global error quoted by the authors
  for the other points.}
\label{kim}
\end{figure}

In Sect.~3.3.2 magnitude and colour vary according to a well defined
sinusoidal pattern which can be attributed to the geometry of the galaxy
as well as to the presence of differential extinction. Here, we
compare this pattern with the spatial variation of different measures
of the distance to the galaxy.  Figure \ref{kim} shows the
distribution of the difference between each distance measurement and
their mean as a function of position angle regardless of their
distance from the centre. Error bars correspond to the value estimated
for the final distance moduli. In the case of Kim et al. (\cite{km02})
it is obtained from averaging the ten measurements, and are shown here
just as a guideline.  The real error bar of each individual point is
smaller than the one indicated.  The sinusoid derived earlier is
overplotted and shows a good agreement, especially with the
measurements by Kim et al.~(\cite{km02}), which were homogeneously
analysed. This suggests that a sinusoidal variation is likely although
current differences from the zero line are not pronounced.  Note that
we did not correct for the different extinction values used by the
different authors and this is partly responsible for the scatter of
the distances obtained.

McConnachie's~(\cite{thesis}) measurement of the tip of the RGB in the
$I$ band differs by $0.13$ magnitudes from the measurement of Tiede et
al.~(\cite{ti04}), both authors studied fields at the same distance
from the centre but at diametrically opposite directions. According to
our study a variation of $0.06$ mag is expected between fields
opposite to the centre. Considering that Brooks et al.~(\cite{br04})
by observing an area very similar to that of Tiede et
al.~(\cite{ti04}) derived a tip of the RGB $0.05$ mag fainter as well
as other uncertainties in the location of the tip of the RGB, the
sinusoidal pattern derived in this study explains the Tiede--McConnachie
difference between their tip of the RGB measurements.  On the
  other hand, McConnachie (\cite{thesis}) explains the variation in
the measured values of the tip of the RGB as due to differential
reddening within the M33 disc.

What is the main cause of the sinusoidal pattern? If this difference
were entirely due to differential extinction a value as large as
$E(B-V)=0.17$ would be obtained assuming $A_{K_{\mathrm s}}=0.06$ and
using the Glass et al.~(\cite{glass}) extinction law. It is more
likely that at infrared wavelength the extinction plays a small
role. On the contrary, an effect due to the orientation of the galaxy
would affect each magnitude equally. Therefore, the sinusoidal pattern
derived in this study despite including a variation of the stellar
population as well as of extinction is predominantly caused by the
geometry of the galaxy.

 In a rotationally supported spiral galaxy of known geometry, like
 M33, the near side is the West side. In Sect. 3.3.2 the sinusoid
   traced by AGB stars has a rather flat minimum and maximum which
   suggests a large uncertainty in the determination of the far versus
   near side of the disc. Moreover, the disc of M33 is warped.
 Corbelli \& Schneider (\cite{co97}) have modeled the HI gas
 distribution across the galaxy. In their study it is clear that the
 SW and the NE regions are peculiar because this is where the warp
 sets in, i.e. they mark the beginning of the deviation from the inner
 (more flat) parts. It is possible that hte analysis of the AGB
   distribution is affected by the presence of the warp but this
   suggestion needs to be confirmed.

\subsection{Confirmed Long-Period Variables}

\begin{figure}
\resizebox{\hsize}{!}{\includegraphics{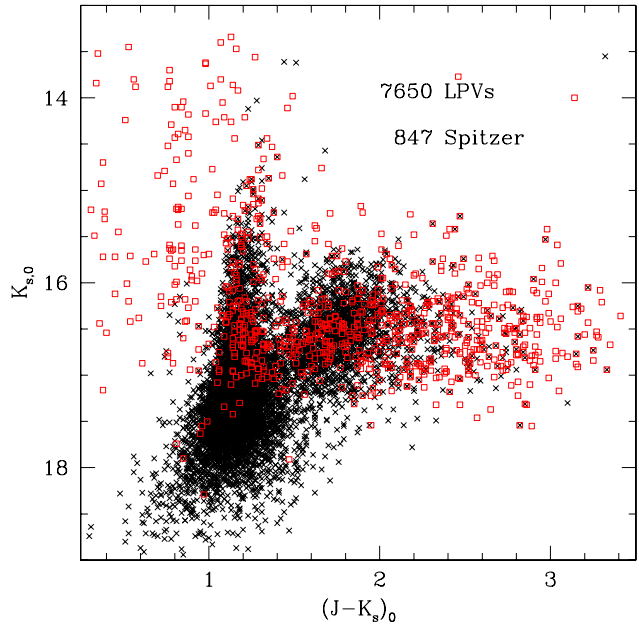}}
\caption{Colour-magnitude diagram of the near-infrared sources matched with the
  LPVs (crosses) from Hartman et al.~(\cite{hart}) and the variable
  stars detected by Spitzer (squares) from McQuinn et
  al. (\cite{spi}). A colour version of this figure is available in
  the electronic version of the paper.}
\label{var}
\end{figure}

In this section we compare the distribution of candidate AGB stars
selected using the colour-magnitude criteria of Sect.~3.1 with the
distribution of confirmed Long-Period Variables (LPVs) from the
catalogue by Hartman et al.~(\cite{hart}). We cross-identified our
catalogue of near-infrared sources with the LPVs candidates located in
a specific region of the (r-i$^{\prime}$) versus i$^{\prime}$
colour-magnitude diagram (Hartman et al.~\cite{hart}, their
Fig.~7). We found that $7650$ candidates have a near-infrared
counterpart within $1^{\prime\prime}$. The histogram of the distance
between near-infrared and LPV matches has a narrow peak at
$0.15^{\prime\prime}$ with a FWHM $=0.25^{\prime\prime}$;  these
values were obtained after correcting the Hartman et al. (\cite{hart})
coordinates by a systematic shift in right ascension of
$0.4^{\prime\prime}$ (the shift in declination is negligible and
amounts to $\sim 0.05^{\prime\prime}$). The distribution of these
sources is shown in Fig.~\ref{var}. The branch of C-rich AGB stars
departing to red colours as well as the vertical branch of O-rich AGB
stars are clearly distinct supporting the selection criteria presented
in Sect.~3.1. The table with the near-infrared photometry from our
study and the optical photometry by Harman et al.~(\cite{hart}), of
which an extract is given in Table \ref{irlpv}, is available
electronically; coordinates are from UKIRT data.

Block et al.~(\cite{bl04}) claim to have observed the brightest
unresolved C star population using 2MASS observations as deep as $1$
mag below the nominal survey limit which results in
$K_s\sim16$. Moreover, in their Fig.~$1$ ({\it top right}) a partial
ring around the galaxy is delineated by sources with
$0.5<J-K_s<1.5$. However, at their sensitivity and within the above
colour range these sources are not C stars but M stars
(cf.~Fig.~\ref{var}). Therefore, what Block et al.~observed is a
ring-like structure traced by the brightest O-rich AGB stars very
similar to the old and metal-poor ring shown in Fig.~\ref{combm}. This
does not exclude that C stars are present in this region but their
colour is redder. In fact recently the authors confirmed
spectroscopically $7$ C stars with $J-K_s>2$ which they selected from
further near-infrared observations down to $K_s\sim17$ (Block et
al. \cite{bl07}).

An outer ring dominated by M instead of C stars has implications on
the age and metallicity distribution. Bright O-rich AGB stars,
especially those brighter than the brightest C star observed, are on
average more massive ($3-5$ $M_{\odot}$) and younger ($\sim0.1$ Gyr
old) than C-rich AGB stars having a mass of $1-3 M_{\odot}$ and an age
of $0.6-2$ Gyr (Cioni et al. \cite{ci03}, Vassiliadis \& Wood
\cite{va93}). This result supports further the conclusion by Block
et al. (\cite{bl07}) that these stars formed recently by gas infall
tied to the HI warp. On the other hand, it shows the uncertainty
of the Block et al. (\cite{bl04}) technique to characterise the outer
halo of galaxies using unresolved stars. 

The most accurate way to determine age and metallicity of a stellar
population is to resolve turn-off main sequence stars. Alternatively,
if only much brighter stars are reached, the $K_s$ method, developed
by Cioni et al. (\cite{lf}) and used here, provides a means to
indicate average values as well as investigate relative
differences. The next more detailed approach is to fit the observed
near-infrared colour magnitude diagrams, dominated by late-type stars,
with synthetic diagrams obtained using newly published and
continuously improving isochrones (Marigo et
al.~\cite{ma07}). Obtaining age and metallicity from unresolved
stellar populations may hide non-negligible information about the
spatial distribution of these quantities.

\begin{table}
\caption{Photometry of candidate Long-Period Variables}
\label{irlpv}
\[
\begin{array}{ccccccc}
\hline
\noalign{\smallskip}
\alpha \mathrm{(deg)} & \delta \mathrm{(deg)} & i^{\prime} &
r-i^{\prime} & J & H &  K_{\mathrm s} \\
23.914114 & 31.014139 & 20.85 & 1.54 & 18.60 & 17.76 & 17.47 \\ 
23.915800 & 30.954123 & 21.14 & 0.96 & 19.07 & 18.01 & 17.74 \\ 
23.917891 & 31.093290 & 21.40 & 0.99 & 18.99 & 17.84 & 17.28 \\ 
23.933479 & 31.087179 & 21.16 & 2.29 & 18.45 & 17.59 & 17.24 \\ 
23.937286 & 31.108717 & 20.77 & 1.75 & 18.51 & 17.68 & 17.27 \\ 
23.948526 & 31.005262 & 20.82 & 1.75 & 18.47 & 17.57 & 17.37 \\ 
23.952553 & 31.029049 & 20.79 & 0.75 & 18.89 & 17.95 & 17.53 \\ 
23.964951 & 31.150126 & 20.76 & 1.52 & 18.09 & 17.24 & 16.89 \\ 
23.786018 & 31.104441 & 21.23 & 1.94 & 18.71 & 17.72 & 17.57 \\ 
23.786890 & 31.061749 & 21.09 & 1.75 & 18.99 & 18.03 & 17.75 \\
...&...&...&...&...&...&...\\
\noalign{\smallskip} 
\hline
\end{array}
\]
\end{table}

\subsection{Spitzer variables}

Very recently McQuinn et al. (\cite{spi}) presented the analysis of
multi-epoch infrared observations of M33 using the Spitzer Space
Telescope. Figure \ref{var} shows the cross-identified sources between
our near-infrared photometry and their Table 3 of variable point
sources. We found that out of $2923$ Spitzer variables $847$ have a
near-infrared counterpart within $1^{\prime\prime}$. The histogram of
the distance between near-infrared and Spitzer matches has a well
defined peak at $0.32^{\prime\prime}$ with a
FWHM$=0.2^{\prime\prime}$;  systematic shifts in both coordinates
are $\le 0.05^{\prime\prime}$. These obscured variables distribute
along the bright half of the branches occupied by O-rich and C-rich
AGB stars as well as populating the region of extreme AGB stars (with
$J-K_s>2$) and supergiants ($J-K_s<1$). The fact that sources with
$J-K_s>2$ are confirmed to vary at infrared wavelengths marks their
nature as LPVs, despite the incompleteness of the Hartman et
al. (\cite{hart}) catalogue, supporting further our selection
criteria. The table with our near-infrared photometry and the Spitzer
photometry by McQuinn et al. (\cite{spi}), of which an extract is
given in Table \ref{irspi}, is available electronically; coordinates
are from UKIRT data.

\begin{table}
\caption{Combined Spitzer and near-infrared photometry}
\label{irspi}
\[
\begin{array}{cccccc}
\hline
\noalign{\smallskip}
\mathrm{Spitzer ID} & \alpha \mathrm{(deg)} & \delta \mathrm{(deg)}& J
& H &  K_{\mathrm s} \\
J013201.90+302603.3 &  23.007860 & 30.434172 & 18.62 & 17.45 & 16.53 \\
J013203.89+302507.9 &  23.016174 & 30.418797 & 17.66 & 17.13 & 16.87 \\ 
J013204.84+302759.1 &  23.020119 & 30.466366 & 19.35 & 18.06 & 16.96 \\ 
J013206.01+302658.2 &  23.024990 & 30.449432 & 18.42 & 17.16 & 16.47 \\ 
J013206.80+303843.4 &  23.028303 & 30.645323 & 18.45 & 17.36 & 16.65 \\ 
J013209.90+302909.6 &  23.041250 & 30.485947 & 15.59 & 14.77 & 14.44 \\ 
J013210.16+304138.9 &  23.042223 & 30.694136 & 18.38 & 17.34 & 16.87 \\ 
J013211.71+302059.1 &  23.048784 & 30.349678 & 15.54 & 14.95 & 14.84 \\ 
J013214.11+304423.9 &  23.058659 & 30.739985 & 18.44 & 17.17 & 16.38 \\ 
J013214.13+302913.7 &  23.058855 & 30.487104 & 17.90 & 16.88 & 16.32 \\ 
...&...&...&...&...&...\\
\noalign{\smallskip} 
\hline
\end{array}
\]
\end{table}

\subsection{The C/M ratio and the mean metallicity of the stellar
  population across M33} 

There are essentially two ways to statistically select C-type and
M-type AGB stars from photometric observations. In this work we use
the ($J-K_{\mathrm s}$, $K_{\mathrm s}$) colour-magnitude diagram
(Sect.~\ref{sel}) while Rowe et al. (\cite{ro05}) use a colour-colour
diagram obtained from the combination of broad-band $VI$ and
narrow-band CN-TiO filters (their Fig.~$10$).  The main advantage of
the first criterion is that near-infrared observations are not
affected by interstellar extinction and produce a more complete sample
of stars. In fact, Rowe et al.~(\cite{ro05}) observations in the
spiral arms suffer from dust extinction which reduces their star
counts because faint stars will fall below the detection limit. If the
dust extinction is as severe as in the centre of the Milky Way
near-infrared observations would suffer a similar effect but this is
not the case in M33. On the other hand, narrow-band observations
target those molecular features that are typical of C-type and M-type
AGB stars and produce a more reliable selection of stars especially at
faint magnitudes where the contamination of RGB stars is stronger.

Rowe et al. (\cite{ro05}) find $7936$ C stars but their sample is
incomplete towards the centre. Their overall AGB distribution drops at
$\sim30^{\prime}$ when the contamination by foreground dwarfs becomes
severe (see their Fig.~9). Depending on the selection criteria we find
$9522/7404$ C stars and judging from Fig.~\ref{fig1} the selected AGB
stars are almost negligibly affected by foreground stars.  A detailed
comparison between our C/M ratio distribution (Fig.~\ref{cm72}) and
the distribution obtained by Rowe et al. (\cite{ro05}; their
Fig.~$18$) confirms that the C/M ratio is high along a ring-like
structure surrounding a central region of low ratio. Our observations
and analysis reveal a much more complex structure because of the high
penetrating power of near-infrared observations which also cover the
whole extent of M33 while Rowe et al. (\cite{ro05}) observations
covered only $2/3$ of the galaxy.  The northern enhancement in the
C/M ratio may also correspond to the stellar arc found by Block et
al. (\cite{bl04}) but the lack of axis labels in their Fig. $1$ makes
it hard to secure such a correspondence.

How well does the C/M ratio alone trace metallicity? M33 presents a
metallicity gradient such that the metallicity ([Fe/H]) decreases
linearly with galactocentric radius from $-0.6$ to $-0.9$ dex (Kim et
al.~\cite{km02}, Barker et al.~\cite{ba07b}) although with
[$\alpha$/Fe]$=0.0$ the metallicity is $0.4$ dex higher (Barker et
al.~\cite{ba07a}); this gradient extends out to $\sim50^{\prime}$
($13$ kpc).  Brooks et al. (\cite{br04}) observed a SE halo field and
derived a metallicity of [Fe/H]$=-1.24$ dex while Tiede et
al. (\cite{ti04}) and Davidge (\cite{da03}) in fields approximately in
the same direction but closer to the galaxy centre derive
[Fe/H]$=-1.0$ dex.  The latter is typical of disc stars rather than
halo stars of M33. Both McConnachie et al.~(\cite{co06}) and
Sarajedini et al.~(\cite{sara}) showed that RGB and RR Lyrae stars,
respectively, belong to two populations: one associated with the halo
([Fe/H]$\sim-1.4$ dex) and the other with the disc ([Fe/H]$\sim-0.8$
dex) of M33. There is a third component in the outer SW region with
the same metallicity as the disc which McConnachie et
al.~(\cite{co06}) attribute to a stellar stream.  The metallicity
spread that we obtain from the distribution of the C/M ratio
reproduces the spread of the above values and the existing gradient
rather well. Both Fig.~\ref{cm72} and \ref{cm36} show a more detailed
structure/substructures than is derived from a gradient representation
of the metallicity. However, we fail to recover a metallicity as high
as [Fe/H]$=-0.26$ dex (Stephens \& Frogel \cite{stfr}) in the nuclear
region.  According to Mouhcine \& Lan\c{c}on (\cite{mola}) the C/M
ratio traces the metallicity of a population older than about a
Gyr. In fact, also across the Magellanic Clouds Cioni et al.
(\cite{lf}, \cite{lfs}) derived that the SFH does affect the C/M if
the population is younger than a few Gyr across the LMC while across
the Small Magellanic Cloud (SMC) the C/M ratio traces only the
intermediate-age epoch of formation of AGB stars. The latter might be
influenced by the spatial orientation of the SMC which was not
corrected for.

The distribution of metallicity shown in Fig.~\ref{combc} and
\ref{combm} has many features in common with the distribution of the
C/M ratio shown in Fig.~\ref{cm72} or \ref{cm36}. There is a broad
metal-poor outer ring, this metallicity may extend to the whole galaxy
if the resolution of the C/M map increases (cfg. Fig.~\ref{cm36} and
\ref{cm72}) which may explain why the metallicity derived from C stars
(Fig.~\ref{combc}) is low in the whole disc. Although the C/M ratio
suggests a metallicity which is slightly higher inside a ring-like
structure it fails to recover a high metallicity in the centre of the
galaxy (see above) which is instead clearly obtained from the analysis
of the magnitude distribution of both C and M stars (Fig.~\ref{combc}
and \ref{combm}). This points to the importance of the age of the
stellar population in the interpretation of the C/M ratio distribution
which perhaps traces, for this galaxy, only the metallicity at the
epoch of formation of AGB stars. On the other hand, an almost flat
metallicity gradient has been found by Magrini, Corbelli \&
Galli~(\cite{magri}) from the observation of various elements in a
limited sample of young stars, HII regions, PNe and RGB stars. Their
results support an evolutionary scenario where M33 is constantly
accreting gas and forming stars at a slowly decreasing rate with
time. The rather smooth distribution of metallicity shown in
Fig.~\ref{combc} would agree with this interpretation.

The absolute values of age and metallicity obtained using the $K_s$
method are model dependent and this means that they may be affected by
systematic differences. Given that the metallicity in the outer disc
of M33 has been inferred to be $\sim1$ dex, it means that the value
obtained from C stars is off by $0.6$ dex. However, if C stars were
 tracing the halo instead of the disc the derived metallicity would
be off by just $0.2$ dex. On the contrary, the disc-like metallicity
obtained from M stars is too high by $0.5$ dex with respect to the
expected disc metallicity. The average of the C and M stars would
produce the same metallicity obtained by other authors. It is possible
that C stars are not just sampling the disc but the halo of the
galaxy. It remains unexplained why both C and M stars trace a
population with a similar mean age but very different mean
metallicities.

\subsection{Mean-age of the stellar population across M33}

Focusing on the most significant distributions of mean age obtained
 from C (Fig.~\ref{combc}-{\it top}) and M (Fig.~\ref{combm}-{\it
 bottom}) stars, M33 appears overall older than the Magellanic Clouds
 (Cioni et al. \cite{lf}, \cite{lfs}). Here, we compare our results of
 the mean age distribution with studies in the literature that cover a
 similar area. This excludes the recent work by Barker et
 al.~(\cite{ba07b}) on the outer M33 regions where they detect an
 increasing age (from $6$ to $8$ Gyr) with increasing radius. Our
 outermost ring-like structure, which does not overlap Barker et al.'s
 fields, is consistent with a mean age of $\sim6$ Gyr; perhaps this is
 the same population connecting to Barker et al.'s inner field. This
 outer mean age is also in agreement with the results by Li et
 al.~(\cite{li04}). These authors derived the age distribution across
 M33 by comparing observations obtained in several narrow bands from
 about $350$ nm to $1000$ nm with synthetic spectral energy
 distributions produced using the PEGASE code. Although they do not
 provide enough information on the size of the three regions outlined
 in their Fig.~$4$ a comparison with our Fig.~\ref{combc} shows that,
 apart from the agreement in the outer region, there is a clear
 disagreement in the mean age of the inner galaxy. Li et
 al.~(\cite{li04}) suggest that the central regions are older than the
 outer regions, except for the spiral arms where the mean age is the
 youngest. In Fig.~\ref{combc} the spiral arm region is not
 distinguishable from the outer region (this is perhaps due to the low
 resolution imposed by the area subdivision into sectors of rings --
 Fig.~\ref{fig1}) but the difference between Li et al's mean age and
 our mean age is $\sim 1$ Gyr. On the contrary, both Fig.~\ref{combc}
 and Fig.~\ref{combm} suggest a population of a few Gyr younger in the
 central regions. This result is consistent with the distribution of
 young stars (super giant stars) confined to the central region almost
 enclosing the nucleus of the galaxy (Rowe et al.~\cite{ro05}), see
 also Fig.~\ref{surf}. These authors show that both AGB and
 main-sequence stars trace the extended structure of M33 where the
 latter is clumpier due to localised regions of massive star
 birth. This cospatial distribution suggests a very similar mean age
 which supports the smooth outer ring derived in Fig.~\ref{combc}.

\section{Conclusions}
\label{con}
In this paper we present wide-field near-infrared observations of M33
obtained with WFCAM at UKIRT. These data reveal a large population of
AGB stars which we have used to determine the distribution of age and
metallicity across the galaxy as well as to constraint its orientation
in the sky.

C-rich and O-rich AGB stars have been selected from the ($J-K_s$,
$K_s$) colour-magnitude diagram using two criteria: one based on a by
eye evaluation of their favorite location in the diagram, often used
in the literature, and one based on stellar evolutionary tracks.  The
selection of the two samples is confirmed via the cross-identification
with the catalogue of candidate LPVs by Hartman et al. (\cite{hart})
and with the recent publication of the list of sources detected by
Spitzer (McQuinn et al. \cite{spi}) in the same region. While the
variable sources delineate well the branches populated by un-obscured
AGB stars, the latter extend to red colours where stars have thick
circumstellar dusty envelope. Both tables containing the
cross-identification between our near-infrared photometry and the
sample of variable stars and of Spitzer detections are available
electronically. 

The confirmed location of C-rich and O-rich AGB stars in the
near-infrared colour-magnitude diagram has shown that the metal-poor
arcs surrounding the galaxy suggested by Block et al.~(\cite{bl04})
are formed by O-rich AGB stars, contrary to C-rich AGB stars, and
support an LMC-type metallicity as well as an old stellar population.
The distribution of the C/M ratio confirms a metallicity gradient
corresponding to a spread of [Fe/H] $=0.6$ dex and shows substructures
in the inner and in the outer parts of the galaxy. A high ratio, or a
low metallicity follows the major spiral arms of the galaxy which
appears more metal rich in the centre.  The peak magnitude and colour
of C stars, but also of M stars, describe a sinusoidal pattern 
  which may explain previous inconsistencies on the determination of
the distance to the galaxy from sample stars at different locations
with respect to the centre, but could also be affected by
  differential extinction and the disc warp. 

We have interpreted the $K_{\mathrm s}$ magnitude distribution of both
C-type and M-type AGB stars using theoretical distributions
constructed from stellar evolution models spanning a range of SFRs and
metallicities. This is the same procedure adopted to study the SFH
across the Magellanic Clouds (Cioni et al. \cite{lf},
\cite{lfs}). Maps showing the distribution of metallicity as a
function of mean age of the stellar population of M33 indicate a metal
poor disc/halo and a metal-rich core as well as metal-rich clumps in
the inner part of the galaxy which change location with time. This is
perhaps tracing the temporal evolution of the galaxy and suggests that
although M33 is more or less isolated in the sky the star formation
proceeded inhomogeneously with time. Contrary to the metallicity value
derived from C and M stars which differ by $\sim 1$ dex the
distribution of age commonly suggests a broad outer ring $\sim 6$ Gyr
old (constant SFR) which surrounds a younger central region, perhaps
with the exception of the core which is also metal rich.

Summarizing, this study deconvolves the effect of reddening,
structure, metallicity and age on the luminosity function (magnitude
distribution in the $K_{\mathrm s}$ band) of the stellar population of
M33 as follows. Assuming that near-infrared observations of the M33
stellar population are negligibly affected by differential reddening,
and that there are no variations in age and metallicity for AGB stars
in restricted range of colours, their peak magnitude traces distances
throughout the extent of the galaxy. Correcting the observed
$K_{\mathrm s}$ magnitude distribution for this geometrical effect,
the fits obtained using theoretical distributions provide the best
measurement of mean age and metallicity across the system where
relative variations are much more significant than absolute values. By
considering the C/M ratio as an independent whilst indirect indicator
of metallicity, the most probable star formation rate (or mean age),
out of $25$ perhaps not realistic simple stellar population templates,
is revealed as a function of position within the galaxy.

Despite the strong links of the $K_{\mathrm s}$ method to the specific
stellar evolutionary models employed and their treatment of the AGB
phase, which is still uncertain (Marigo et al.~\cite{ma07}), this
method provides a satisfactory approach to the interpretation of
variations of age and metallicity across galaxies, complemented by the
C/M ratio as an indirect indicator of iron abundance. These studies
are particularly important in distant systems where only the brightest
stars, e.g. AGB stars, are observed and therefore it is not possible
to determine an accurate star formation history and its spatial
variations from classic indicators like the main-sequence turn-off.

\end{document}